\documentclass[useAMS,usegraphicx,usenatbib]{mn2e}
\usepackage{txfonts,amssymb,hyperref,subfigure,epsfig} 
\newcommand{\Mearth}{M_{\oplus}}
\newcommand{\Mstar}{M_{\star}}
\newcommand{\fracbrac}[2]{\left(\frac{#1}{#2}\right)}

\def \aap{A\&A}

\def \aj{AJ}

\def \apjl{ApJ}

\def \apj{ApJ}

\def \mnras{MNRAS}

\def \nat{Nat}

\def \pasp{PASP}

\title[Dynamical Simulations of HD69830]{Dynamical Simulations of the Planetary System HD69830}
\author[M.~J.~Payne]{Matthew~J.~Payne$^{1}$
  \thanks{E-mail:~\href{mailto:mpayne@ast.cam.ac.uk}{mpayne@ast.cam.ac.uk}
    (MJP)}, Eric~B.~Ford$^{2}$, Mark~C.~Wyatt$^{1}$ and Mark~Booth$^{1}$\\
  $^1$Institute of Astronomy, University of Cambridge, Madingley Road,
  Cambridge CB3 0HA\\
  $^{2}$Department of Astronomy, University of Florida, 211 Bryant Space Science Center, PO Box 112055 Gainesville, FL, 32611-2055, USA}

\begin{document}

\date{Accepted -------. Received -------; in original form -------}
\pubyear{2008}

\maketitle
\label{firstpage}

\begin{abstract}
The star HD 69830 exhibits radial velocity variations attributed to
three planets as well as infrared emission at $8-35\mu$m attributed to
a warm debris disk.  Previous studies have developed models for the
planet migration and mass growth \citep{Alibert_et_al_05,Alibert_et_al_06} and the
replenishment of warm grains \citep{Wyatt_et_al_07a}.  In this paper, we
perform n-body integrations in order to explore the implications of these models for:
1) the excitation of planetary eccentricity, 
2) the accretion and clearing of a putative planetesimal disk, 
3) the distribution of planetesimal orbits following migration, and 
4) the implications for the origin of the infrared emission from the HD 69830
system.  

We find that:
i)  It is not possible to explain the observed planetary eccentricities ($\sim 0.1$) purely as the result of planetary perturbations during migration unless the planetary system is nearly face-on. However, the presence of gas damping in the system only serves to exacerbate the problem again.
ii) The rate of accretion of planetesimals onto planets in our n-body simulations is significantly different to that assumed in the semi-analytic models, with our inner planet accreting at a rate an order of magnitude greater than the outer ones, suggesting that one cannot successfully treat planetesimal accretion in the simplified manner of \citet{Alibert_et_al_06}.
iii ) We find that the eccentricity damping of planetesimals does \emph{not} act as an insurmountable obstacle to the existence of an excited eccentric disk: All simulations result in a significant fraction ($\sim$15\%) of the total planetesimal disk mass, corresponding to $\sim 25 \Mearth$, remaining bound in the region $\sim$1-9 AU, even after all three planets have migrated through the region
iv) This swarm of planetesimals has orbital distributions that are size-sorted by gas drag, with the largest planetesimals ($\sim 1,000km$), which may contain a large proportion of the system mass, preferentially occupying the highest eccentricity (and thus longest-lived) orbits. Although such planetesimals would be expected to collide and produce a disk of warm dust, further work will be required to understand whether these eccentricity distributions are high enough to explain the level of dust emission observed despite mass loss via steady state collisional evolution.
\end{abstract}

\section{Introduction}

Radial velocity surveys have discovered over 300 extrasolar planets
and roughly 30 multiple planet systems (Butler et al.\ 2006;
\mbox{http://www.exoplanets.org}, \mbox{http://www.exoplanet.eu}).
The multiple planet systems are particularly valuable for gaining
insights into the processes of planet formation (e.g.,
\citet{Lee02,Ford05}).  Among the known multiple planet systems, the
{\em Spitzer} Legacy Program FEPS \citep{Meyer06} has identified four
systems that emit an excess of infrared radiation
(relative to that expected from the stellar photosphere), most likely
due to reradiation of starlight absorbed by a dust disk.  In these
systems, the properties of the dust disk can provide additional
constraints on the system's formation and dynamical evolution (e.g.,
\citet{Moro-Martin07}).  

One of these systems is HD 69830, a bright K0V star located 12.6 pc away \citep{Perryman97}. In addition to possessing 3 Neptune-mass planets \citep{Lovis_et_al_06}, this exceptional system was the \emph{only} system out of 84 FKG stars studied by \citet{Beichman_et_al_05} which was found to display an excess of emission at 24$\mu$m, indicative of \emph{warm} dust (other systems were observable only at 70$\mu$m, indicative of much cooler ($T<100$K) dust).

All three known planets orbit at less than 1 AU, so the putative dust disk is exterior to the known planets, but still close enough that gravitational perturbations from the planets might be effective at stirring a disk of planetesimals, resulting in collisions that could generate dust that would give rise to the observed IR excess.  Thus the HD 69830 system is particularly interesting due to a unique combination of radial velocity and IR excess observations that have revealed multiple planets and a close-in warm dust disk.

\subsection{Planets: Radial Velocity Observations and Theoretical Models}\label{Planets}

Based on HARPS radial-velocity measurements of HD 69830,
\citet{Lovis_et_al_06} identified planets ($M\sin{i}=10.2$, $11.8$ \&
$18.1M_{\oplus}$ respectively) orbiting close-in to the central star
(semi-major axes of 0.08, 0.19 \& 0.63 AU).  The planets'
eccentricities ($0.10\pm 0.04$, $0.13\pm 0.06$ \& $0.07\pm 0.07$ respectively,
\citet{Lovis_et_al_06}) are large by solar system standards, but
modest when compared to those of extra-solar planets
\citep{2007astro.ph..3163F}.  The host star is somewhat cooler
($T_{\rm eff}\simeq$ 5385 K) and less massive than the Sun
($M_{\star}\simeq 0.86 M_\odot$).  Otherwise, it is quite similar to
the Sun, with a nearly solar metallicity ([Fe/H]= $-0.05\pm 0.02$) and age (4-10
Gyr; \citet{Lovis_et_al_06}).  Since the known planets produce modest
radial velocity perturbations ($2.2 - 3.5\textrm{m s}^{-1}$) just large enough for detection, similar
planetary systems could have often eluded detection by broad radial
velocity planet searches on account of an insufficient number of
velocity observations and/or velocity precision due to photon noise
and/or radial velocity ``jitter''.  Recent results from the HARPS search for southern extra-solar planets and its discovery of a planetary system with 3 Super-Earths \citep{Mayor08} bolsters the view that similar planetary systems might be quite common.

These radial velocity observations alone reveal an interesting planetary system.  Most of the known planets in multiple planet systems have masses roughly comparable to Jupiter (most likely due to detection biases). The unusual intermediate planetary masses revealed in the HD 69830 system have already inspired several theoretical investigations.  As a result, theorists have developed a detailed model for the mass growth and orbital migration of the observed planets through a series of papers \citep{Alibert_et_al_05,Alibert_et_al_06,Lovis_et_al_06}.  In particular, \citet{Alibert_et_al_06} found that the observed characteristics could be reproduced by starting with a gas disk in which the surface density, $\Sigma$, is related to the disk radius, $a$, by $\Sigma \propto a^{-3/2}$ and normalized to $800 \textrm{g/cm}^2$ at 5 AU (This surface density is around 4 times greater than the minimum mass solar nebula, giving $M_{disk} = 0.07M_{\odot}$ ($0.07\textrm{AU} \rightarrow 30 \textrm{AU}$)). The disk has a dust-to-gas ratio of 1/70. They then inserted 3 planetary embryos of mass $0.6 M_{\oplus}$ at initial semi-major axes of 3, 6.5 \& 8 AU and used a semi-analytic model to migration and grwoth of the embryos due to their interaction with the disk over the course of the disk lifetime, $\tau_{disk} = 2$Myr. After accretion and migration, the 3 planets in the model were found to have shifted inwards to the observed semi-major axes, have total masses consistent with the minimum observed masses and possess \emph{core} masses of $\sim 10$, $\sim 7.5$ \& $\sim 10 \Mearth$ respectively.

While the \citet{Alibert_et_al_06} model does an impressive job of matching the observed planet masses and semi-major axes, it is a 1-D semi-analytic model, using numerous assumptions and approximations to model the interaction between the protoplanetary disk and the growing and migrating planetary cores. This raises questions regarding the effects of the planets on any planetesimal disk and the feedback effects of disk evolution and ``shepherding'' on planetesimal accretion rates. In addition, the model does not address the eccentricity evolution of the planets and planetesimals.  In this paper, we build on this work, as we continue the quest to understand the formation and orbital evolution of this fascinating system by applying n-body methods to follow planetesimal evolution within the context of the semi-analytical model of \citet{Alibert_et_al_06}.

\subsection{IR Observations}\label{IRobs}

Spitzer observations of the HD 69830 system reveal a strong infrared
excess (relative to the stellar photosphere) between 8 \& $35\mu
m$, but \emph{no} significant excess at $70\mu m$
\citep{Beichman_et_al_05}.  This combination suggests that the system
contains a disk of warm ($\sim 400K$), small ($<1\mu m$) crystalline
silicate grains orbiting close to the star ($\sim 1 \textrm{AU}$).
For these parameters, the collisional timescale (for $\mu$m grains at
$\sim1$AU from HD 69830) is $\sim400$ yr, while the Poynting-Robertson
drag timescale is $\sim700$ yr \citep{Beichman_et_al_05}.  Thus, collisions in such a disk would grind down the grains
until they became small enough to be removed via radiation pressure.
Unless we happen to be observing the system at a very special time,
this implies that the grains are replenished on a timescale
$<1000$ yrs so as to sustain the observed IR excess.

Several models \citep{Beichman_et_al_05,Lisse_et_al_07,Wyatt_et_al_07a} have been proposed
to explain the origin of the dust responsible for the IR excess, including:
\begin{enumerate}
\item A massive cometary population,
\item The capture of a super-comet onto a circular orbit at $\sim 1$
AU,
\item The steady-state evolution of a planetesimal belt at $\sim 1$ AU,
\item A recent, large collision in a planetesimal belt at $\sim 1$ AU, and
\item Recent dynamical instability which results in planetesimals from
an outer belt being thrown inwards.
\end{enumerate}

In the first scenario, the observed dust would be released by pristine
comets entering the inner region of the planetary
system. \citet{Beichman_et_al_05} noted the observed spectral
signature was similar to that of comet Hale-Bopp, but that reproducing
the emission using cometary ejecta similar to that of Hale-Bopp would
require $\sim10^6$ such comets per year to be delivered to the inner
regions of the system.  If the putative comets were of the same mass
as Hale-Bopp and the dust were to be sustained for $\sim10^7$ yr, then
this would imply $\sim900M_{\oplus}$ of comets entering the inner
regions of the planet system, unfeasibly large for a residual
Kuiper-Belt.

In the second scenario, \citet{Beichman_et_al_05} suggested that
a single object of the size of a large Kuiper Belt Object (e.g.,
Sedna) composed of ice and rock may have been scattered inwards to
$\sim 0.5$AU.  They estimate that such an object would have an
evaporation timescale of $\sim2$ Myr, allowing a reasonable chance of
observation.  Given the masses and orbits of the three known planets,
a Sedna-like object at $\sim0.5$AU could not be dynamically stable,
unless it were on a low eccentricity orbit. In order to have a
reasonable chance of observing the IR excess from such a body, the dynamics of the system must be such that a Sedna-like object could have recently been perturbed into the inner system (presumably on a highly eccentric orbit) and then had its orbit circularized.

For the third scenario, \citet{Wyatt_et_al_07a} considered the possibility that the dust originates in a primordial
planetesimal belt that is coincident with the dust and which has evolved in a quasi-steady state. They found that collisional processing would have removed most of the belt's mass over the $>2$ Gyr age of the system and that the observed levels of dust are incompatible with this interpretation for a similar reason. One possible resolution to this is that planetesimals (and dust) may orbit with significant eccentricities, thus prolonging their survival, meaning that the emission at 1 AU could come from the inner edge of an extended disk when the planetesimals and dust are at pericentre (Wyatt et al, in preparation).

Similarly, for the fourth scenario \citet{Wyatt_et_al_07a} showed that an {\em in situ} planetesimal belt would be extremely unlikely to have undergone the single recent collision that would give rise to the observed IR excess. The results of \citet{Lohne08} may slightly increase the probability of a collision having occurred, as their model predicts higher remnant masses at late times, but a significant increase in probability would seem to require unfeasibly high initial disk masses.

Finally, in the fifth scenario, \citet{Wyatt_et_al_07a} have proposed a model in which the current IR excess is due to a delayed epoch of orbital instability among planetesimals in an outer disk, perhaps similar to that of the Late Heavy Bombardment (LHB) in our own solar system \cite{2005Natur.435..466G}. Clearly this model then raises further issues such as how such a delayed instability might have been triggered and why there has been no detection of the source population of planetesimals.

All of these scenarios have unresolved problems, but their unifying feature is that they all depend to some extent on the existence of an extended distribution of planetesimals beyond the observed planets in the system. Fortunately the architecture of the planetary system sets constraints on how it formed, which in turn has implications for the planetesimal population remaining following planet formation. By combining planetary formation models (described in \S \ref{Planets}), with the N-body techniques described in \S \ref{Method}, we aim to develop an understanding of the remnant disk structure likely to populate the systems at the end of the planet formation process.

\subsection{Outline}
We perform dynamical simulations to investigate the formation and current state of the HD 69830 planetary system.  Using numerical methods described in \S\ref{Method}, we look in \S\ref{PlanetExcite} at the implications of the \citet{Alibert_et_al_05} model for the planetary eccentricities and accretion rates.  In \S\ref{PlanetesimalExcite}, we perform additional simulations to model the interaction between the planets and a putative protoplanetary distribution of planetesimals in order to model the predicted planetesimal distribution after the migration and growth of planets has been completed. In \S\ref{Problems} we consider possible variations in the formation and orbital evolution of the HD 69830 planetary system.  In \S \ref{Discussion} we consider the planetesimal distributions resulting from our simulations and discuss their implications for explaining the observed dust emission. Finally, we summarize our conclusions and future prospects in \S\ref{Summary}.

\section{Methodology}\label{Method}

We adopt the MERCURY n-body package of \citet{Chambers99} for the
basis of our simulations, allowing us to follow the evolving orbits of a population of planets and planetesimals. We model the planets as massive bodies, but treat the planetesimals as massless bodies (test-particles), adding in various additional routines to allow the modelling of additional physical forces and effects such as
i) Forced-Planet Migration, ii) Gas Drag on bodies of arbitrary size,
iii) Dynamical Friction and iv) Mass Growth. All of these effects
are implemented within MERCURY within the ``mfo\_user.for''
subroutine, thus ensuring that they are called sufficiently frequently
that perturbations in mass and / or semi-major axes are adiabatic.

To produce forced migration in a planetary system we implement the
required $\dot{a}$ in MERCURY using the constant-migration-rate model of \citet{Wyatt03}, such that the change in velocity, $\dot{{\bf v}}$ is given by
$\dot{{\bf v}} = 0.5 \kappa \sqrt{GM_{\star}/a^3}\frac{{\bf v}}{/v/}$ , where $\kappa$ is a constant and $\Mstar$ is the stellar mass, giving a constant migration rate of \\$\dot{a}=2\dot{{\bf
v}}\sqrt{a^3/GM_{\star}}=\kappa$. This method allows us to arbitrarily
select a constant migration rate or to adopt a time dependent
migration rate using as an input the data from previous work,
e.g. \citep{Alibert_et_al_06}.  For the latter, we take a list of
specified semi-major axes as a function of time and calculate from
this list the resultant rate-of-change of semi-major axis, $\dot{a}$
at any given time. This is then used as a time dependent $\kappa$
input to the afore mentioned migration model. We implement a similar
mechanism to increase the planetary masses, where an $\dot{M}$ term
can be specified and implemented as a function of time.

Given that much of our work involves migration rates dictated by the results of \citet{Alibert_et_al_06}, the planets are already subject to a force which gives rise to inward migration. We take this to specify the relevant timescale, and then apply the simplified model for a, e and i damping described in \citet{Zhou07}, in which any bodies of mass $M$ orbiting in a gas disk background can be taken to suffer a gravitational tidal drag force, where to leading order in e \& i, the semi-major axis evolution can then be linked to the eccentricity and inclination damping rates via:
\begin{eqnarray}
\frac{1}{T_{Tidal}}  &=& \fracbrac{8}{5e^2 +2 i^2} \frac{<{\dot a}>}{a} \nonumber
\\
\frac{<{\dot e}>}{e} &=& \frac{1}{T_{Tidal}} \nonumber
\\
\frac{<{\dot i}>}{i} &=& \frac{1}{2 T_{Tidal}} \nonumber
\end{eqnarray}
We then implement the eccentricity and inclination damping in MERCURY in the $mfo\_user.for$ subroutine via direct damping of the relevant elements.

We implement gas drag on the planetesimals using a model similar to that used by \citet{Mandell07}. We consider a three-dimensional gas disk model where the gas density, $\rho_g$, and the vertical scale height, $z_0$, are given by \citep{Mandell07}, 
\begin{eqnarray}
\rho_g &=& \rho_{g0} \left(\frac{r}{1AU}\right)^{-\epsilon}e^{-z^2/z_0^2} e^{-t/\tau_{disk}}\label{EQN:rho}
\\
z_0 &=& 0.0472 \left(\frac{r}{1AU}\right)^{\gamma} AU. \label{EQN:z}
\end{eqnarray}
where $\epsilon = 11/4$, $\gamma = 5/4$ and the gas disk is taken to decay exponentially on a timescale $\tau_{disk}=2 Myr$. Note that the model of \citet{Alibert_et_al_06} assumes that the $0.6\Mearth$ embryos will have taken $\sim 0.93$ Myr to grow. As such, our simulations are taken to start at $t=0.93$ Myr, and hence any gas disk in the simulation will have dissipated by an amount $e^{-0.93/\tau_{disk}}$ at the start of the n-body simulation.

Large bodies orbiting in such a fluid will be subject to a deceleration \citep{Rafikov04}
\begin{eqnarray}
\frac{d{\bf v}}{dt} &\approx& -\frac{3C}{4\pi}\frac{\rho_g v_r}{\rho_p r_p}{\bf v_r},
\end{eqnarray}
leading to a drag timescale of 
\begin{eqnarray}
\tau_{GD} &\approx& 5\frac{\rho_p r_p}{\rho_g v_r}\label{EQN:GDTS}
\end{eqnarray}
where $ {\bf v_r}$ is the velocity of the particle w.r.t. the local
gas velocity and the subscript, $_p$, refers to the particle in
question and we have assumed that planetesimals $\ge 100$km will have a drag coefficient $C\approx 1$.

The sub-keplerian local gas velocity, $ {\bf v_g}$, is calculated by considering only the horizontal component of the stellar gravity in calculating the circular velocity and combining this with a further reduction due to internal pressure support, giving
\begin{eqnarray}
{\bf v_g} &=& \fracbrac{x^2+y^2}{r^2}(1-\eta){\bf v_K}, \nonumber
\end{eqnarray}
where {\bf $v_K$} is the standard Keplerian circular velocity, \\ $\eta = \fracbrac{\pi}{16}\fracbrac{z_0^2}{r^2}(\epsilon - \gamma + \beta)$ and $\beta = 1/2$ sets the disk temperature structure, $T \propto r^{-\beta}$.

The above formalism makes it clear that a highly eccentric and/or inclined planetesimal will pass through regions of the disk significantly more removed from the centre and/or midplane of the gas disk, thus experiencing significantly lower gas densities and associated drag forces.

Unless otherwise specified, we apply gas drag as though the
planetesimals have a characteristic radius $\sim 100km$. We note that
this planetesimal size will cause the escape velocity from the
planetesimals to be comparable to the velocity dispersion of the
planetesimals for $<2i>\approx<e>\approx 0.05$, hence giving a
quasi-equilibrium state \citep{Lufkin06}. We take the initial distribution of our planetesimals to be a Rayleigh distribution with $<2i> = <e> = 0.05$

Whilst the addition of non-conservative forces to the MERCURY routine prevents the traditional accuracy checks on conservation of energy and momentum, we did confirm that, (i) in the \emph{absence} of migration, using MERCURY in hybrid mode to simulate non-migratory planets embedded in a planetesimal swarm would typically result in fractional energy changes of $\sim 10^{-10}$ and fractional ang. mom. changes of $\sim 10^{-14}$, and (ii) when applying our gas damping routines to both non-migratory and migratory planets in a static (non-dissipating) disk, we observed the damping timescales for a, e and i to scale as expected with mass, semi-major axis, eccentricity, inclination and gas density.

Using the above algorithms in conjunction with the basic MERCURY
package allows us to investigate a wide variety of physical effects in
the HD69830 system.

As a simplified test of the model, we conduct simulations similar to those of \citet{Lufkin06}, in which a Jupiter-mass planet migrates inwards at a rate of $10^{-3}$ AU per year through a planetesimal disk distributed between 0.5 \& 4.5 AU. Using the rapid hybrid-symplectic (H-S) algorithm  we typically find that the inward migration of the planet will trap those planetesimals at smaller semi-major axes into mean motion resonances (MMRs) and then sweep them into more and more highly eccentric orbits, but that as the migration continues, many of the planetesimals are scattered out of the resonance into highly eccentric bound orbits, with an average eccentricity of 0.5. Repeating the simulation using the slower but more accurate Bulirsch-Stoer (B-S) algorithm results in a system in which the planetesimals remain tightly confined to resonance, being excited to higher and higher eccentricities as the planet shepherds them inwards until finally, many of the planetesimals are ejected from the system, leaving \emph{fewer} of the planetesimals surviving in the system, and those that do survive have a \emph{lower} average eccentricity ($e\sim 0.1$).

The difference appears to stem from the way in which the two algorithms resolve the orbits of highly eccentric planetesimals trapped in the MMRs ahead of the planet. The B-S algorithm, resolving the encounter in greater detail, results in much greater numbers of planetesimals being confined within the resonance until they finally either (i) directly impact the planet, or (ii) are given a large kick and ejected from the system entirely. In contrast, the H-S algorithm tends to use a fixed timestep which is too large to properly resolve the fast moving, small pericentre planetesimals. This results in many planetesimals ``leaking'' from the resonance, not being shepherded to such high eccentricities, and subsequently tending to only suffer high impact-parameter scattering and thus remain on bound (but excited) orbits within the system. 

We note that the results originally reported by \citet{Lufkin06} are intermediate between the results of our B-S \& symplectic tests, with few of the particles in \emph{their} simulations being excited to the point of ejection. We thus suggest that the use of approximate integration techniques (such as the MERCURY H-S integrator and the PKDGRAV Runge-Kutta 4th order integrator) can, in cases of high eccentricity excitation, lead to anomalous results, and should be avoided when studying cases where significantly eccentric planetesimal populations arise.

We also tested MERCURY on simulations similar to those of \citet{Edgar04} in which a planet migrates from 6 to 0.5 AU, scattering planetesimals initially distributed in an annulus between 3 and 4 AU. These simulations result in excitation of the planetesimals, but to a much lower level than in the previous test, with little or no ejection from the system occurring. We find that using MERCURY in both H-S and pure B-S mode gives essentially identical results, and that these results follow exactly the same dependencies on planetary mass and migration rate as observed by \citet{Edgar04} (although we note that the absolute values of the eccentricity excitations we observe are always off-set slightly below the values found by \citet{Edgar04} using the Runge-Kutte based PKDGRAV integrator).

Having tested the integrators in this manner, we felt confident in the behaviour and accuracy of the H-S algorithm \emph{only} when the planets are at large semi-major axes and have excited little eccentricity in the planetesimal population. For the purposes of speed, we therefore used the H-S algorithim when the planets are at large semi-major axes (when we were sure that the fixed timestep would be able to resolve the smallest pericentres) but then switch permanently to the B-S algorithm as soon as any of the planets migrate inside 1 AU and start to excite significant eccentricities and small pericentres.

\section{Dynamical Modelling of Planetary Eccentricity Excitation}\label{PlanetExcite}
Here we consider a MERCURY model containing just 3 planets.

\subsection{Alibert Model}\label{AlibertPlanet}

We consider the initial conditions as adopted in the model of \citet{Alibert_et_al_06}. We then force the three planetary embryos to grow in mass and decrease in semi-major axis according to the output of their model (kindly provided in numerical format by Yann Alibert). This allows us to assess the planetary eccentricities excited by mutual perturbations during inward migration along the track proposed by \citet{Alibert_et_al_06}.

\begin{figure*}
\centerline{
  \psfig{figure=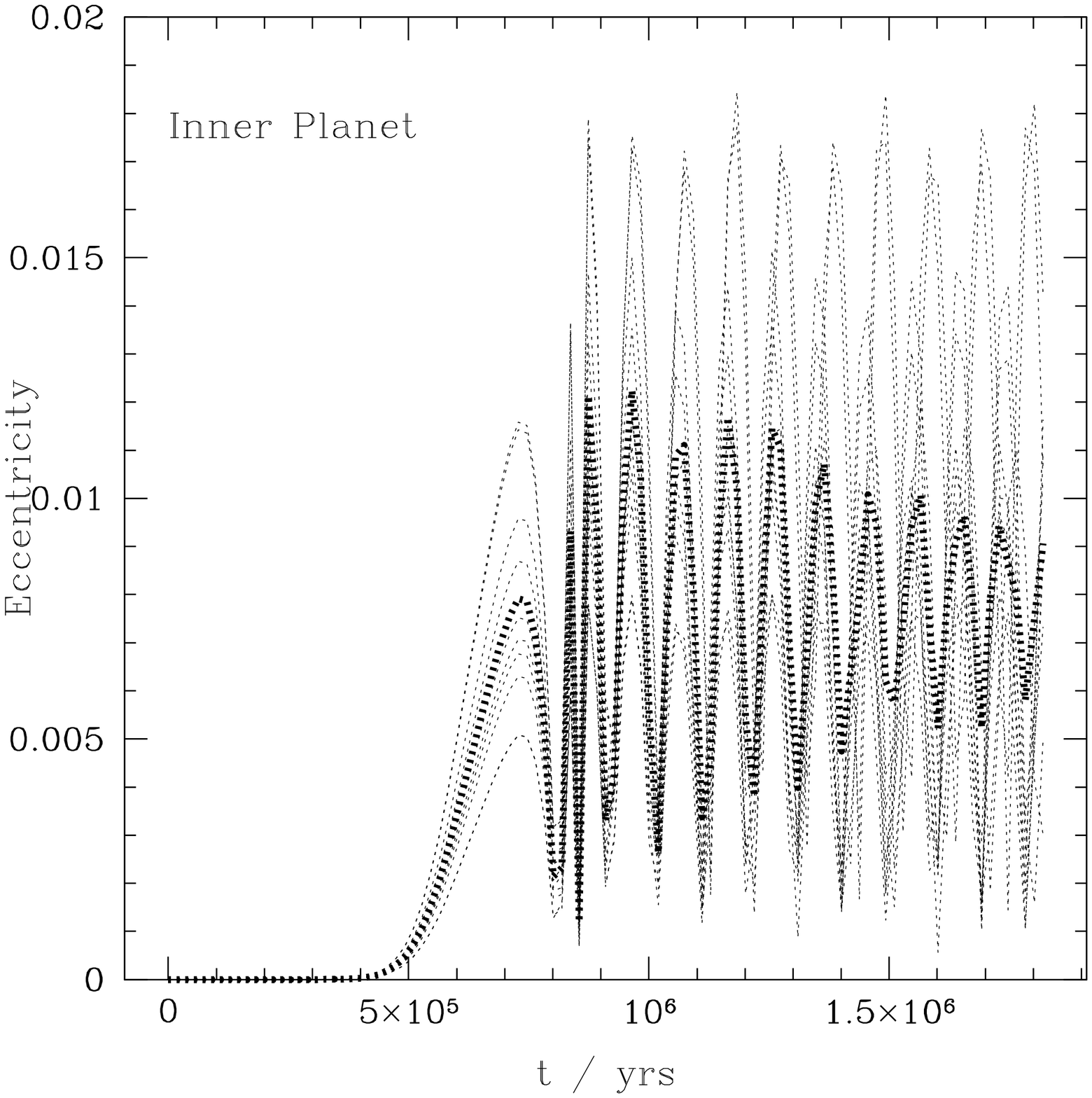,width=0.33\textwidth}
  \psfig{figure=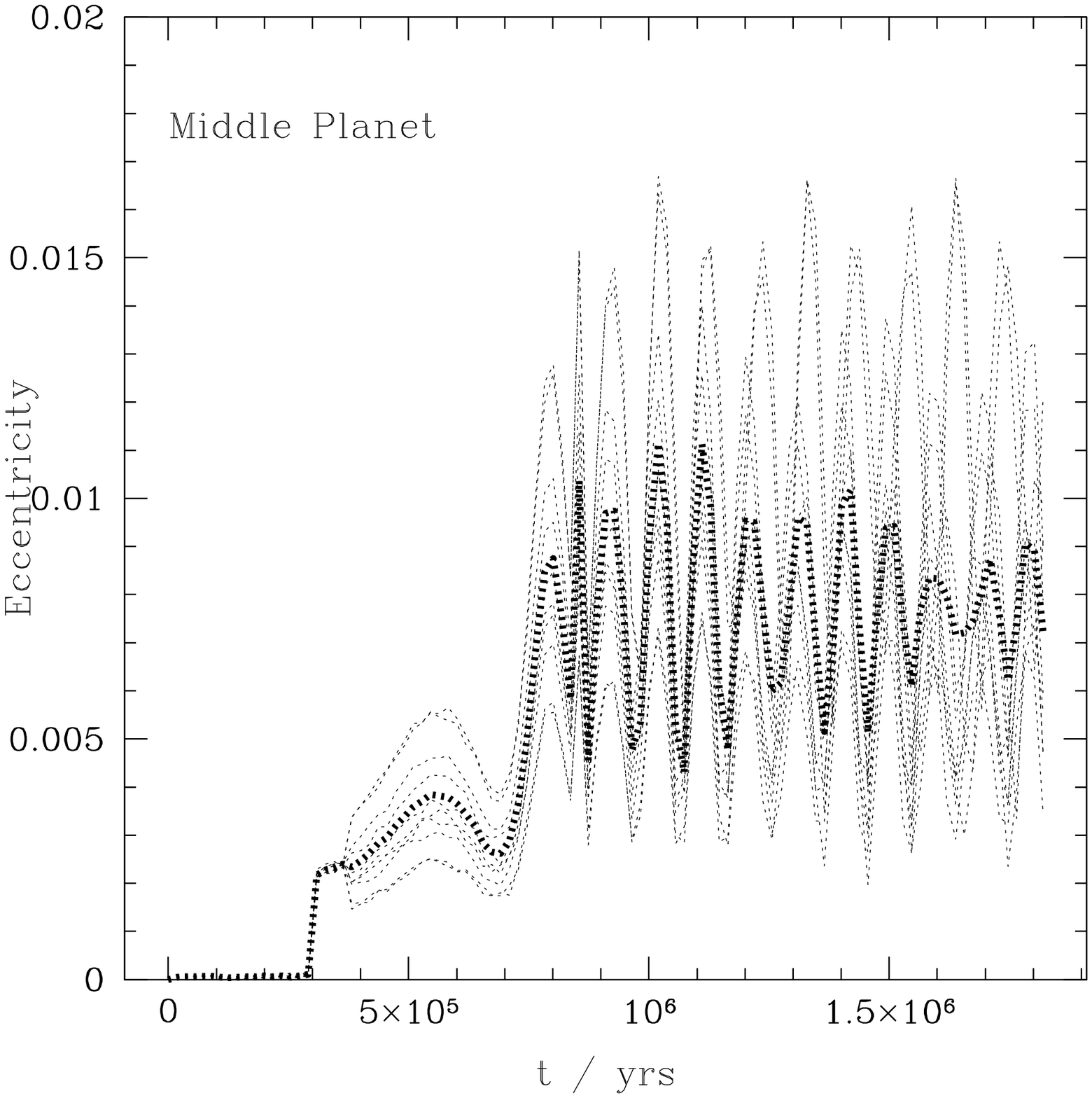,width=0.33\textwidth}
  \psfig{figure=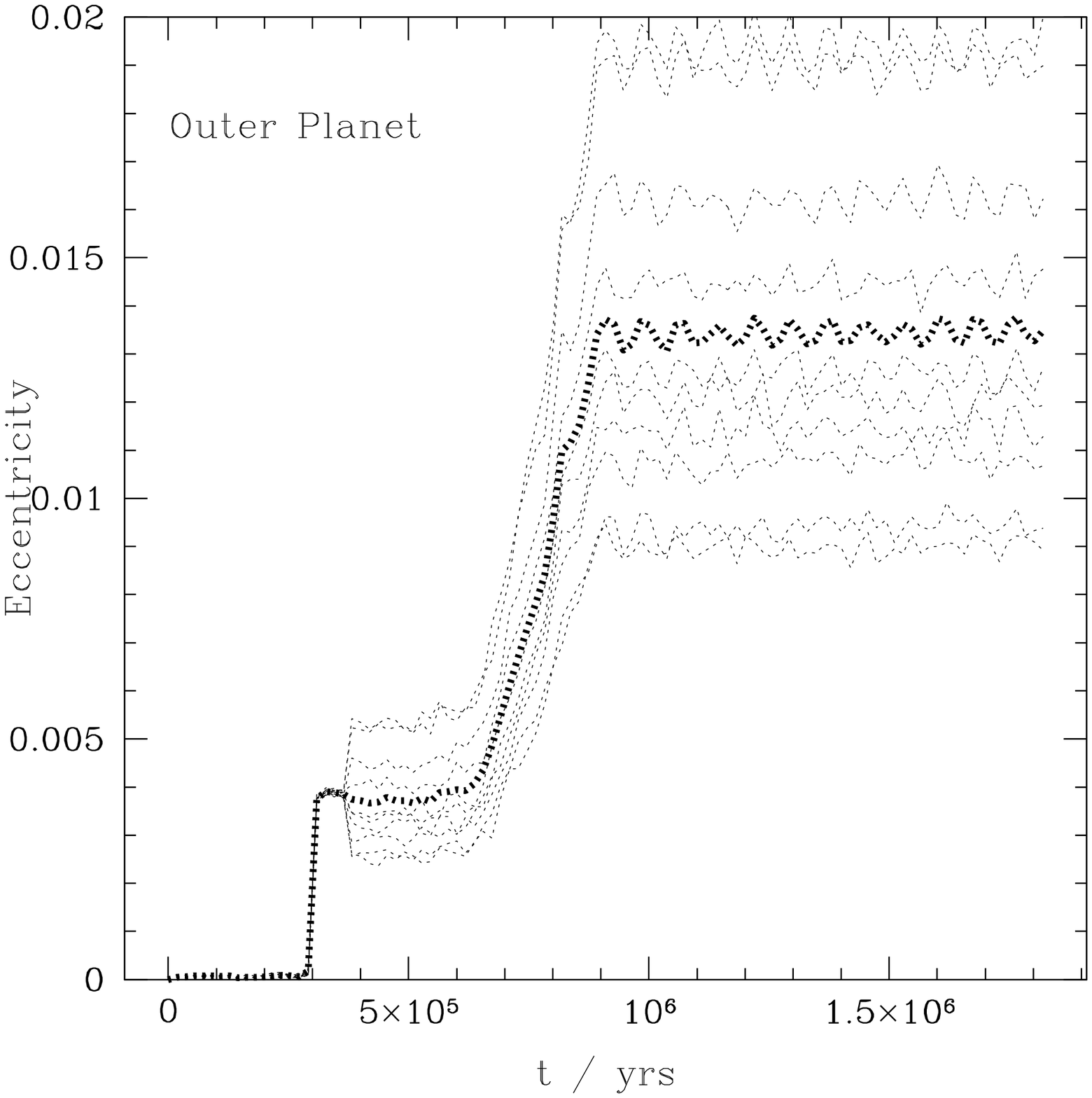,width=0.33\textwidth}
          }
          \caption{Eccentricity evolution for the three planets in the HD69830 system. Ten simulation runs are superimposed. The individual runs are plotted in grey, whilst the overall average is plotted in black. The planets grow and migrate according to the minimum mass Alibert model, with no gas damping operating.}
  \label{FIG:750_759e}
\end{figure*}

We conduct a large number of such simulations, each simulation
starting with the planets on orbits having zero eccentricity, small ($< 1$
degree) randomised inclinations and different randomised mean
anomalies. A sample of some typical results is shown in
Fig. \ref{FIG:750_759e}, combining and averaging the results from 10
simulations. As the planets migrate inwards, we find that the
average planetary eccentricities excited along this forced migratory path are 0.007, 0.008 and 0.013 respectively. We note that there is a large, fast increase in eccentricity visible in the middle and and outer planets in Fig \ref{FIG:750_759e}. In contrast with \citet{Alibert_et_al_06}, we find that the crossing of the 3:2 MMR by the outer two planets is the source of this excitation.

We note that the observed eccentricities and their 1-$\sigma$ errors were $0.10\pm 0.04$, $0.13\pm 0.06$ \& $0.07\pm 0.07$ respectively, so the simulation results are an order of magnitude below the mean observed values. Our own Markov-Chain Monte-Carlo (MCMC) analysis of the radial velocity data using the methodology described in \citet{Ford06} is summarised in Table \ref{TABLE:MCMCEccentricities}. It again suggests that actual planetary eccentricities are likely to be significantly higher than those seen in our simulations. However, we caution that there can be significant uncertainties in the radial velocity observations.  In particular, best-fit orbital solutions can overestimate the eccentricity, $e$, of a nearly circular orbit, particularly for planets with small velocity amplitudes (Shen \& Turner 2008).  Zakamska et al.\ (in prep) find that the summary statistic $\hat{e}\equiv\sqrt{\tilde{h}^2+\tilde{k}^2}$ is significantly less biased than several other estimators, where $h=e\cos \omega$, $k=e\sin \omega$, $\omega$ is the argument of periastron, and the tilde denotes the median value from a Bayesian posterior sample.  Using the same MCMC simulations as reported in Table 1, we find $\hat{e}_b = 0.076$, $\hat{e}_c = 0.080$, and $\hat{e}_d = 0.021$, suggesting that the bias is modest for the inner two planets.  In addition to the statistical errors reported in Table 1, eccentricities can be overestimated due to model misspecification (e.g., the presence of an additional planet that was not included in model fitting).  Thus, we caution that further radial velocity observations are warranted to define how non-circular the orbits are.

\begin{table}
\caption{MCMC Analysis of Planetary Eccentricities}
\label{TABLE:MCMCEccentricities}
\begin{tabular*}{\columnwidth}{lccccc}
\hline
\multicolumn{1}{|c|}{Planet} & 
\multicolumn{5}{|c|}{Percentiles} \\
                     &  2.5  &   16.3  &    50    &    83.7  &   97.5    \\
\hline
HD69830-b            & 0.010 &   0.045 &    0.088 &    0.131 &   0.173   \\
HD69830-c            & 0.007 &   0.042 &    0.106 &    0.174 &   0.241   \\
HD69830-d            & 0.003 &   0.020 &    0.066 &    0.141 &   0.240   \\
\hline
\end{tabular*}
\end{table}

\subsection{More Massive System}\label{mass}
Given that the \citet{Alibert_et_al_06} model is calibrated to produce
the \emph{minimum} masses for the HD69830 system, any relative
inclination between the system and the plane of the sky would mean that
the actual planetary masses could be significantly above this
minimum. Greater planetary masses during the migratory stage could
allow the planets to self-excite to much greater
eccentricities. Therefore, we run a new suite of simulations to look
at the effect of scaling-up the initial embryo masses, as well as the subsequent planetary mass at each time-step in the simulations. For each mass scaling we run ten sets of
simulations that differ only by the initial mean anomalies, giving an
approximate measure of the variability of the potential evolutionary
scenarios and outcomes. We stress that these increased mass models are only an approximation (directly increasing the mass in the \citet{Alibert_et_al_06} model would change various relative growth and migration timescales), but they do allow us to gain some insight into the behaviour of a more massive system.

\begin{figure*}
\centerline{
  \psfig{figure=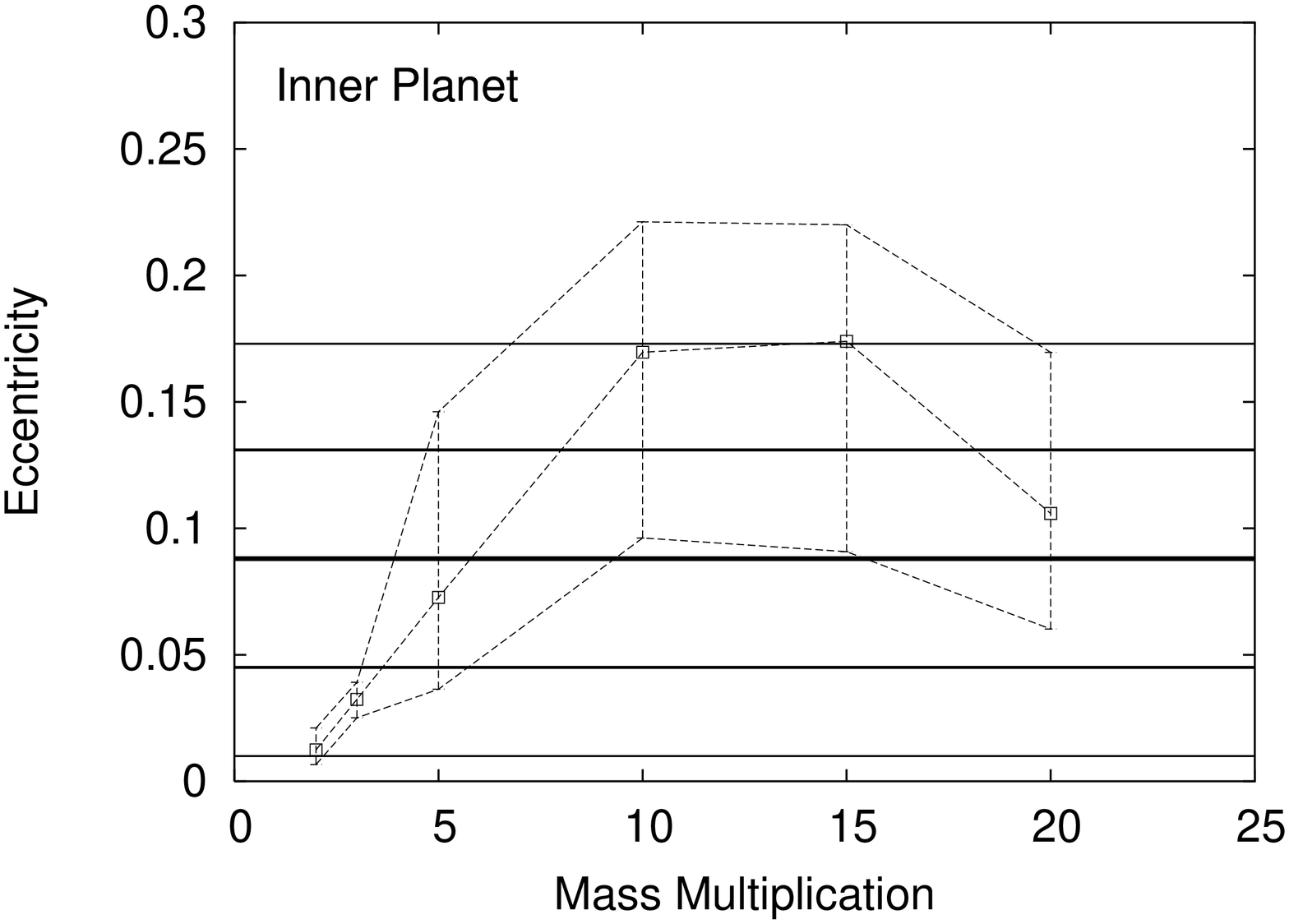,angle=-90,width=0.33\textwidth}
  \psfig{figure=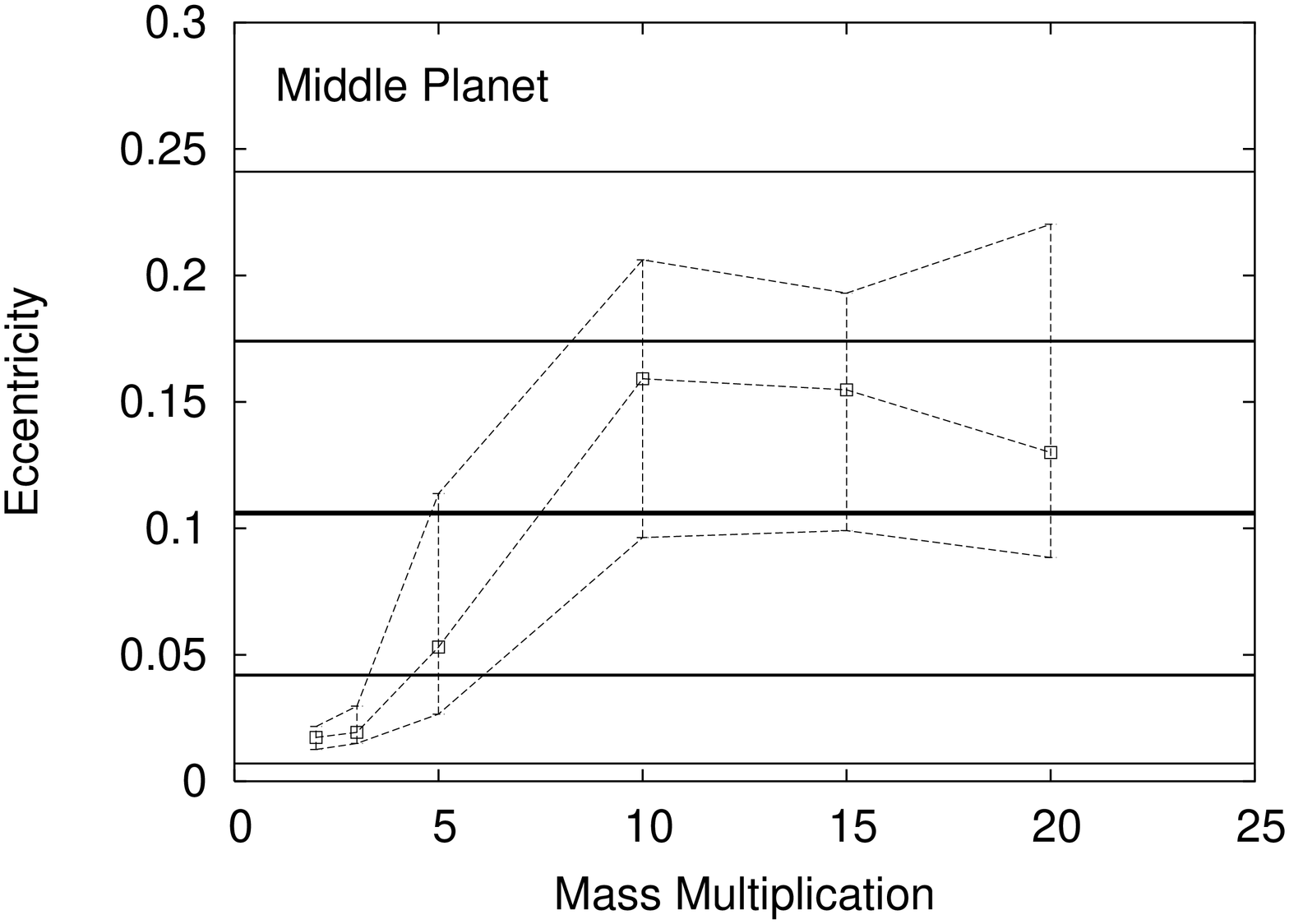,angle=-90,width=0.33\textwidth}
  \psfig{figure=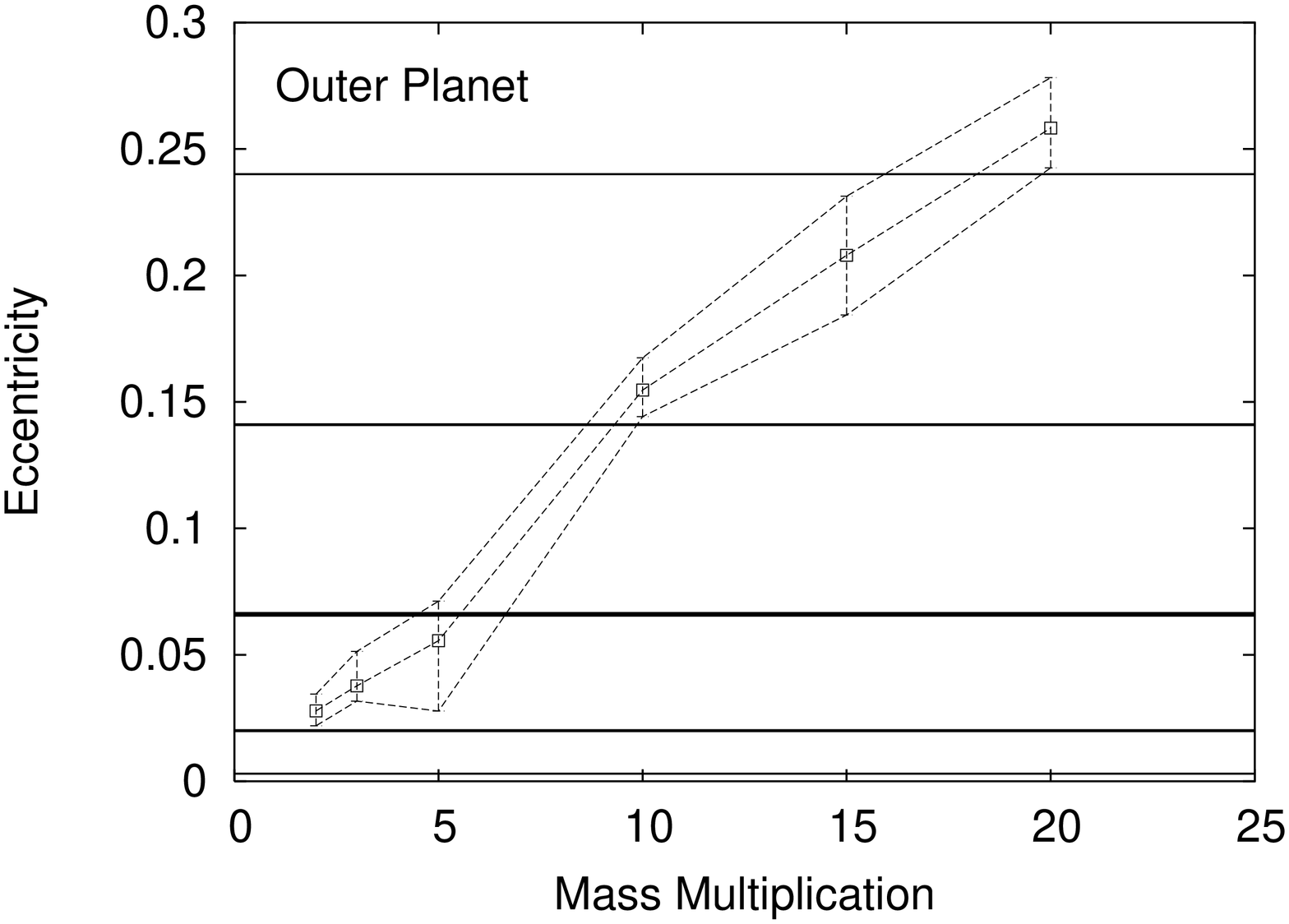,angle=-90,width=0.33\textwidth}
          }
\caption{The dashed lines and data points give the eccentricities for the three planets as functions of the mass-scaling (x-axis) in our simulations, each point gives the mean eccentricity over 10 identical runs, along with the upper and lower quartile bounds. The solid lines give the MCMC 2.5, 16.3, 50, 83.7 \& 97.5 percentile values for fits to the R.V. observations (50th percentile in bold, 16.3 \& 83.7 in medium and 2.5 \& 97.5th in light).}
  \label{FIG:MeanMass}
\end{figure*}

We plot in Fig \ref{FIG:MeanMass} the mean, upper-quartile and lower-quartile eccentricities from simulations for the three planets as we vary both the initial planetary masses and the subsequent mass growth rates. Over the top of this we also plot (solid lines) the observationally inferred eccentricities which result from our MCMC analysis of the best fit orbital elements to the radial velocity data. 

If we focus on the results for the inner two planets, then we can see that (i) values of $e\approx 0.01$ (as seen in Fig \ref{FIG:750_759e}) are approximately coincident with the 2.5th percentile line, and (ii) the lower mass scalings ($2\times$ \& $3\times$ minimum mass) for both planets fall below the 16.3th percentile line. So, whilst it is not impossible that the observations are consistent with zero eccentricity, the likelihood is that it is significantly above this, suggesting that if one wishes to ascrib ethe observed planetary eccentricities to mutual planet-planet excitation during planetary migration, then the planets would be expected to have masses over $5\times$ the minimum mass from the Alibert models. Turning to the results for the third planet, and performing a similar analysis, we find that we expect from the MCMC fits a mass scaling \emph{less} than 10 times the minimum mass. Taken together, these results suggests that we concentrate on mass scalings 5 - 10 times higher than the minimum mass models. Such a large increase in the masses would demand a nearly face-on system ($4 < i < 9$ degrees for $\frac{1}{15} < \sin{i} < \frac{1}{5}$), the probability of which is rather low, as well as the implication that the initial disk mass and surface density would be challengingly high.

\subsection{Gas Damping}\label{DampPlanets}
\begin{figure*}
\centerline{
  \psfig{figure=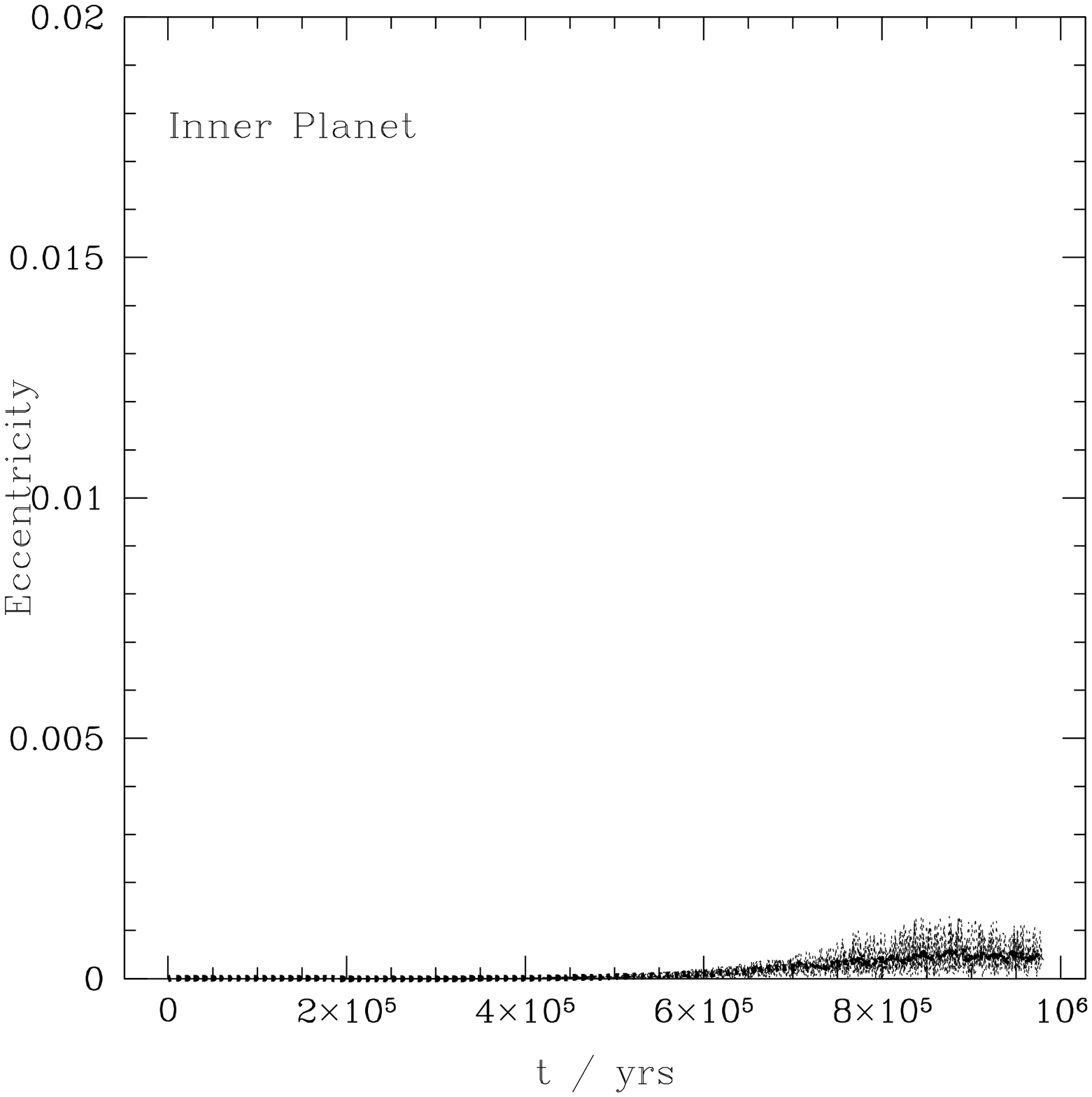,width=0.33\textwidth}
  \psfig{figure=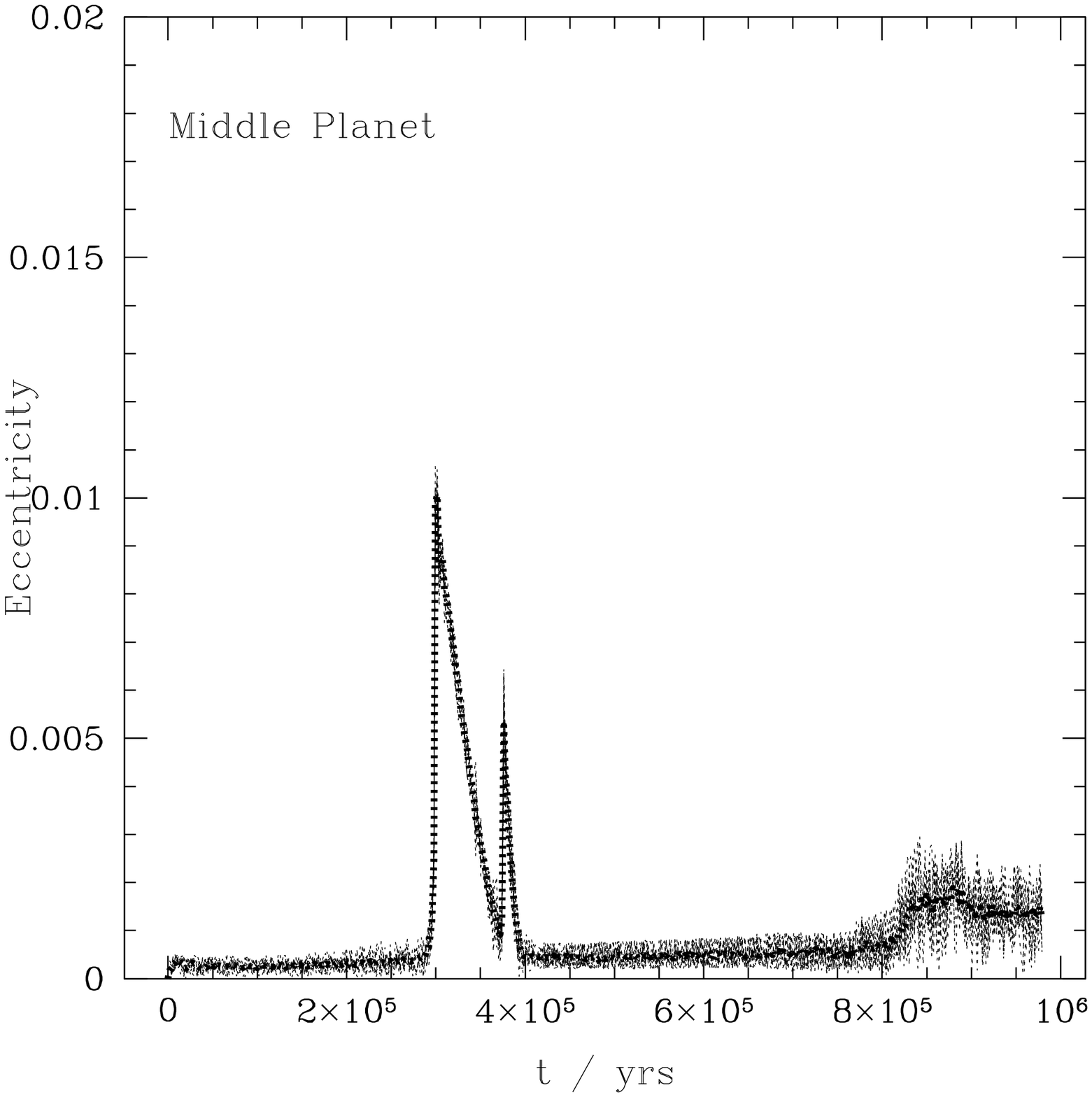,width=0.33\textwidth}
  \psfig{figure=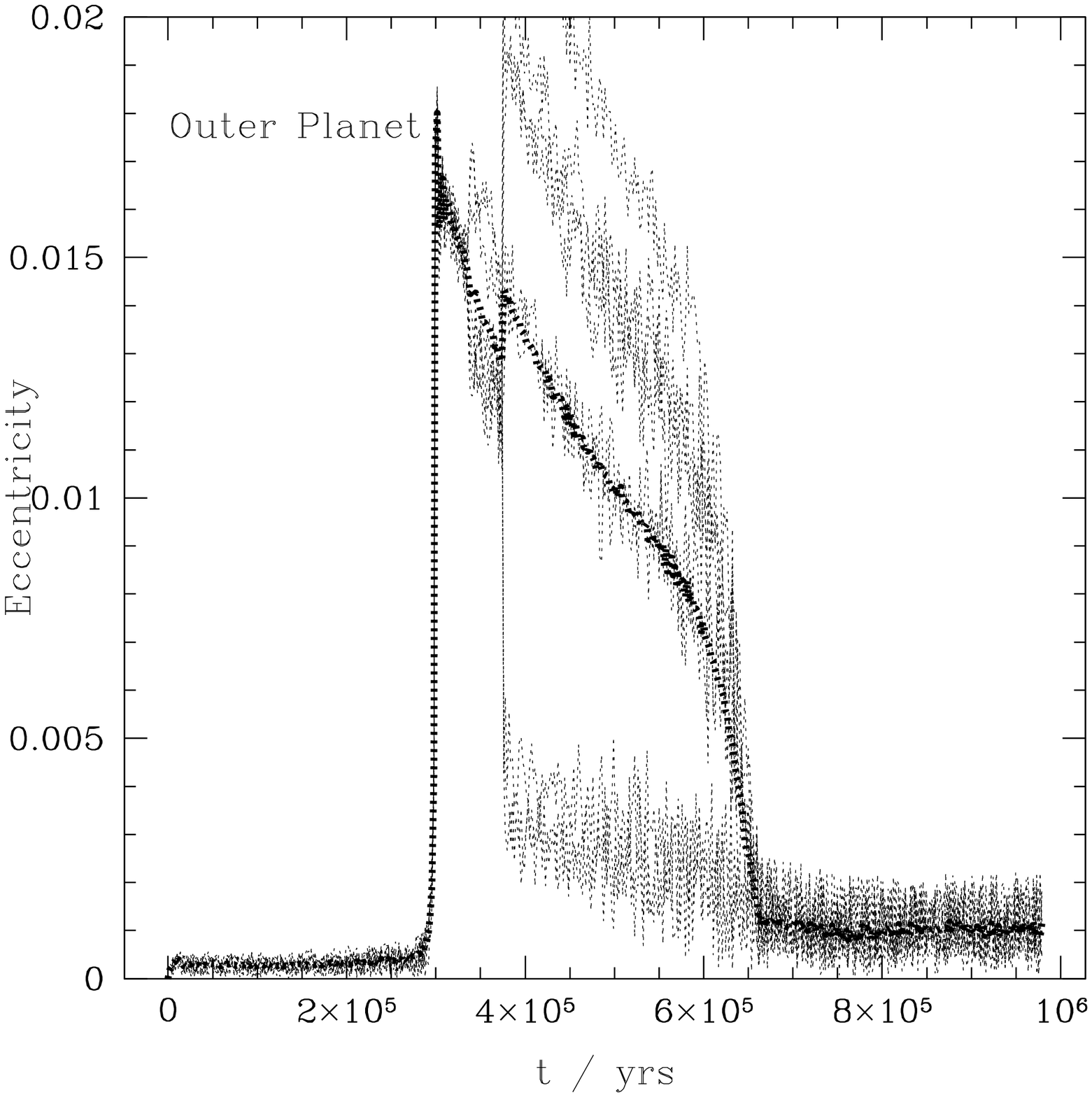,width=0.33\textwidth}
          }
          \caption{Eccentricity evolution for the three planets in the HD69830 system. Planets are $5\times$ the minimum observed mass and are subject to gas damping. Ten simulation runs are superimposed, the individual runs plotted in grey, the overall average plotted in black.}
  \label{FIG:GD5}
\end{figure*}

Given that the planets are growing and migrating within a gas disk, one should include gas-induced eccentricity damping in the system simulations. We model this using the planetary damping model described in \S \ref{Method} and plot the results in Figure \ref{FIG:GD5}. Whilst the eccentricities of the middle and outer planets are again clearly still excited by the 3:2 MMR resonance crossing at $\sim 3\times10^5$ yrs, this is damped back down on a timescale of a few $\times 10^5$ years.

We find that this rate of damping is far too high to allow any significant eccentricity to be preserved in the system. For both the standard mass simulations (not shown) and for the simulations with 5 times greater masses (Fig \ref{FIG:GD5}) we find that the eccentricities for all of the planets are damped to values lower than $10^{-3}$, two orders of magnitude below the observed values.

These results illustrate the problem, e.g. \citet{Papaloizou01,Chatterjee07}, of how to generate an extra-solar planetary
system in which there is sufficient gas present in the disk to cause
mass growth and semi-major axis decay, whilst simultaneously ensuring
that the gas present in the system does not effectively circularise
the planetary orbits. Given the difficulties of the \citet{Alibert_et_al_06}
model, another form of eccentricity excitation must be
posited following the decay of the gaseous disk.

\section{Excitation of Planetesimal Eccentricities}\label{PlanetesimalExcite}
There are a number of papers \citep{2005A&A...441..791F,
2007arXiv0710.3730F, 2007A&A...461.1195F, 2007A&A...472.1003F,
Mandell07} which focus on the formation of terrestrial
planets during and after the migration of a Jupiter-mass planet
through the inner system. They typically find that the passage of the
giant planet does \emph{not} clear the inner system of solids, but
rather initially shepherds the material inwards before exciting it and finally scattering/expelling it to exterior orbits. If this
situation can be replicated in simulations of the HD69830 system, it
could provide a means of producing a scattered disk external to HD69830d which may result in a long-lived population of hot dust as observed.

In this section we add massless planetesimals to out MERCURY simulations of planetary migration. To investigate the scattering of planetesimals in this model, we perform a number of different simulations to understand the effects of various physical phenomena on the final distribution of planetesimals after the conclusion of planetary migration. We vary the planetary masses, the gas drag acting on the planets and the gas drag on the planetesimals. The details of these simulation are recorded for convenience in Table \ref{TABLE:Simulations}.

Unless otherwise stated, all simulations have an initial distribution of planetesimals which follows the minimum mass Solar nebular model surface density profile of $\Sigma\propto a^{-3/2}$, with the initial semi-major axes limited between 0.1 and 9 AU, and with the eccentricities and inclinations drawn from a Rayleigh distribution with $2<i> = <e> = 0.05$.

In general, we perform composite simulations consisting of $m$ different simulation runs, each simulation containing $n$ planetesimals and 3 planets. This allows us to gain a speed benefit in finding the distribution of $m\times n$ planetesimals by running simulations in parallel, but more importantly allows us to include the range of different \emph{planetary} eccentricity excitations seen in \S \ref{AlibertPlanet} by using random initial planetary mean anomalies in each of the parallel runs, thus giving us a view of the probabilities of different outcomes for a given set of starting conditions.

\begin{table*}
\begin{minipage}{1.0\textwidth}
\caption{Summary of the main planetesimal scattering simulations.}
\label{TABLE:Simulations}
\begin{tabular}{lcccccccccccc}
\hline
Simulation & Number of     & Planet          & \multicolumn{2}{|c|}{Gas Damping } & $\%$ Ejected   & \multicolumn{3}{|c|}{$\%$ Accreting onto } & $\%$ Surviving & \multicolumn{3}{|c|}{Surviving}    \\
           & Planetesimals & Mass            & \multicolumn{2}{|c|}{on ...}       &        or      & \multicolumn{3}{|c|}{planets}      &    with        & \multicolumn{3}{|c|}{with $q > 1$} \\
           &               & ($\times$ Min.) &    Planets   &  Planetesimals      &  Hit Star      &                   b & c & d                &    $q < 1$     &       $\%$ & $<e>$ & $<q>$         \\
\hline
\hline\\
A      & 10,000 & 1 & No  & No           &  4 & 65 &  3  &  1  & 12  & 15 & 0.30 & 3.55 \\ %Standard Simulation, 3700_4098
A+     &        &   &     &              & 11 & 66 &  5  &  2  &  1  & 15 &      &      \\ %Extended Simulation of A, 8501_8578
B      & 2,500  & 5 & No  & No           & 43 & 37 &  1  & 0.2 &  3  & 15 & 0.47 & 3.86 \\ %Increased Mass, 5200_5299
C1     & 2,500  & 1 & Yes & No           &  4 & 65 &  4  & 0.3 & 12  & 14 & 0.29 & 3.66 \\ %Standard Planets + Gas Drag on Planets. 5600_5699
%C2     & 2,500  & 5 & Yes & No           & 13 &  2 & 0.1 & 0.3 & 72  & 12 & 0.44 & 4.46 \\ %Increased Mass Planets + Gas Drag on Planets. 6000_6099  - - - - - - Problem, planets seem to turn around!!!
D1     & 2,000  & 1 & No  & Yes, 100km   &  0 &  3 &  1  & 0.2 & 80  & 15 & 0.13 & 3.75 \\ %Standard Planets + Gas Drag on Planetesimals, 100km. 4202_4399
D2     & 2,000  & 5 & No  & Yes, 100km   & 12 &  1 & 0.5 & 0.1 & 74  & 13 & 0.28 & 4.32 \\ %Increased Mass Planets + Gas Drag on Planetesimals, 100km. 8000_8199
E1     & 1,000  & 1 & No  & Yes, 1000km  &  0 &  6 &  3  & 0.6 & 69  & 23 & 0.25 & 2.92 \\ %Standard Planets + Gas Drag on Planetesimals, 1000km. 8400_8499
E2     & 1,000  & 5 & No  & Yes, 1000km  & 34 &  5 & 0.6 & 0.3 & 41  & 19 & 0.46 & 3.97 \\ %Increased Mass Planets + Gas Drag on Planetesimals, 1000km. 8200_8299
\hline
F(i)   & 2,300  & 1 & No  & No           & 17 & 56 & 2  & 0.3 & 11   & 14 & 0.29 & 3.53 \\ %Additional Planets, Collision with HD69830(c), 8800-8822
F(ii)  & 300    & 1 & No  & No           & 18 & 57 & 2  & 0   &  9   & 14 & 0.29 & 3.47 \\ %Additional Planets, Collision with HD69830(d), 8830-8832
F(iii) & 700    & 1 & No  & No           & 24 & 49 & 3  & 0   & 11   & 13 & 0.27 & 3.97 \\ %Additional Planets, Scattered to outer system, 8840_8846
F(iv)  & 100    & 1 & No  & No           & 44 & 34 & 1  & 1   &  9   & 11 & 0.30 & 4.20 \\ %Additional Planets, Ejected from system, 8850
\hline
G      & 2,500  & 1 & No  & No           & 40 & 38 & 2   & 0   &  9  & 11 & 0.41 & 3.83 \\ %Migration after growth. 6402_6499
\hline
H      & 1,250  & 1 & No  & Yes, 100km   & 0 & 0.3 & 0.8 & 1.2 & 84  & 13 & 0.15 & 3.59 \\ %Different Initial Semi-major axes 7600_7649
\hline
\end{tabular}
\end{minipage}
\end{table*}

\subsection{Planetesimal Scattering in the 3-Planet Model}\label{SimpleScatter}
Given the results of \S \ref{PlanetExcite}, we initially examine the
case of planetesimal excitation in an undamped system consisting of
three planets following the growth and migration paths of \S
\ref{AlibertPlanet}: This is our simulation set A.

\begin{figure*}
\centerline{
  \begin{tabular}{cccccc}
    \multicolumn{2}{|c|}{\subfigure[]{\psfig{figure=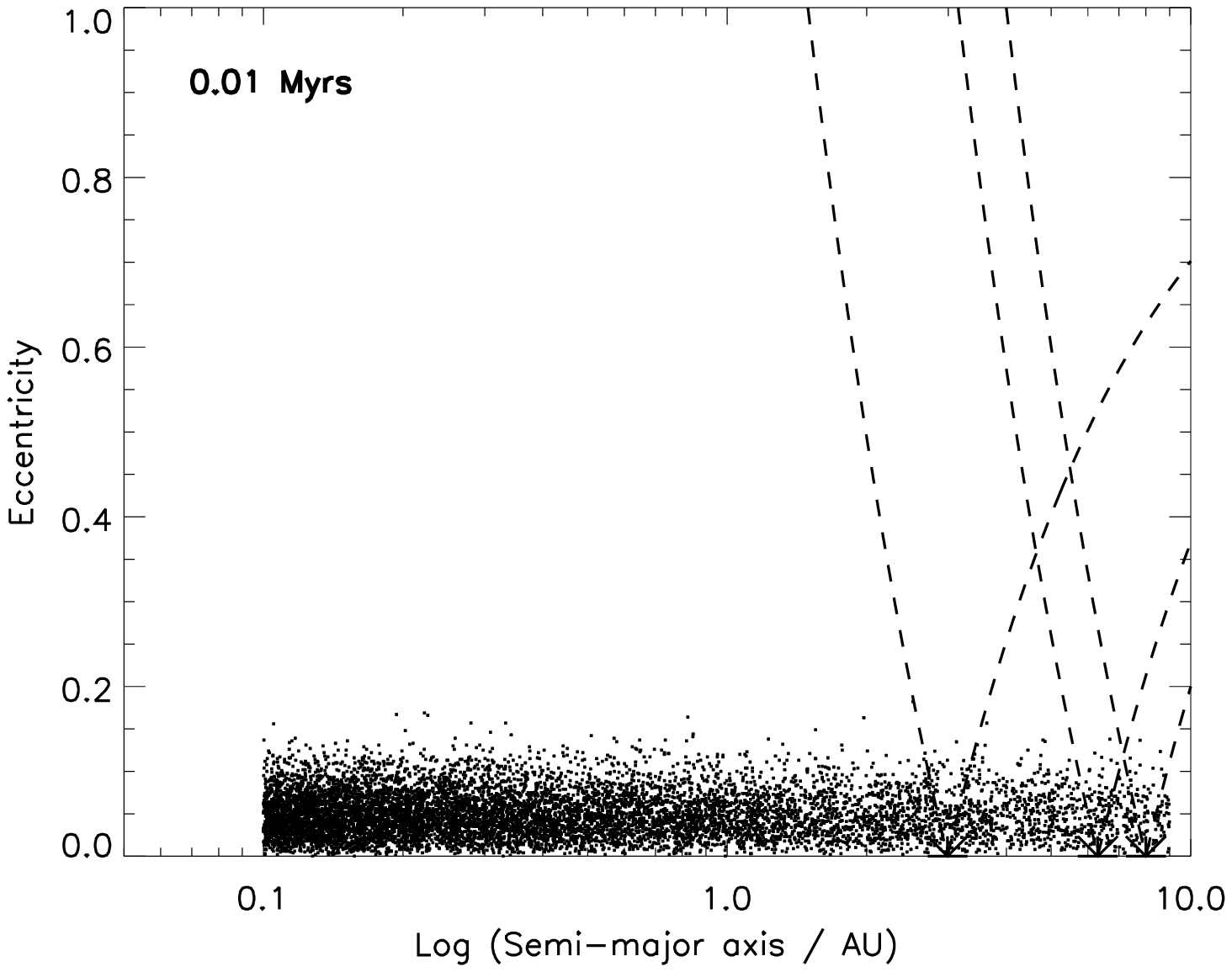,width=0.33\textwidth }}}&
    \multicolumn{2}{|c|}{\subfigure[]{\psfig{figure=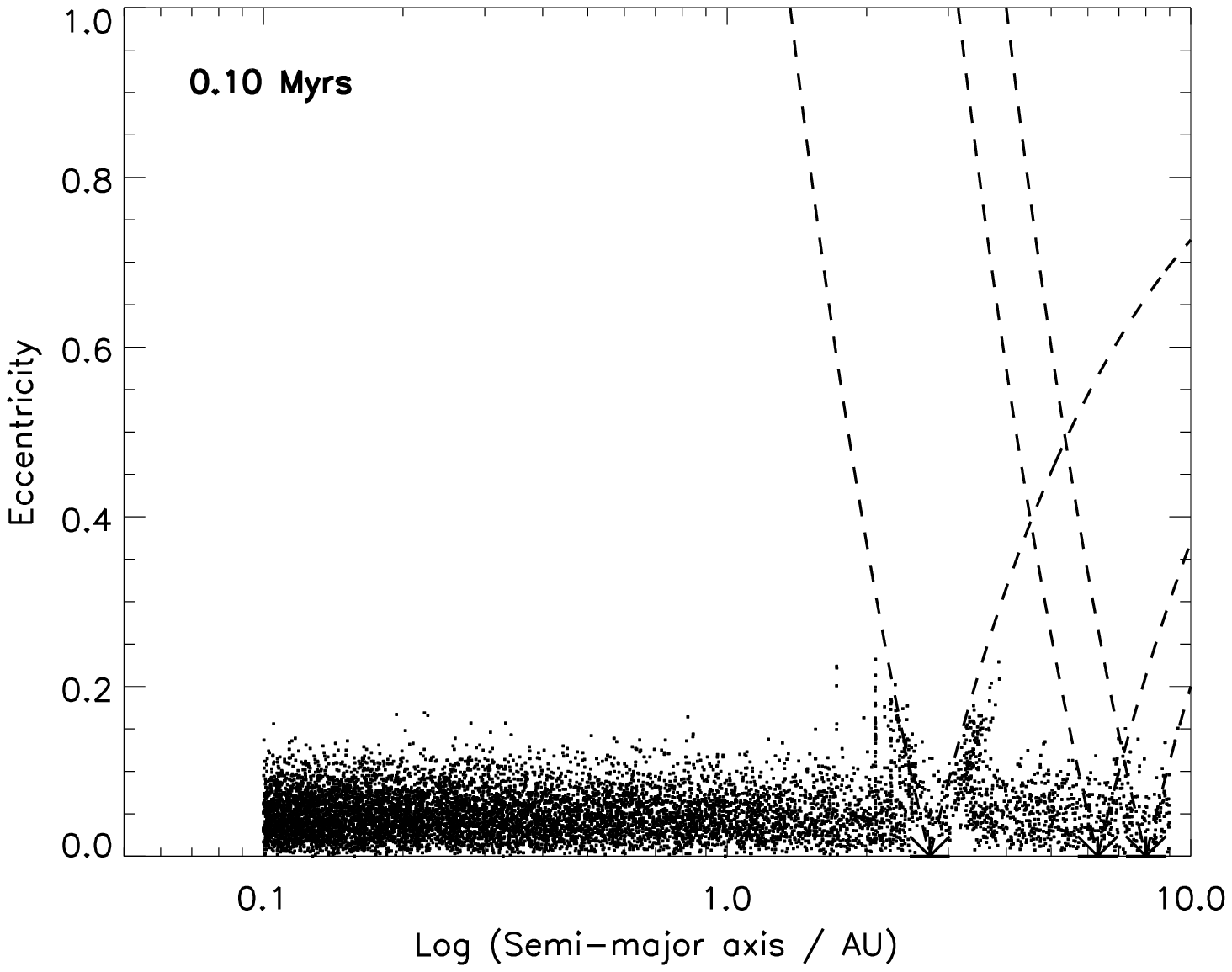,width=0.33\textwidth}}}&
    \multicolumn{2}{|c|}{\subfigure[]{\psfig{figure=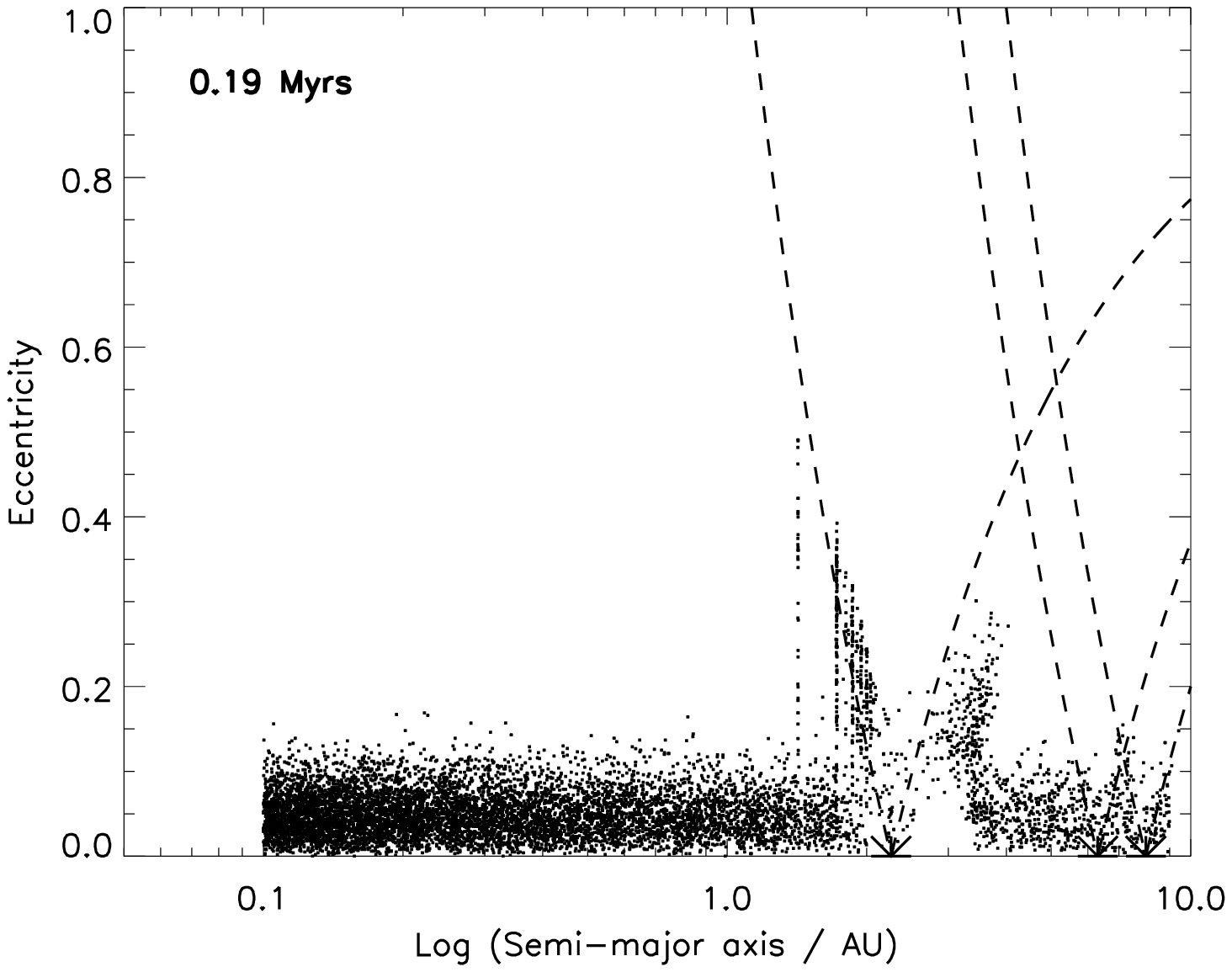,width=0.33\textwidth}}}\\
    %\vspace{-0.3cm}
    \multicolumn{2}{|c|}{\subfigure[]{\psfig{figure=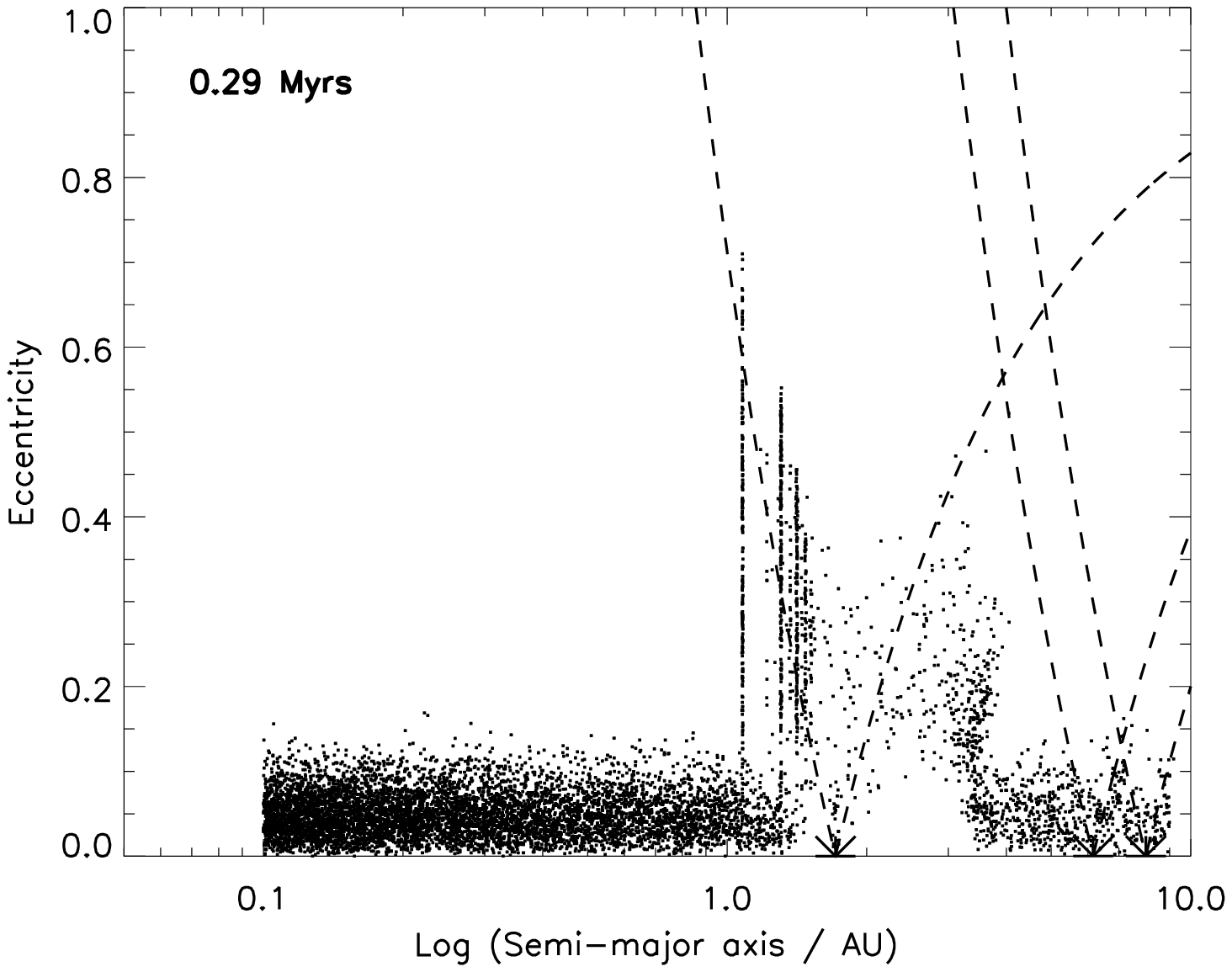,width=0.33\textwidth}}}&
    \multicolumn{2}{|c|}{\subfigure[]{\psfig{figure=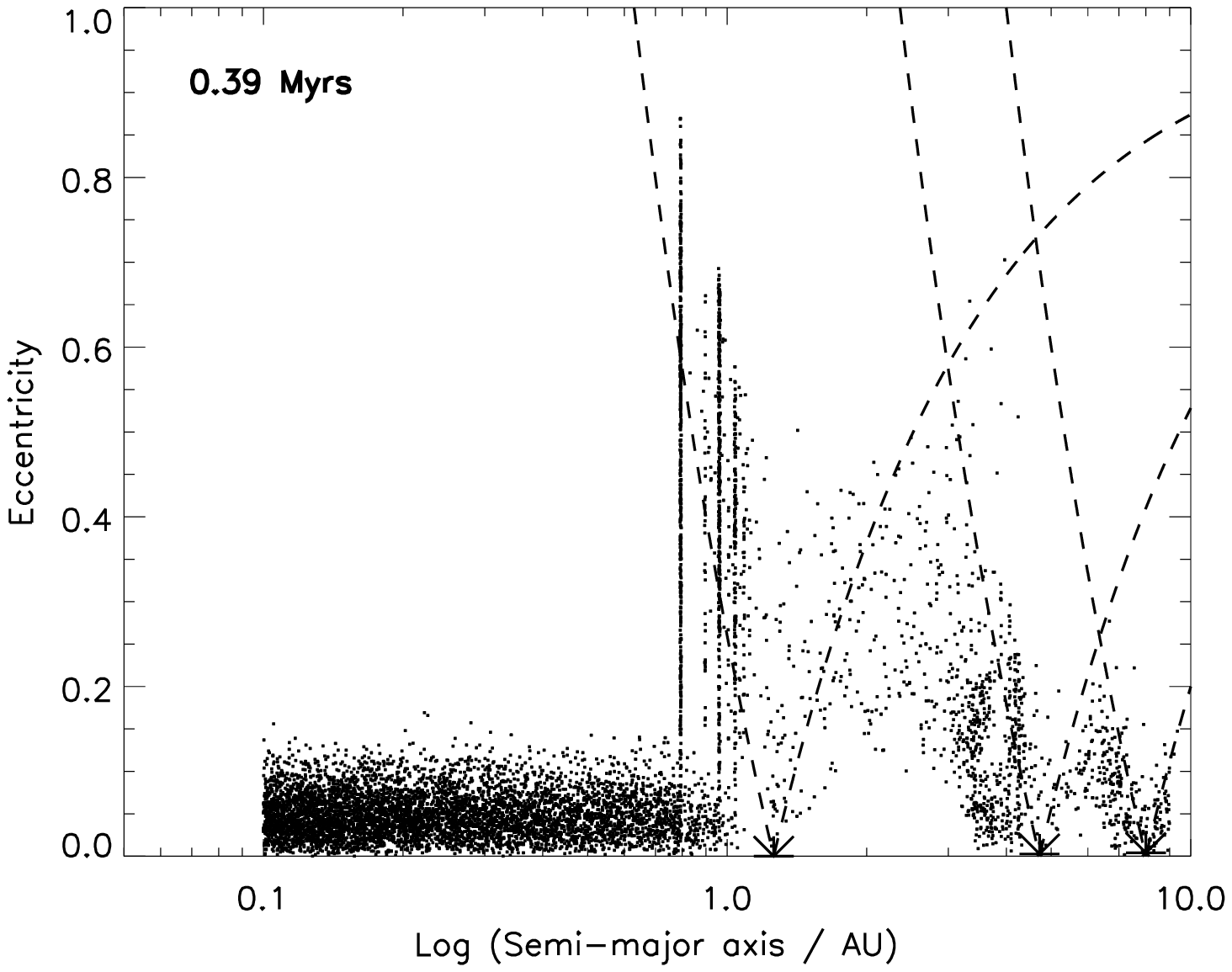,width=0.33\textwidth}}}&
    \multicolumn{2}{|c|}{\subfigure[]{\psfig{figure=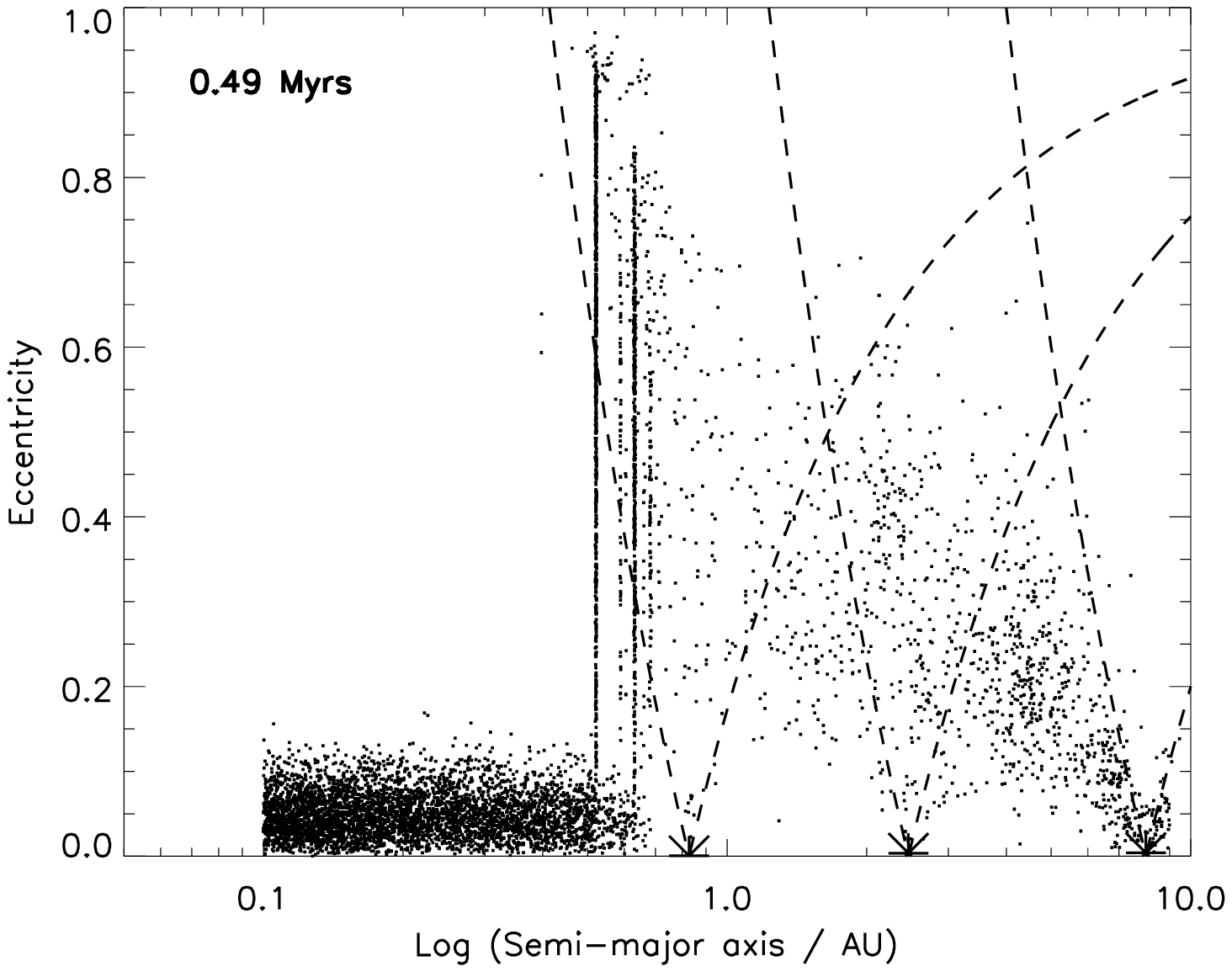,width=0.33\textwidth}}}\\
    %\vspace{-0.3cm}
    \multicolumn{2}{|c|}{\subfigure[]{\psfig{figure=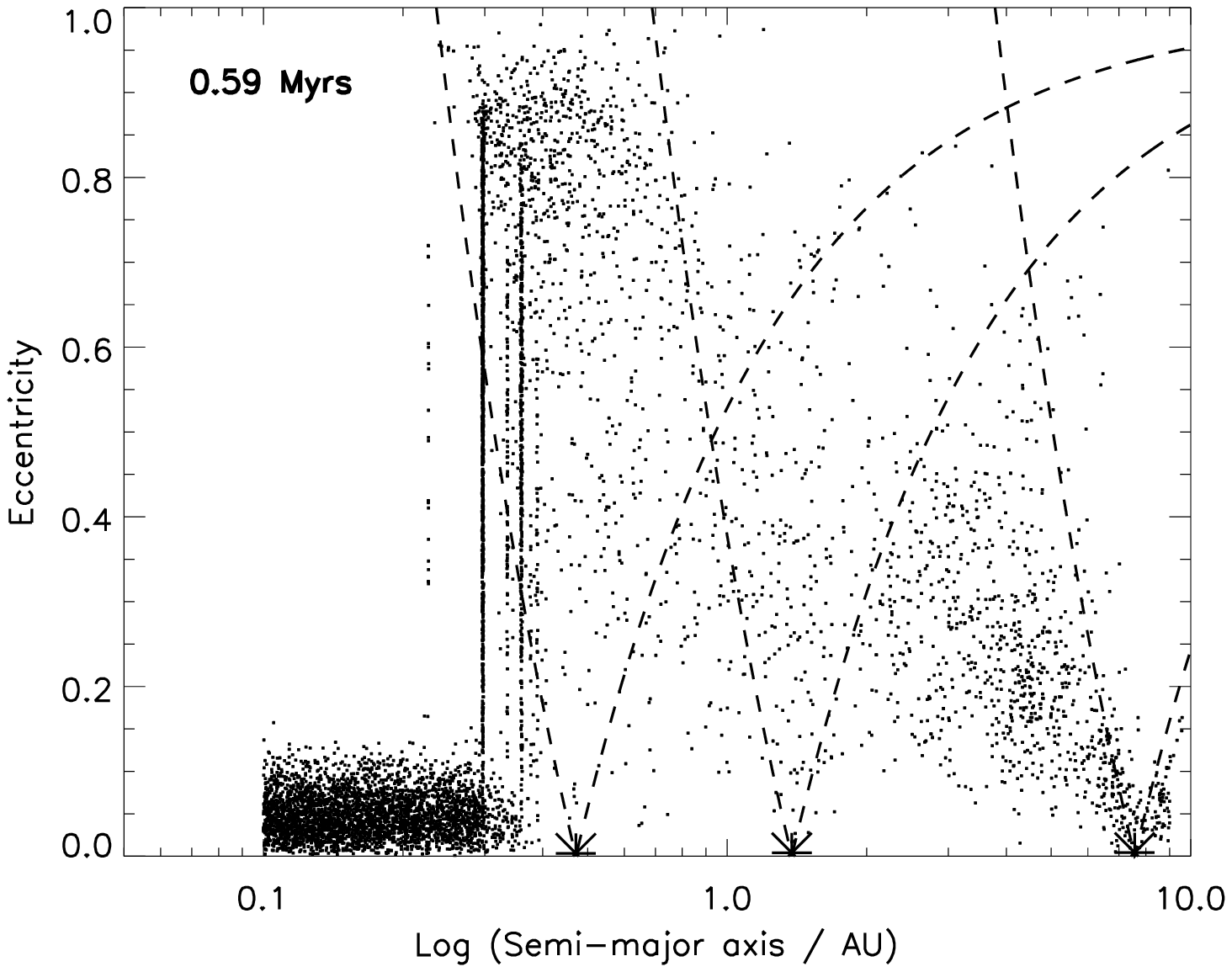,width=0.33\textwidth}\label{t0.5}}}&
    \multicolumn{2}{|c|}{\subfigure[]{\psfig{figure=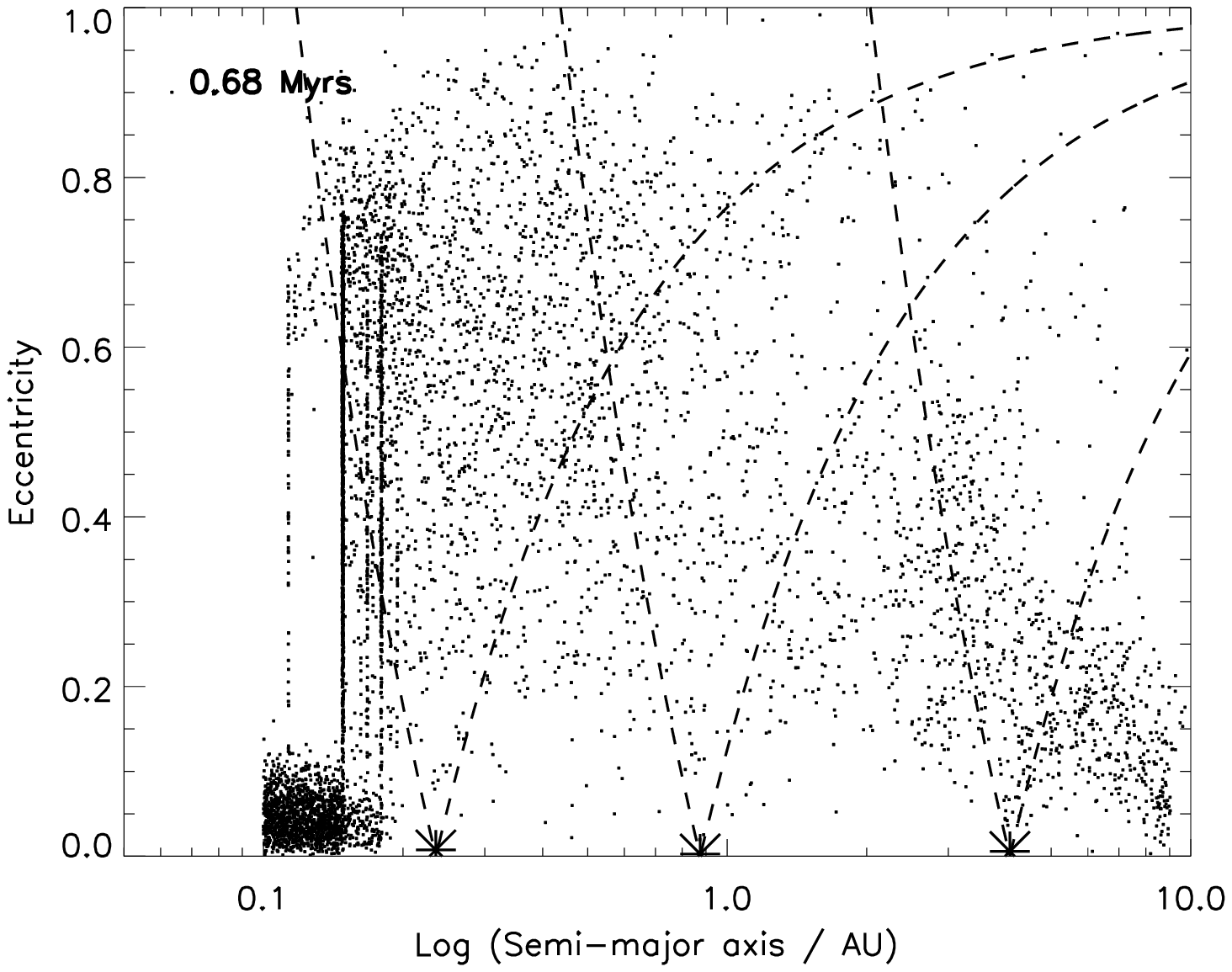,width=0.33\textwidth}}}&
    \multicolumn{2}{|c|}{\subfigure[]{\psfig{figure=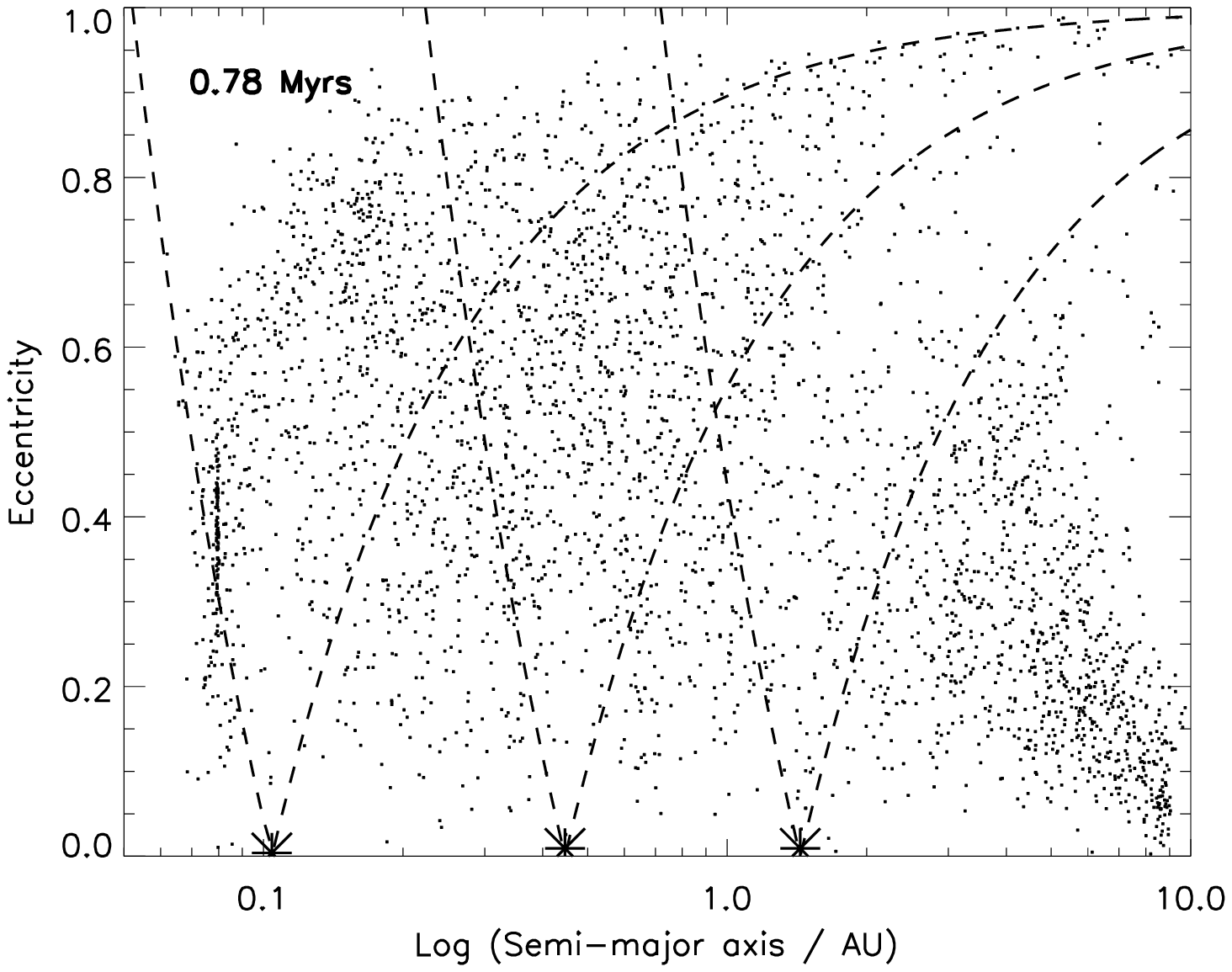,width=0.33\textwidth}}}\label{t0.8}\\
    %\vspace{-0.3cm}
    \multicolumn{2}{|c|}{\subfigure[]{\psfig{figure=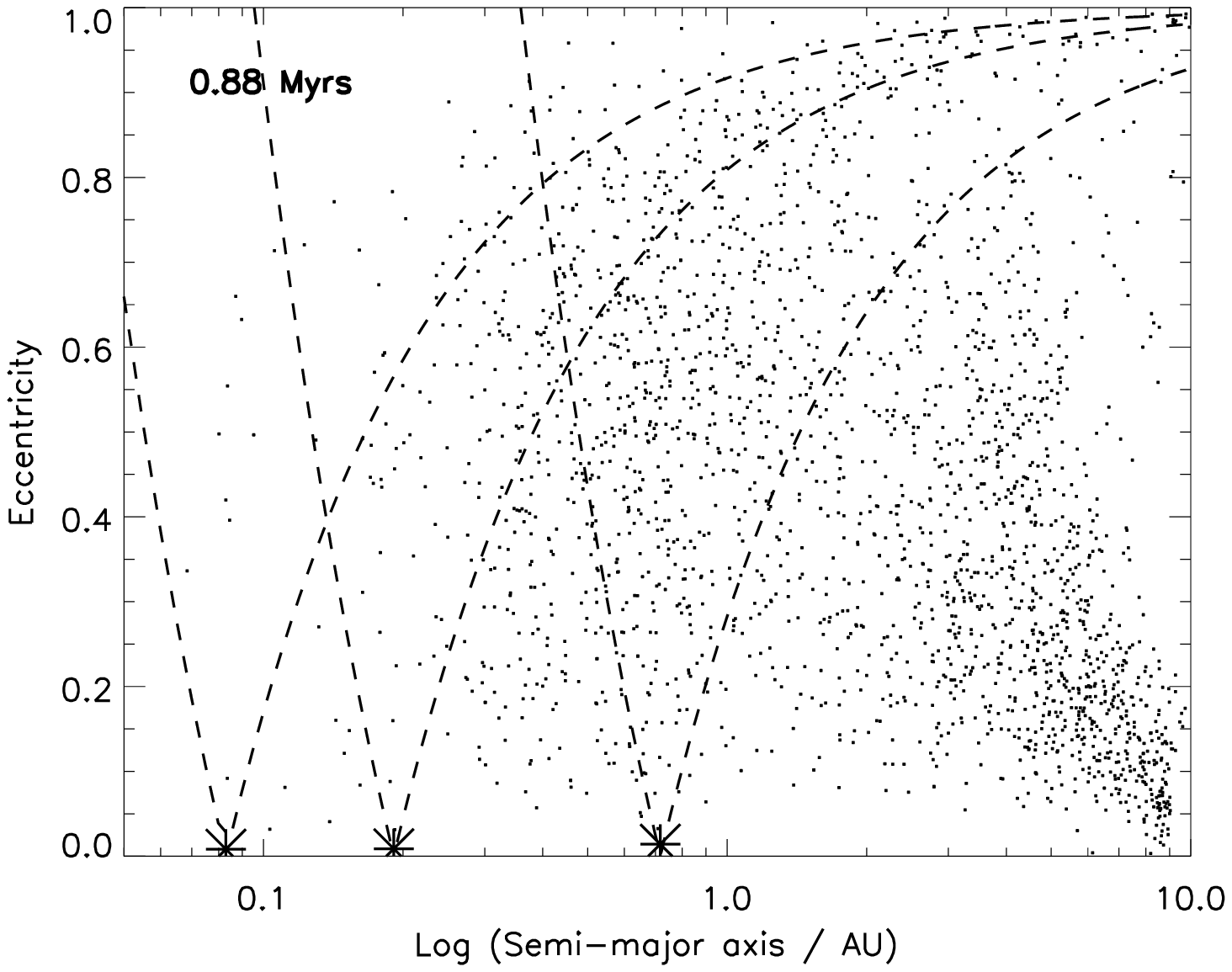,width=0.33\textwidth}}}&
    \multicolumn{2}{|c|}{\subfigure[]{\psfig{figure=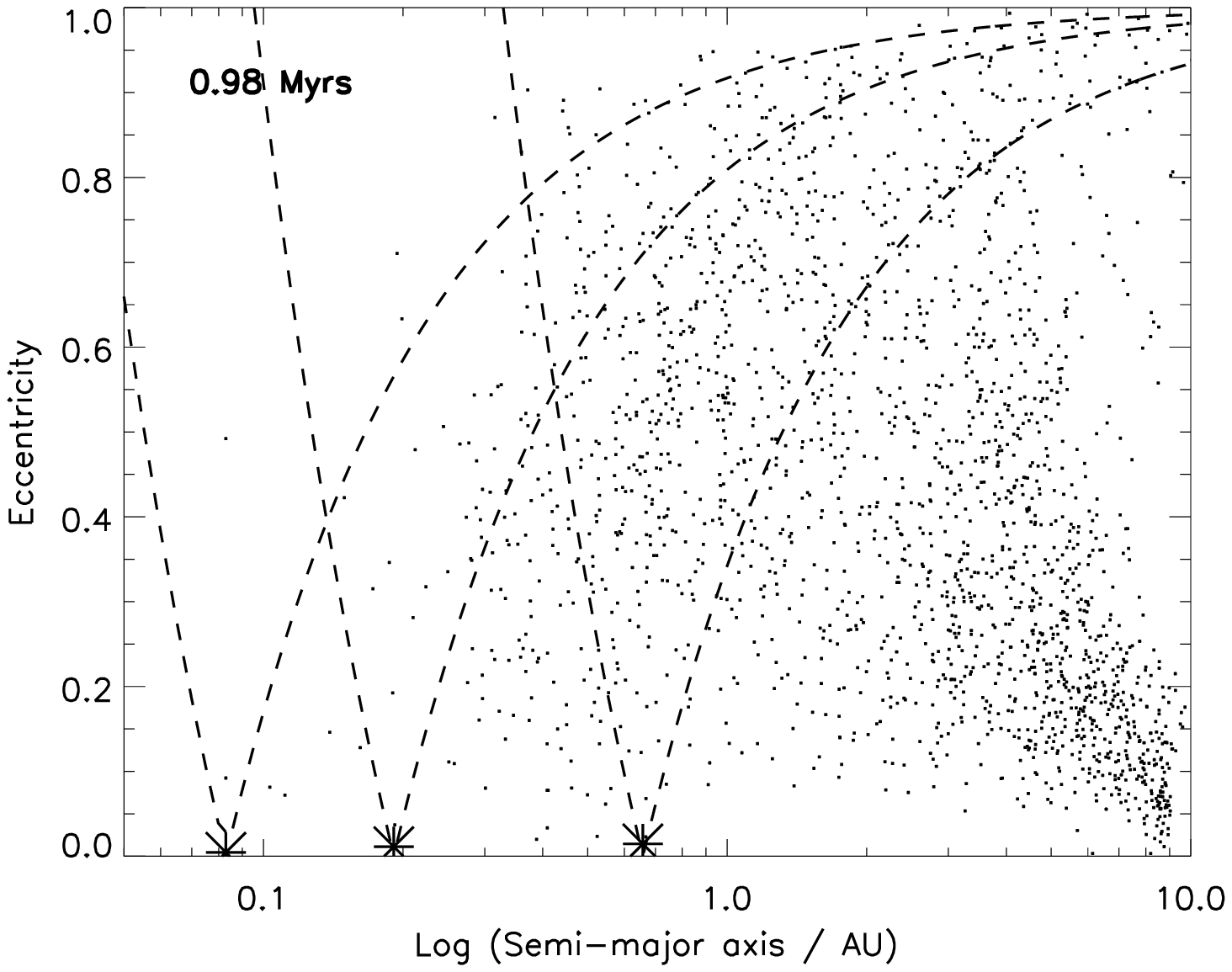,width=0.33\textwidth}}}
  \end{tabular}
}
\caption{Evolution of eccentricity - semi-major axis for the planetesimals distribution due to the forced migration of planets. Results of Simulation Set A - Minimum mass planets with no gas damping present in the system.}
\label{FIG:Evolution_ae}
\end{figure*}

The shepherding effect on the planetesimals is clearly visible in the sample illustrations given in Fig \ref{FIG:Evolution_ae}. Here we see that as the planets migrate inwards, numerous planetesimals are forced to migrate ahead of the planets, subsequently rising up a number of the MMRs interior to the planetary semi-major axes, with the vast majority being shepherded inside the innermost planet. From around $0.5$ Myr onwards, we start to see (Fig 4(g)) planetesimals being scattered out of the MMRs, some colliding with planets, some more being ejected from the system, whilst others are scattered onto high eccentricity orbits at intermediate semi-major axes. By $\sim 0.8$ Myr, we find that the resonance structures interior to the innermost planet have now been almost completely destroyed, leaving a large swarm of planetesimals effectively populating the majority of the parameter-space $0.1<a<10$, $0<e<1$.
%
%%%%%%%%%%%%%%%%%%%%%%%%%%%%%%%%%%%%%%%%%%%%%%%%%%%%%%%%%%%%%%%%%%%%%%%%%%%%%%%%%%%%%%%%
%%%% Collection of figures to display the evolution in the planetesimal semi-major-axis - eccentricity plot %%%%

%
%%%%%%%%%%%%%%%%%%%%%%%%%%%%%%%%%%%%%%%%%%%%%%%%%%%%%%%%%%%%%%%%%%%%%%%%%%%%%%%%%%%%%%%%

When the simulations have completed (I.e. the three planets are at the semi-major axes observed) we find that $27\%$ of the planetesimals survive in the system, the rest being either ejected or suffering collisions with the planets or the central star. 

It should be noted that in the interests of brevity, these simulations were terminated after $10^6$ years. If (impractically) the simulations were continued on to billion-year time scales, then it is clear that many of the remaining planetesimals whose orbits cross those of the planets would also be ejected or suffer a collision, thus effectively clearing out most material with pericentre inside $1 AU$, although some material may remain in a belt between planets c \& d \citep{Lovis_et_al_06, Ji07}. However, as shown in \citet{Wyatt_et_al_07a}, any material inside $1 AU$ which was sufficiently close to circularised to avoid planetary collision / excitation would be collisionally ground down by the current age of the system. As such, we therefore concentrate our subsequent analysis on those planetesimals with pericentres outside 1 AU.

\begin{figure*}
  \vspace{-0.9cm}
  \subfigure[tight][Simulation A: Standard Alibert-Mass Simulations after 1 Myr]{
    \centerline{
      \psfig{figure=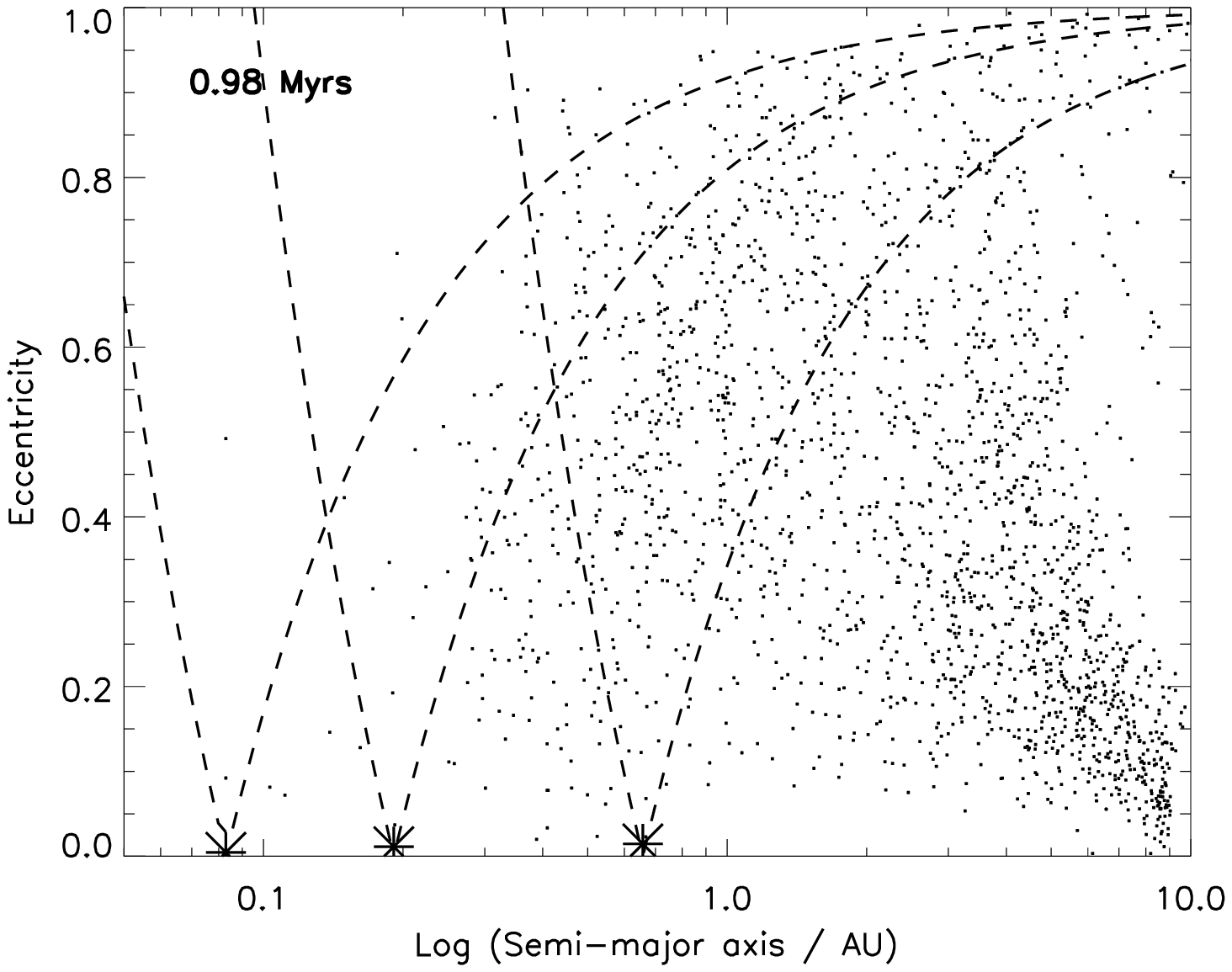,width=0.33\textwidth}
      \includegraphics[trim = 0mm 0mm 0mm 49mm, clip, width=0.33\textwidth]{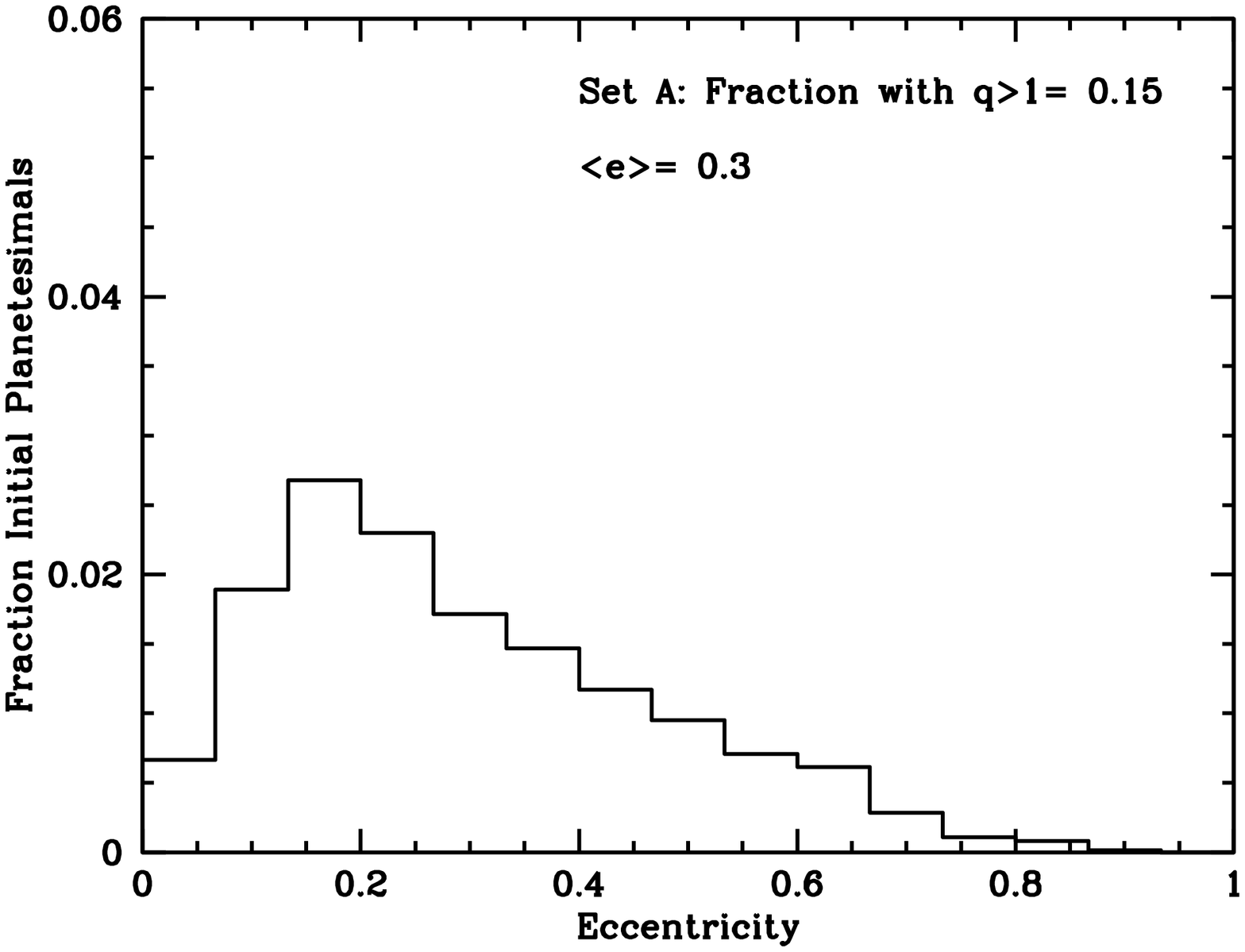}
      \includegraphics[trim = 0mm 0mm 0mm 49mm, clip, width=0.33\textwidth]{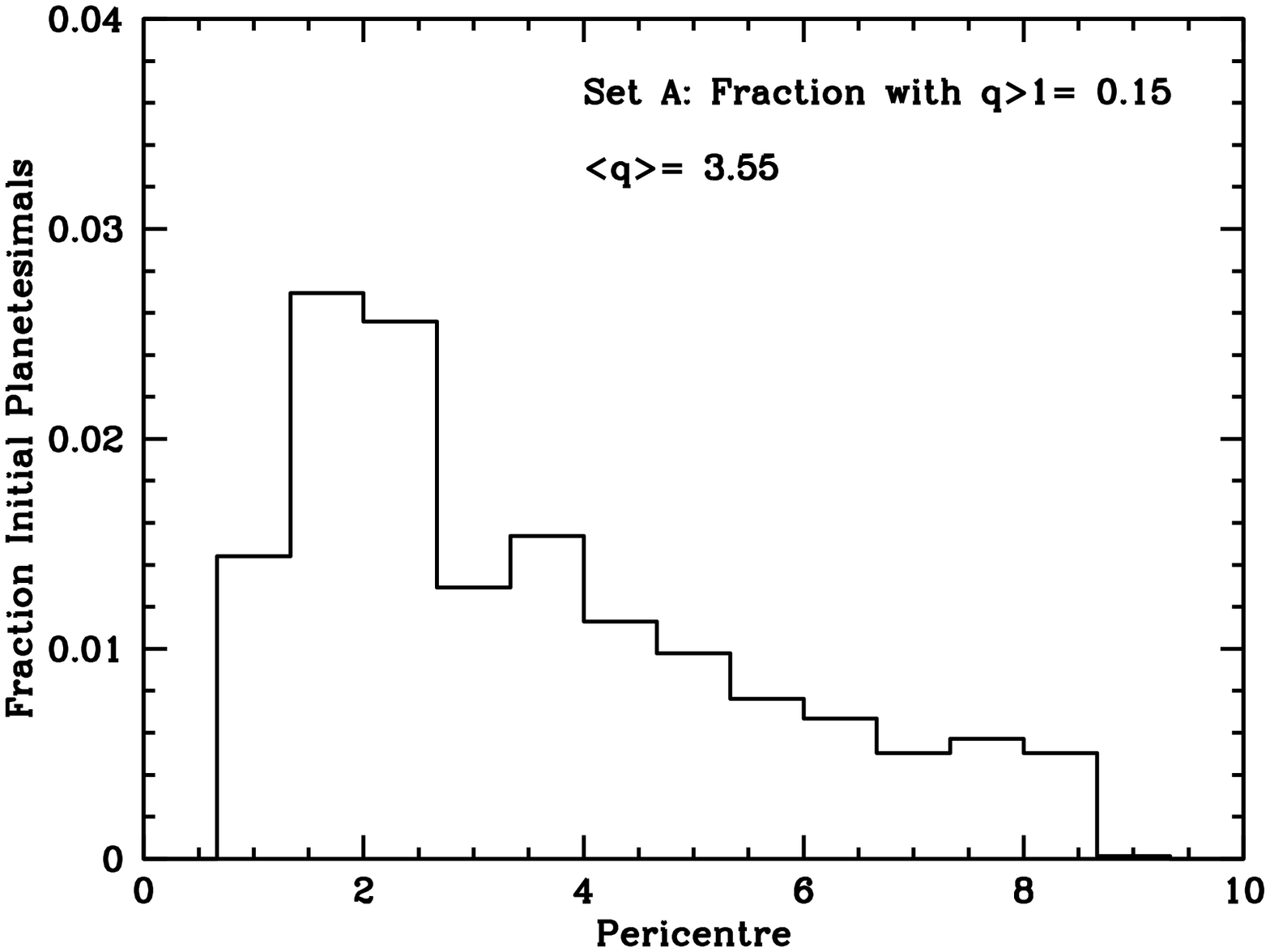}
    }
    \label{FIG:SIMA}
  }
  \vspace{-0.9cm}
  \subfigure[tight][Simulation B: Solid Histograms - Increased Planetary Masses; Dotted Histograms - Standard Masses (Set A)]{
    \centerline{
      \psfig{figure=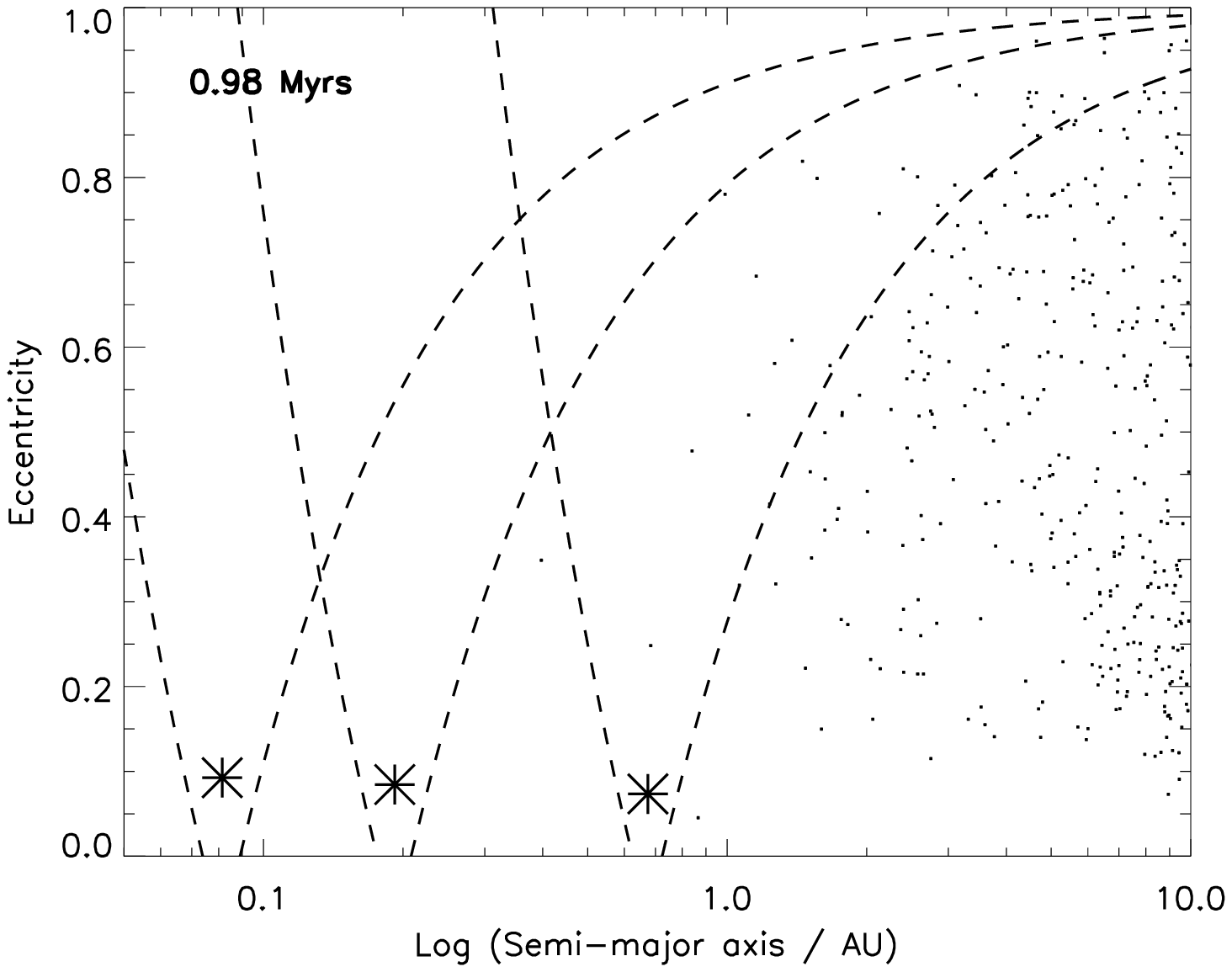,width=0.33\textwidth}
      \includegraphics[trim = 0mm 0mm 0mm 49mm, clip, width=0.33\textwidth]{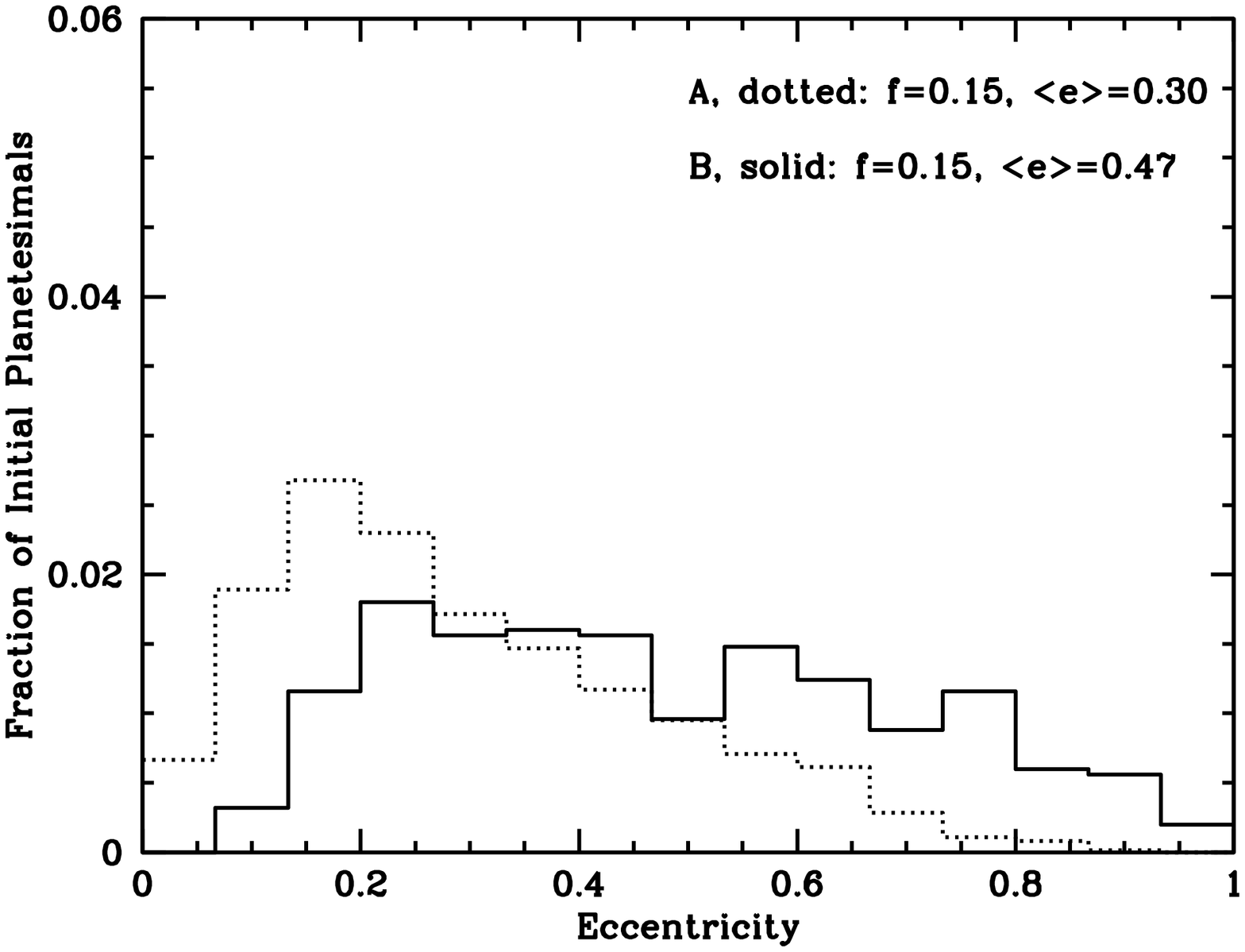}
      \includegraphics[trim = 0mm 0mm 0mm 49mm, clip, width=0.33\textwidth]{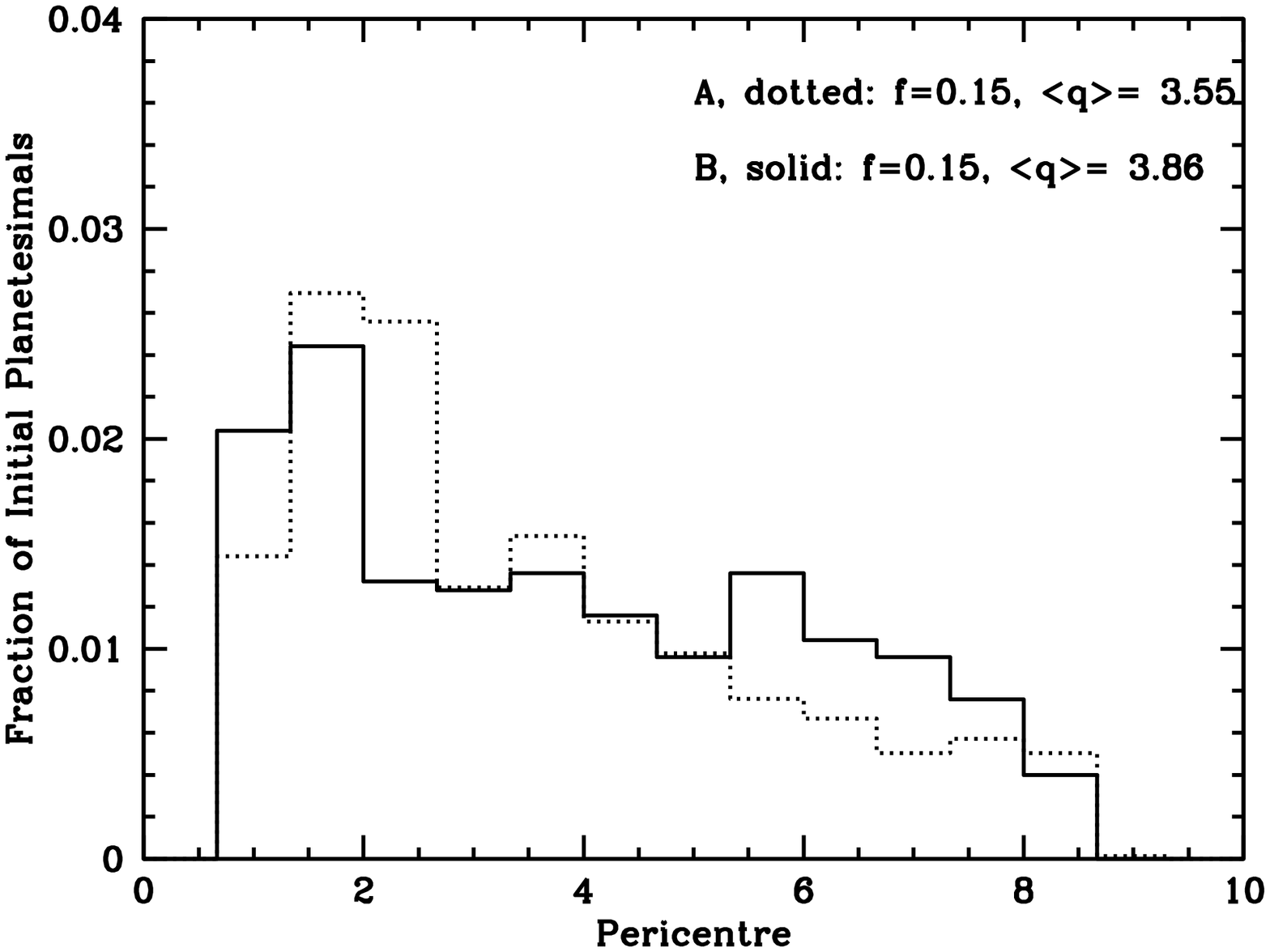}
    }
    \label{FIG:SIMB}
  }
  \vspace{-0.9cm}
  \subfigure[tight][Simulation C: Solid Histograms  - Gas Drag on Planets; Dotted Histograms  - Standard Zero Drag (Set A)]{
    \centerline{
      \psfig{figure=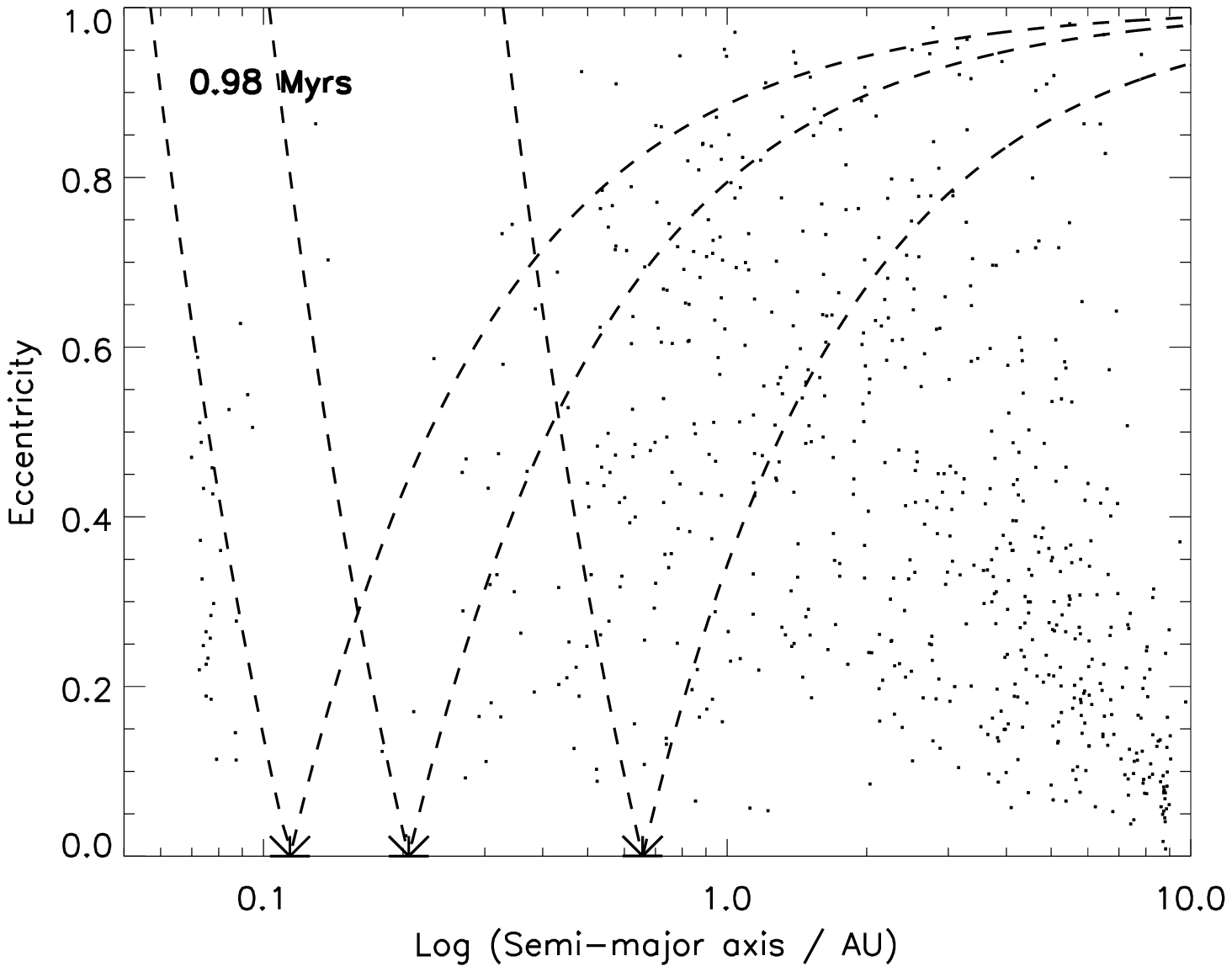,width=0.33\textwidth}
      \includegraphics[trim = 0mm 0mm 0mm 49mm, clip, width=0.33\textwidth]{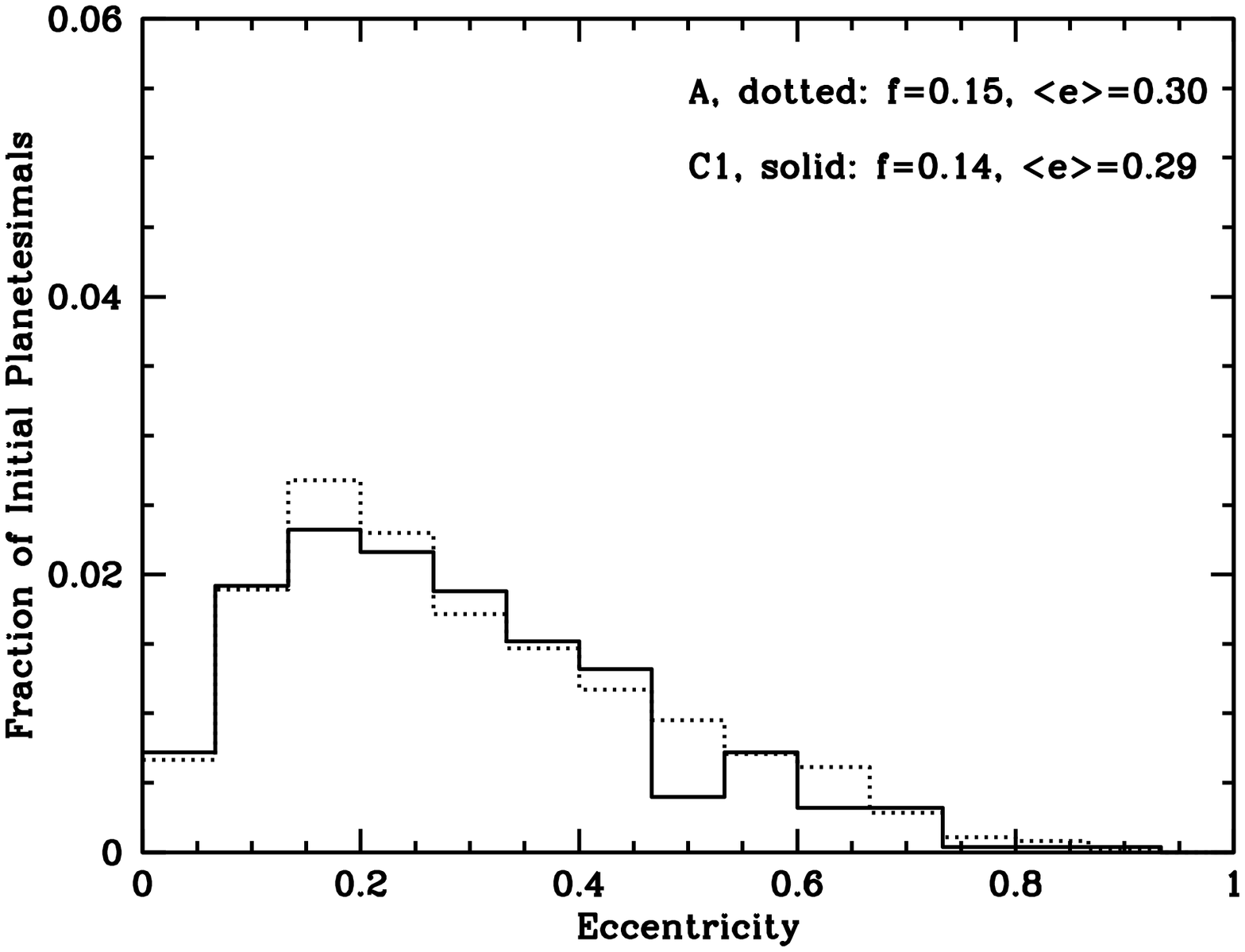}
      \includegraphics[trim = 0mm 0mm 0mm 49mm, clip, width=0.33\textwidth]{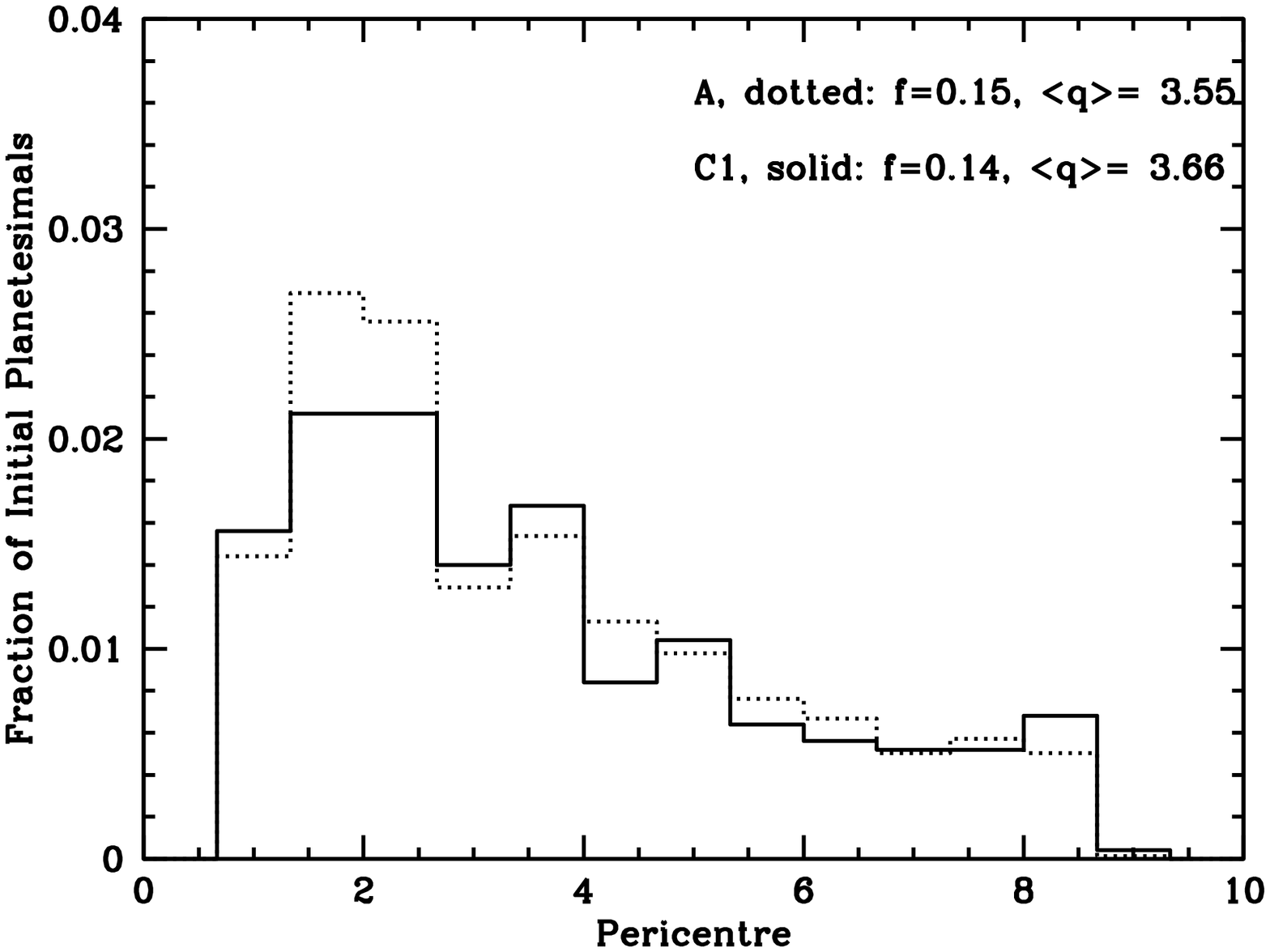}
    }
    \label{FIG:SIMC}
  }
  %\vspace{-0.9cm}
  \subfigure[tight][Simulations D \& E - All have Standard Mass Planets.  LEFT - Eccentricity v. Semi-major Axis for planets and planetesimals for Set D only. CENTRE - Eccentricity Histograms for Gas Drag on 100km Planetesimals (Set D, Solid Histogram), Gas Drag on 1000km Planetesimals (Set E, Dashed Histogram) and Standard Zero Drag (Set A, Dotted Line). RIGHT - Pericentre Histograms, Line-types as for centre plot.]{
    \centerline{
      \psfig{figure=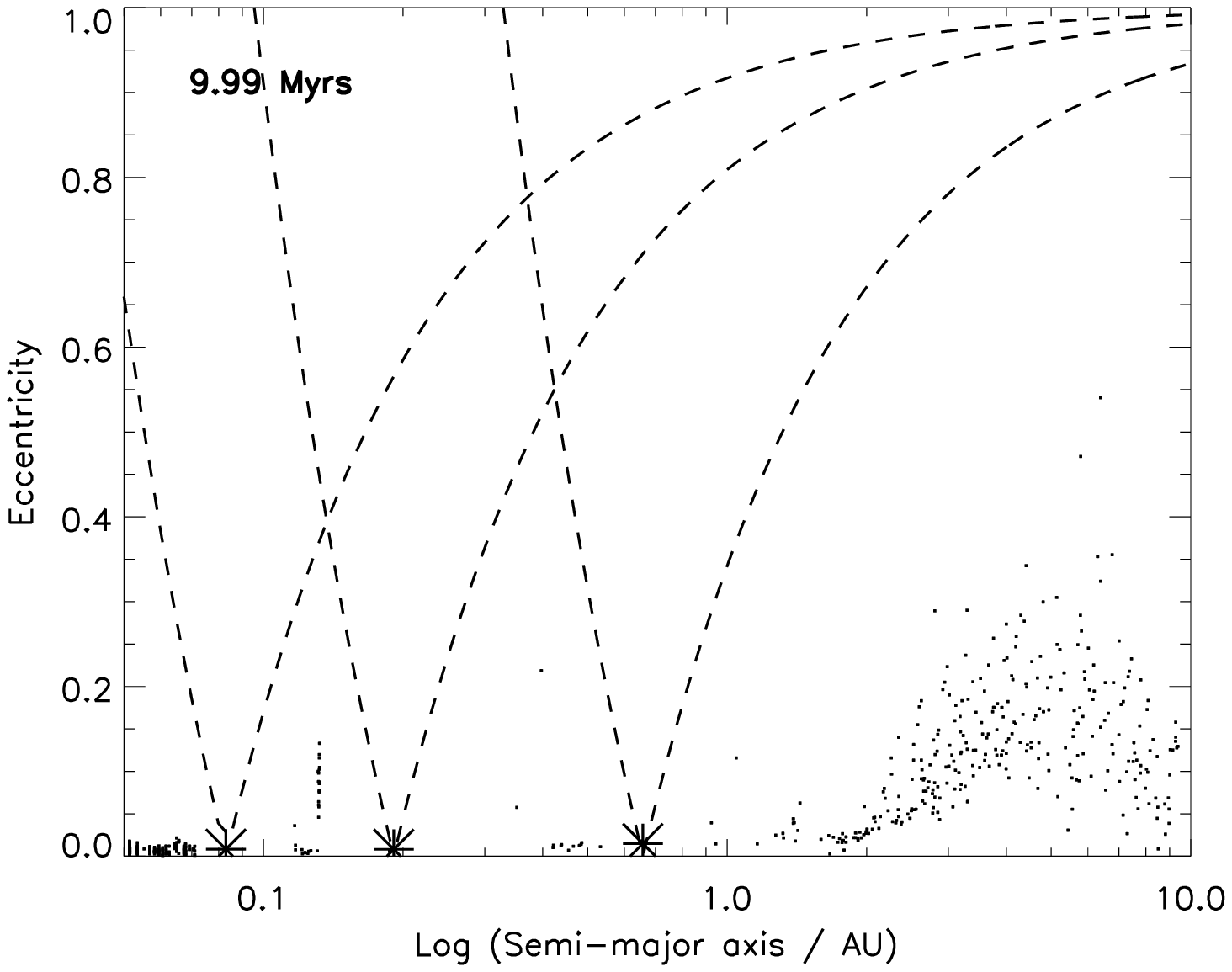,width=0.33\textwidth}
      \includegraphics[trim = 0mm 0mm 0mm 49mm, clip, width=0.33\textwidth]{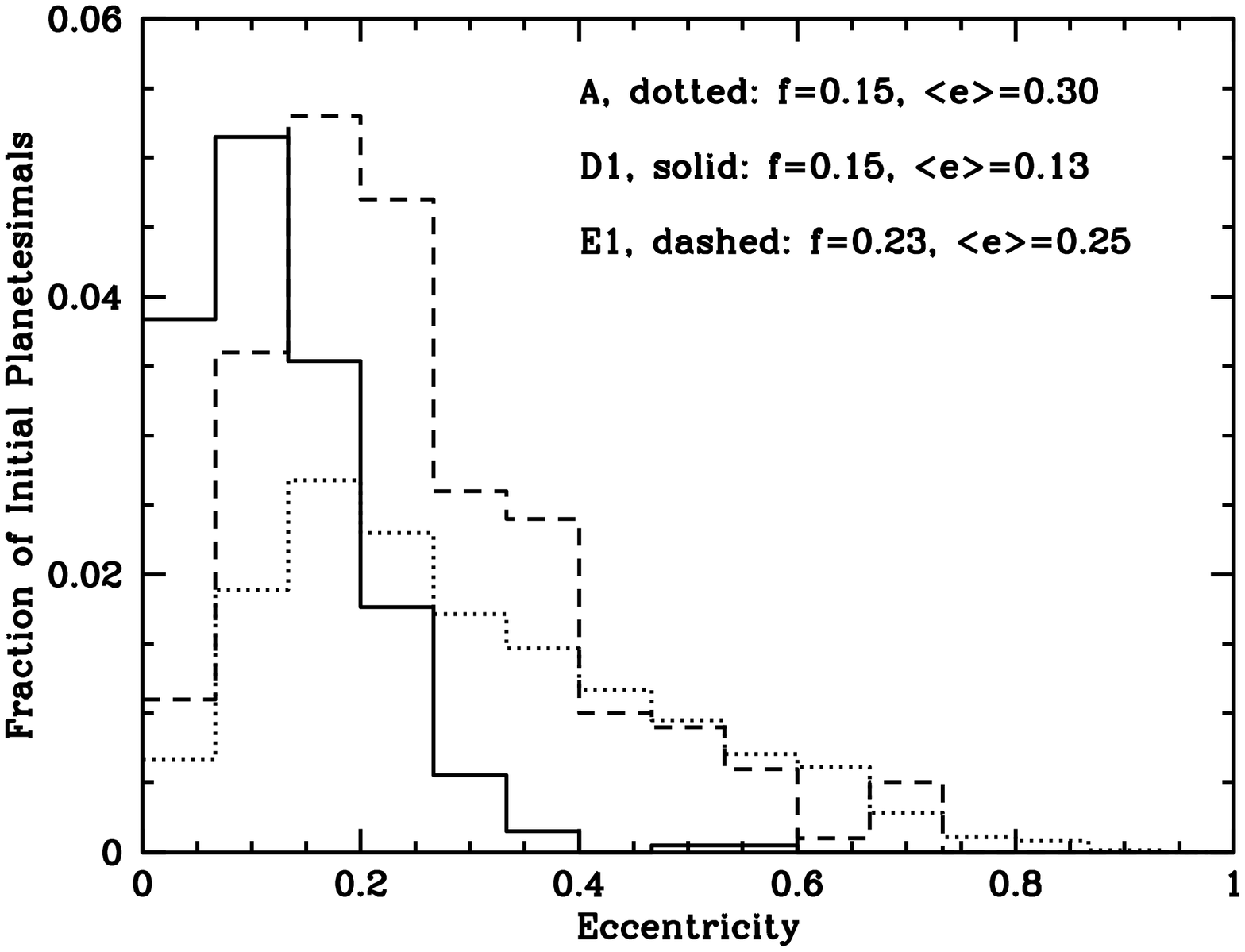}
      \includegraphics[trim = 0mm 0mm 0mm 49mm, clip, width=0.33\textwidth]{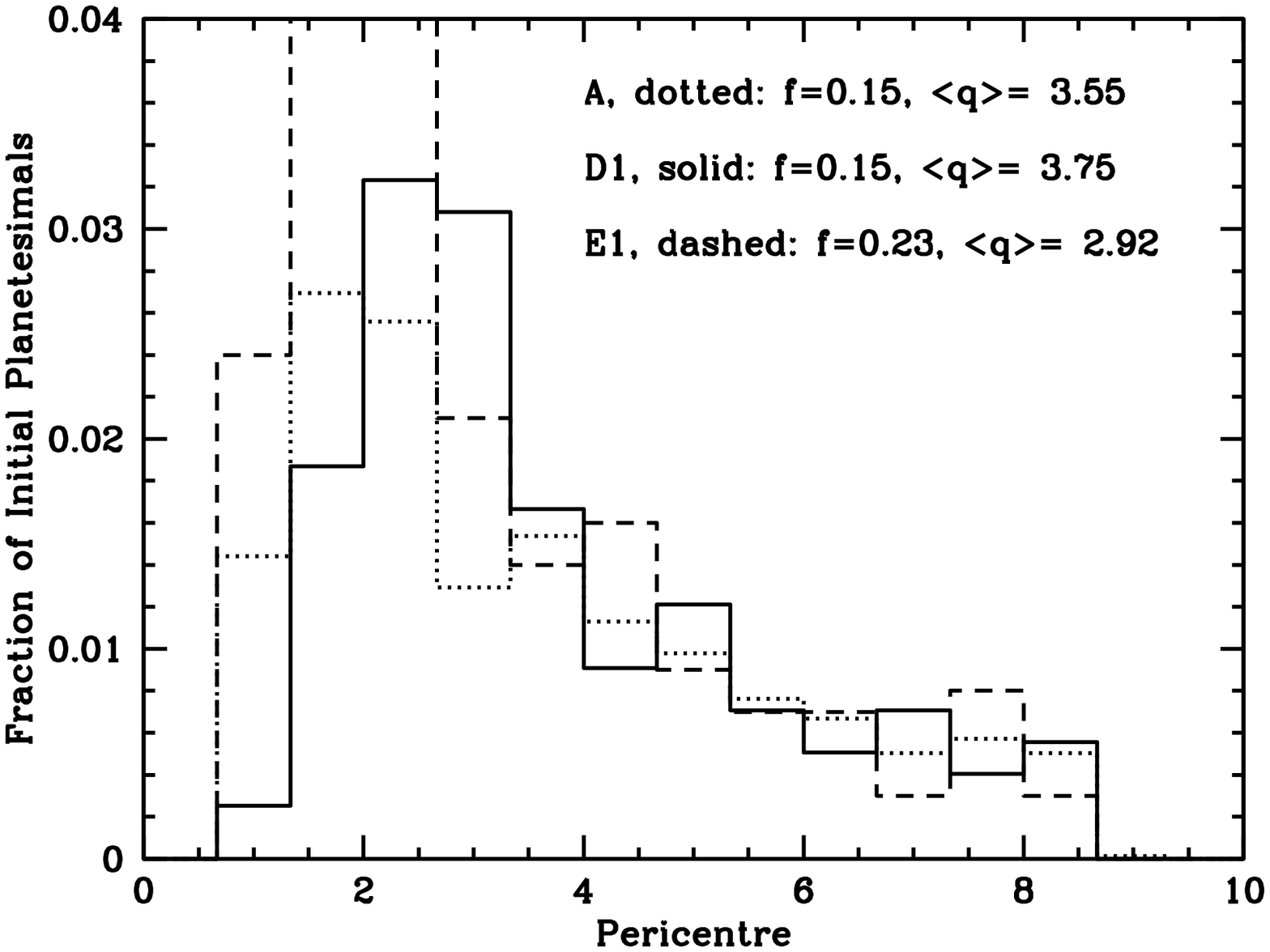}
    }
    \label{FIG:SIMD}
  }  
  \caption{
    Comparative results plots for selected simulations. 
%    \\ 
    Left Hand Column = Eccentricity-v-Semi-Major Axis Plots for Planets \& Planetesimals; 
%    \\
    Central Column = Final \emph{Eccentricity} Histograms for planetesimals (only showing planetesimals with $q<1$); 
%    \\
    Right Hand Column = Final \emph{Pericentre} Histograms for planetesimals (only showing planetesimals with $q<1$).
  }
  \label{FIG:SIMGRID}
\end{figure*}

We justify the neglect of the planetesimals with $q<1 AU$ by now considering the possibility that the planetesimals discarded by this approach (i.e. the planetesimals which survive within the system at 1 Myr with $q<1 AU$) could be scattered out to large semi-major axes / pericentres through interactions with the planets, thus enriching the extended eccentric disk. To investigate this scenario, we continue the integration of Set A from 1 Myr to 100 Myr, focussing solely on the 12\% of planetesimals which have $q<1$ at 1 Myr. We find that over the course of the subsequent 100 Myr integration, over half of these planetesimals collide with the central star, a third collide with the various planets and a negligible percentage are scattered out to the external disk. We summarise this data in the Table \ref{TABLE:Simulations} under the entry labelled A+, where we have assumed that the amount of material with $q> 1AU$ is fixed. Removing those planetesimals whose pericentres are inside 1.0 AU, leaves us with $15\%$ of the original planetesimal material occupying orbits with pericentres greater than 1.0 AU. Note that the initial proportion of bodies in this same region was approximately $21\%$, so this region is depleted only slightly from its initial value. However, the mean eccentricity has been greatly excited to $<e>=0.30$. See Fig \ref{FIG:SIMA} for the eccentricity and pericentre histograms. We note that the number of planetesimals scattered to semi-major axes larger than 10 AU is very small. 
 
We should note that the migration rate in the Alibert model is artificially reduced compared to the analytic estimates of \citet{Tanaka02}, a reduction of 1-2 orders of magnitude being necessary to explain observations of extrasolar giant planets (E.g. \citet{Alibert_et_al_05b,Daisaka06,IL5,Benz08}), and also suggested by various other results in which the migration behaviour is found to be more complex than that suggested by linear type-I theory. In particular, the results of \citet{Nelson04} suggest that migration can become stochastic as a result of disk turbulence. If such an instantaneously fast, stochastic migration was overlaid on a much longer timescale orbital decay, then the overall migration rate could be similar to that used in the Alibert model. However, the instantaneously fast stochastic migration would work to reduce shepherding, since the planetesimal trapping probability is a strong function of migration rate (with lower probabilities for faster migration rates, e.g., \citet{Wyatt03}), and stochastic variations could cause previously trapped planetesimals to fall out of resonance (e.g., \citet{MurrayClay06,Adams08,Rein08}).

%
%%%%%%%%%%%%%%%%%%%%%%%%%%%%%%%%%%%%%%%%%%%%%%%%%%%%%%%%%%%%%%%%%%%%%%%%%%%%%%%%%%%%%%%%
%
%
%%%%%%%%%%%%%%%%%%%%%%%%%%%%%%%%%%%%%%%%%%%%%%%%%%%%%%%%%%%%%%%%%%%%%%%%%%%%%%%%%%%%%%%%

\subsubsection{Surviving Planetesimals}\label{SurvEsimals}
The planetesimals which ``survive'' in the system are those planetesimals which are neither ejected from the system, nor suffer collisions with either the central star or one of the migrating planets (The planetesimals are treated as test particles so no planetesimal-planetesimal collisions take place). To gain some additional insight into the behaviour of the system, we look in detail at the planetesimals that do not survive to find out if and how their ultimate fate is dictated by their initial starting positions.

\begin{figure}
  \centering
  \psfig{figure=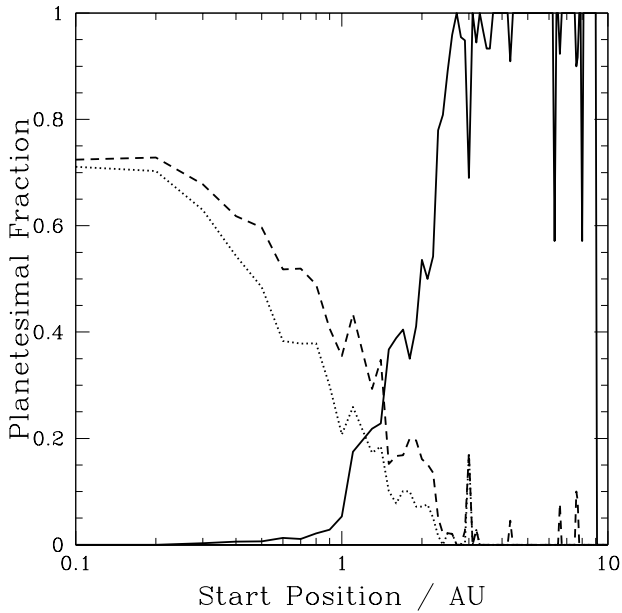,width=\columnwidth}
  \caption{Fate of planetesimals as functions of starting semi-major axes for: (i) Fraction of planetesimals which survive with $q>1.0$ - Solid Line; (ii) Fraction of all planetesimals which do \emph{not} survive - Dashed Line; (iii) Fraction of planetesimals which do not survive becasue they hit the inner planet - Dotted Line. Excludes any gas damping effects.}
  \label{FIG:Surv1}
\end{figure}

We plot the fate of the planetesimals as a function of their starting semi-major axis in Fig \ref{FIG:Surv1}. This makes it clear that the planetesimals which are located well within the initial orbit of the inner planet (3 AU) have a higher chance of being lost from the system ($\sim 70\%$ for the inner-most planetesimals), whilst those that are located outside 2AU are almost certain to survive as bound bodies within the system. In addition, we see that almost all of the planetesimals which start outside 2 AU will end up on an eccentric orbit with a pericenter beyond $\sim~1$AU.  Thus, these planetesimals could be ``useful'' for providing a parent population that explains the origin of the $3-35\mu m$ emission (contingent on their eccentricities and collisional lifetimes ). 

Next, we investigate in detail the fate of the individual planetesimals.  We find that out of the 10,000 planetesimals with which we start the simulation, 65\% hit the inner planet, 3\% hit the middle planet and 0.5\% hit the outer planet. A further 3\% were put into parabolic orbits such that they collided with the central star or were ejected from the system. We note from Fig 4(i) that the majority of the collisions with the inner planet are happening between 0.7 \& 0.8 Myr into the simulation, at which point the inner planet is effectively pushing into and colliding with the planetesimals in the MMRs ahead of it. To eliminate the possibility that these results are unduly influenced by the surface density profile that we have adopted in the simulations, we repeat the analysis for a set of simulations in which the surface density profile is distributed evenly between 0.5 and 9.5 AU and in which we try removing 1 or 2 of the outer planets. These simulations showed that the surface density profile does \emph{not} skew the results. We find a very similar ejection profile to that with a $\Sigma \propto a^{-3/2}$ profile, with the planetesimals which do not survive as free bodies within the system primarily suffering collisions with the inner planet.

\subsubsection{Accretion Rate}\label{AccRate}
\begin{figure}
  \centering
  \psfig{figure=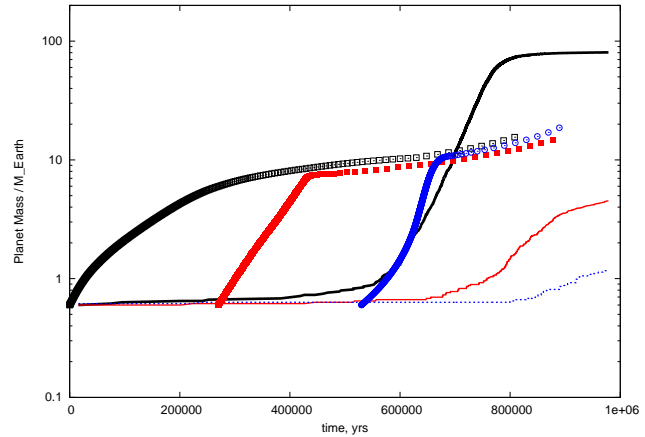,height=\columnwidth,angle=-90}
  \caption{Comparison of planetary growth rates inferred from standard accretion scenario (Set A, thin, solid lines) with those from the semi-analytic growth rates of \citet{Alibert_et_al_06} (Thick, dotted lines). Black plots - Inner Planet(b), Red plots - Middle Planet(c), Blue plots - Outer Planet(d).}
  \label{FIG:MassGrowth}
\end{figure}

Given that the solid disk in the Alibert model is massive, containing $\sim 0.5 M_J$ of solid material in the 0.1 - 9 AU region that we simulate,  an impact fraction of 65\% onto the inner planet would imply that the solid core of the planet would attain a mass of $\sim 0.3 M_J$ or $\sim 100 \Mearth$, far above both the observed minimum masses for the system and the $\sim 10 \Mearth $ cores in the Alibert model. We compare the growth rates for the planets as inferred from the planetesimal accretion rates with those found in the \citet{Alibert_et_al_06} model in Fig \ref{FIG:MassGrowth}.
We find that the inner planet, starting out at 3 AU would collide with many more planetesimals than the outer two planets, thus growing 10 - 100 times more massive, significantly at odds with the observed mass ratios.

The \citet{Alibert_et_al_05} model is a semi-analytic model in which the core accretion is taken to occur smoothly from an annular region extending out to 4 Hill radii either side of the planet, with the disk profile depleting self-consistently, but always maintaining a local surface density profile $\propto a^{-3/2}$. In contrast, our dynamical model makes no such assumptions, simply recording the number of collisions between the planetesimal population and the planets. Fundamentally, we would expect this to give a more accurate representation of the accretion rates, as long as no crucial physics have been omitted. 

To reconcile the results of our n-body simulations with the semi-analytic results would require that the relative number of planetesimal collisions onto each of the planets in the dynamical model be in approximately the same ratios as the core masses of the three planets in the growth model. Given the ubiquitous nature of the shepherding phenomena in n-body simulations, it seems highly unlikely that any simple semi-analytic model (such as that of \citet{Alibert_et_al_05}) which \emph{neglects} this fundamental rearrangement of the solid disk profile can ever be in agreement with n-body simulations, suggesting the necessity of making improvements to these semi-analytic models, in particular the treatment of shepherded low eccentricity planetesimals.

Finally, we note that the accretion rates in our model were calculated directly from the collision rate between the planets and planetesimals, and that this required that we make an assumption about the planetary density ($r \propto M^{1/3} \rho^{-1/3}$). We used a constant density of $1.5 g\ cm^{-2}$ throughout our entire calculation, corresponding approximately to that observed in the Solar-System ice-giants, initially suggesting that our accretion rates at the start of the simulation (when the body is a solid core) will be slightly too high, while at the end of the simulations they should be approximately correct. However, \citet{Fortier07} show that the effect of gas drag in the planetary envelope increases mass growth rates by up to a factor of 2 compared to their standard assumption of a pure solid core with density $3.2 g\ cm^{-2}$. I.e. they require a higher effective radius, or alternatively a \emph{lower} effective density, $\rho_{eff}=3.2\times\fracbrac{1}{2}^{3/2}g cm^{-2} \approx 1.1 g \ cm^{-2}$. Thus the initial growth rate in our simulations may be more accurate than initially implied. However, we emphasise that these slight uncertancies in the accretion rates would apply to \emph{all} planets in our simulations and would \emph{not} therefore help to reconcile the order of magnitude difference in accretion rates that we observe between the inner and outer planets.

\subsection{Planetesimal Eccentricity Excitation as a function of Planetary Mass \& Planetary Eccentricity}
\subsubsection{Planetary Mass}
We demonstrated in \S \ref{mass} that the mass of the planets can have a significant impact on the eccentricity that the planets excite amongst themselves as they migrate inwards. We now look at how the mass of the planets affects the properties of the scattered planetesimal disk. In simulation B, we look at a model in which the initial masses of the planets and their subsequent growth rates are 5 times those of the standard Alibert model.

In Fig \ref{FIG:SIMB} we can compare the distributions of bound planetesimals with pericentres greater than 1.0 AU for simulations A \& B (where the masses are respectively $1$ and $5\times$ those of the standard model). We find that as the mass of the planets is increased, the available fraction of planetesimals remaining bound remains static at $15\%$, but the eccentricity distribution is shifted, with the higher mass planets creating more higher eccentricity planetesimals at the expense of low eccentricity planetesimals, increasing the average eccentricity of this population from 0.30 to 0.47. The pericentre distribution is not as greatly affected, although there is again a slight increase in the number of high $q$ planetesimals at the expense of the number of low $q$ planetesimals. In addition we find that the range of initial semi-major axes which is ejected is \emph{not} changed by an increase in the planetary masses: again, only the planetesimals inside 2 AU are ejected/suffer collisions with any efficiency, whereas outside this region, almost all the planetesimals remain bound to the system as free bodies.

\subsubsection{Planetary Eccentricity}
We saw in \S \ref{DampPlanets} that gas damping significantly reduces the planetary eccentricities. When we look at the effect on the \emph{planetesimals} though, the difference made by damping the \emph{planetary} eccentricities is nowhere near as significant: we see in Fig \ref{FIG:SIMC} the results of simulation C1 in which the planets are damped (but the planetesimals remain undamped). We find that the the distribution of planetesimals is very similar to that found from simulation A. Overall this suggests that the eccentricity of the \emph{planets} has only a small influence on the final distribution of the \emph{planetesimals}, probably because the vast majority of the excitation is caused by the inward-shepherding and excitation, and this takes place undisturbed, irrespective of (modest) planetary eccentricity.

\subsection{Gas Damping of Planetesimal Eccentricity}\label{EsimalDamping}
We simulate gas damping on the planetesimal population using the model described in \S \ref{Method}. We again note that the start point of these simulations with the embryo mass at $0.6 \Mearth$ is taken to correspond to a ``system-time'' of 0.93Myr, i.e. the gas disk has already suffered a significant amount of dissipation. In addition we note that the dissipation of the disk and the growth of the planets is \emph{not} calculated self-consistently. 

The Alibert model assumes a disk lifetime of $2\times 10^6$ years, so we adopt this as our disk dissipation timescale. Since we adopt an exponential model, dissipation will result in a significant low density ``tail'' of gas remaining in the system and thus cause non-trivial damping for a few times this timescale. As such we run our simulations for $2\times 10^7$ years to allow the planetesimals to approach an equilibrium.

\subsubsection{Scattered Disk}
In simulation D1, we apply drag as if the planetesimals had a physical radius of 100km and a density of $1\textrm{g cm}^{-2}$, and assume that the planets have the standard minimum masses and are unaffected by gas drag. We find in Fig \ref{FIG:SIMD} that we still have around $15\%$ of the planetesimals remaining in the useful zone out beyond 1 AU but that this population now has the much lower average eccentricity of 0.13. At this stage in the simulation of the system's evolution, the gas density has decreased to $\sim 10^{-5}$ of its initial value, hence the possibility of any further eccentricity damping can essentially be neglected. In simulation D2 (not plotted) we repeat this gas drag simulation for planets with $5\times$ the standard Alibert masses, and now find that even with this significant gas drag present, the average eccentricities of the remaining planetesimals can have $<e> \sim 0.28$.

In simulations E1 \& E2, we repeat the gas drag simulations of D1 \& D2, but we now model the gas drag on the planetesimals as appropriate for bodies with a physical radius of 1,000km. From Eqn \ref{EQN:GDTS} we see that the gas drag timescale,
\begin{eqnarray}
\tau_{GD} & \propto & r_p,\nonumber
\end{eqnarray}
\begin{figure}
  \centering
  \psfig{figure=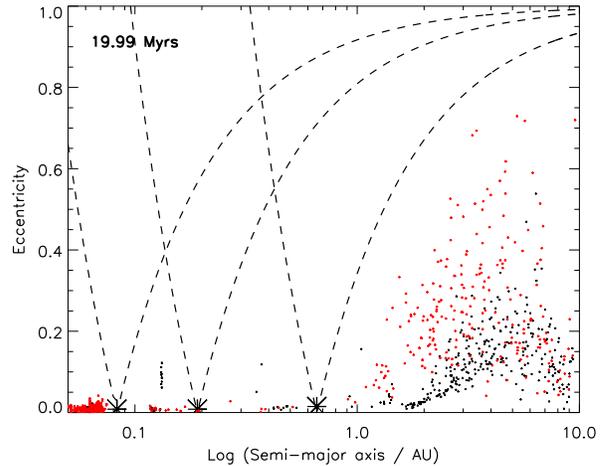,width=\columnwidth}
  \caption{Eccentricity vs. Semi-Major Axis plot after planetary migration for damped planetesimals: BLACK - 100km (Set D1); RED - 1,000km (Set E1).}
  \label{FIG:SIME}
\end{figure}
so the damping timescale for a 1,000km planetesimal will be over twice as long as for a 100km one. The overplotted results in Fig \ref{FIG:SIMD} and Fig \ref{FIG:SIME} show that these larger bodies, less affected by gas drag, have a mean eccentricity of $<e>=0.25$ after the migration of standard mass planets (E1). More massive planets (E2, not plotted) have $<e>=0.46$. So the much larger planetesimals of radius $\sim 1,000km$ will tend to occupy orbits of almost double the eccentricity of the smaller bodies of radius $\sim 100km$. 

We emphasise again the generic implication of the above: Any initially mixed population of planetesimals subject to gas-drag will become size-sorted over time, with the larger bodies tending to occupy the high eccentricity orbits, whilst the smaller bodies (D1 \& D2) will tend to populate more circular orbits.

In summary, we find that the eccentricity damping of planetesimals does \emph{not} act as an insurmountable obstacle to the existence of an excited eccentric disk with pericentres $\geq 1$ AU in the HD69830 system. The amount of available material with pericentre $> 1$ AU remains essentially unchanged at 15\%, but the average eccentricity is reduced to 0.13. This still means that we have more than 10\% of the solid disk mass available in an excited form to potentially act as the source of the $\mu m$ emission. Further work will be required to understand whether the observed distribution can survive collisional processing for the $>2$ Gyr age of the system (see \S\ref{Problems} for further discussion).

\subsubsection{Accretion Rates}
In \S\ref{SurvEsimals} we saw that the collision rate of planetesimals onto the inner planet was far too high to be compatible with the semi-analytic model of \citet{Alibert_et_al_06}. In \S \ref{SurvEsimals} we saw collision rates for the three planets of 65, 3 \& 0.5 \% respectively, whereas with gas damping acting on the planetesimals the problem was somewhat ameliorated: collision rates reduced to 3, 1 and 0.2 \% respectively.

The difference is essentially due to the fact that in the damped case, all of the planetesimals in the MMRs inside of the inner planet are damped to $e\approx 0$, meaning that their orbits do not overlap with that of the inner planet and hence collisions rarely occur. However, even with gas damping on planetesimals, the number of planetesimal collisions onto the inner planet is still an order of magnitude greater than onto the outer planet.
\begin{figure}
  \centering
  \psfig{figure=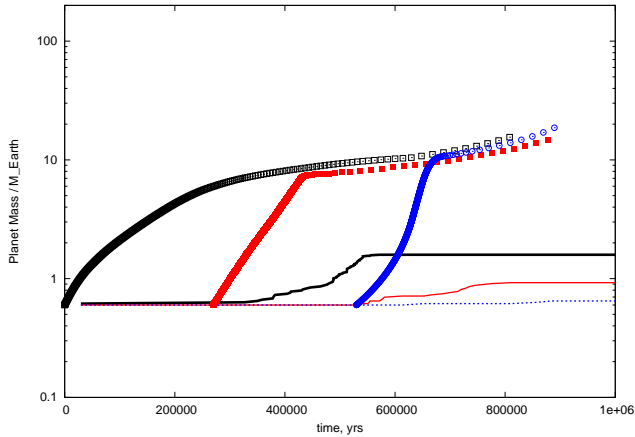,height=\columnwidth,angle=-90}
  \caption{Comparison of planetary growth rates in gas-drag simulations (Set D1, thin, solid lines) with those from the semi-analytic growth rates of \citet{Alibert_et_al_06} (Thick, dotted lines). Black plots - Inner Planet(b), Red plots - Middle Planet(c), Blue plots - Outer Planet(d).}
  \label{FIG:MassGrowth2}
\end{figure}

The growth profile plotted in Fig \ref{FIG:MassGrowth2} makes it clear that we still have an order-of-magnitude difference in planetary masses (accretion rates), but now \emph{all} of the planets are too low in mass (core masses $\sim 10\Mearth$ would require accretion of $\sim 6\%$ of the solid disk).

\section{Outstanding Problems and Potential Solutions}\label{Problems}
The work in sections \ref{PlanetExcite} and \ref{PlanetesimalExcite} has highlighted a number of results and outstanding problems. These include, (i) Lack of excitation and possible over-damping of the planetary eccentricities, (ii) The incompatibility of the collision rates with the core masses from \citet{Alibert_et_al_05} and (iii) The existence of relatively massive and moderately eccentric planetesimal disks. The first two indicate that the \citet{Alibert_et_al_05} model will need tweaking to provide a self-consistent formation model. The third provides significant grounds for optimism in producing a parent population of eccentric planetesimals, but as we do not yet know whether the degree of eccentricity found is sufficient to extend the collisional lifetime of the system past 2Gyr, we wish to take this opportunity to understand whether further eccentricity excitation can be fostered. Below we consider three ways in which the model may be tweaked and consider its implications for planetary eccentricities, planetary accretion rates and planetesimal eccentricity excitation.

\subsection{Additional Embryos}\label{FourEmbryos}
The large initial separation of the three embryos considered thus far and their relatively low masses would suggest that a nascent system in this configuration would have a large number of similar size embryos distributed between them. Whilst it is possible to conceive of this intervening material being accreted in an adiabatic manner, it also seems reasonable to suggest that these ``additional'' bodies could be growing and migrating in parallel with the three bodies considered thus far, opening the door to catastrophic impact and / or ejection scenarios. We briefly investigate this scenario by conducting 100 simulations in which an additional planetary embryo of mass $0.6\Mearth$ is initially randomly placed between 4.1 and 5.2 AU (i.e. in the middle third of the gap between HD69830-b and -c). The standard \citet{Alibert_et_al_06} semi-major axis evolution is applied to the 3 standard planets, whilst a scaled Type-I migration rate ($10\times$ slower than analytic calculations suggest) is applied to the fourth planet. This gives the additional planet a migration rate intermediate between that of HD69830-b \& -c. Gas damping is \emph{not} applied to the planets.

We typically find (in $81\%$ of simulations - See Table \ref{TABLE:AdditionalEmbryos}) that the fourth planet collides with one of the other planets in the system. This tends to result in rather excited systems with mean eccentricity $> 0.1$, rather than the under-excitement that we have seen previously. Reducing the mass of the added planet would, of course, reduce the level of additional excitation given to the remaining three planets.

\begin{table}
\setlength{\tabcolsep}{4.5pt}
\caption{Results of Simulations with an Additional Embryo Present}
\label{TABLE:AdditionalEmbryos}
\begin{tabular}{lccc}
\hline
        & Percentage  & Average      &               \\
        & of          &  Planetary   & Planetesimal  \\
Outcome & Simulations & Eccentricity & Simulations?  \\
\hline
Collision with (b)            & 10 & 0.17 & No       \\ \hline \\
Collision with (c)            & 61 & 0.11 & F(i)     \\ \hline \\ %8680-8689
Collision with (d)            & 10 & 0.27 & F(ii)    \\ \hline \\ %8690-8699
Scatter to Outer System       &  4 & 0.23 & F(iii)   \\ \hline \\ %8650-8669
Ejection from System          &  9 & 0.19 & F(iv)    \\ \hline \\ %8670-8679
Other (Multiple Collisions)   &  6 & 0.45 & No       \\
\hline
\end{tabular}
\end{table}

\begin{figure*}
  \vspace{-0.5cm}
  \subfigure[tight][Simulation F(i): Additional Planet, Collides with HD69830(c)]{
    \centerline{
      \psfig{figure=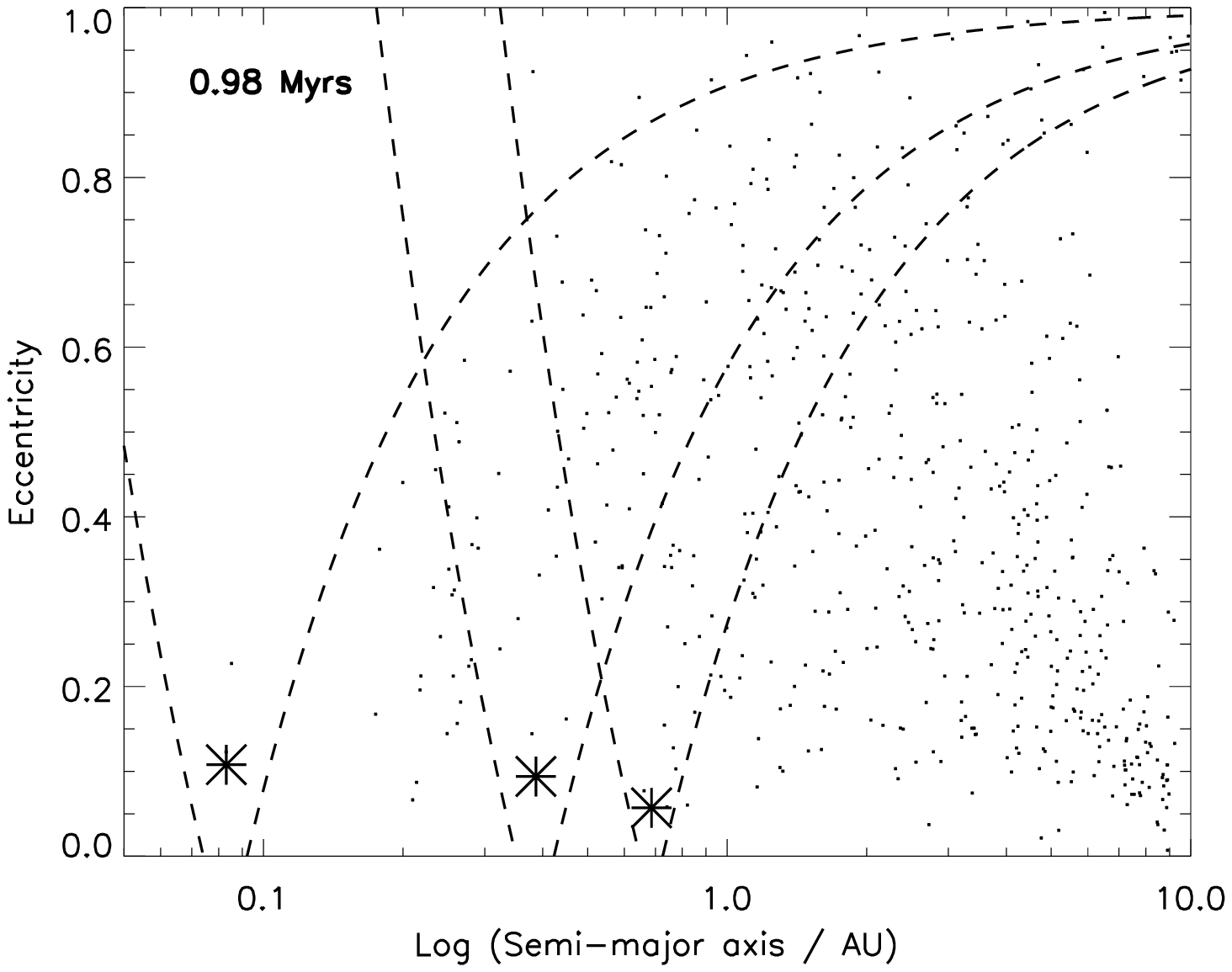,width=0.33\textwidth}
      \includegraphics[trim = 0mm 0mm 0mm 49mm, clip, width=0.33\textwidth]{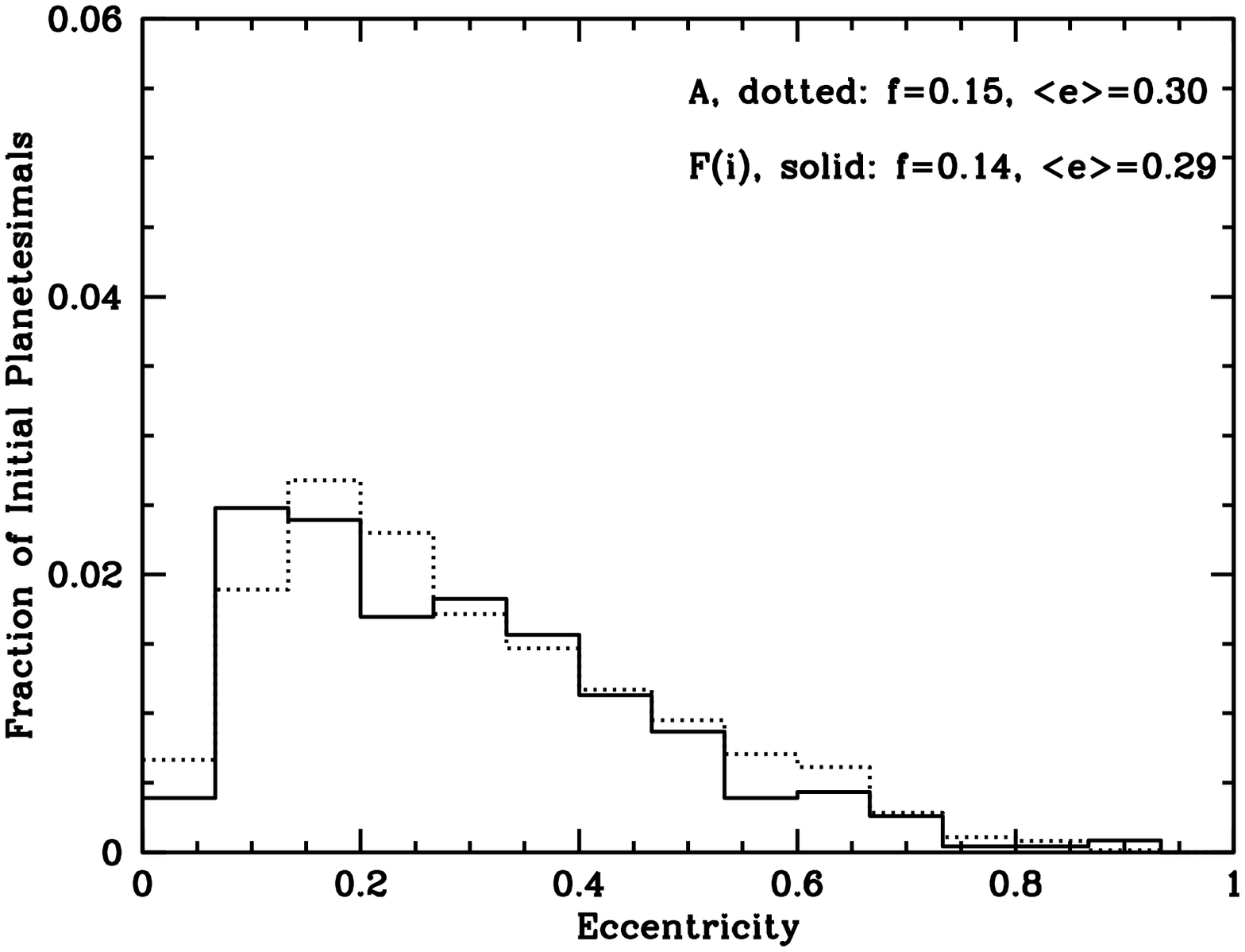}
      \includegraphics[trim = 0mm 0mm 0mm 49mm, clip, width=0.33\textwidth]{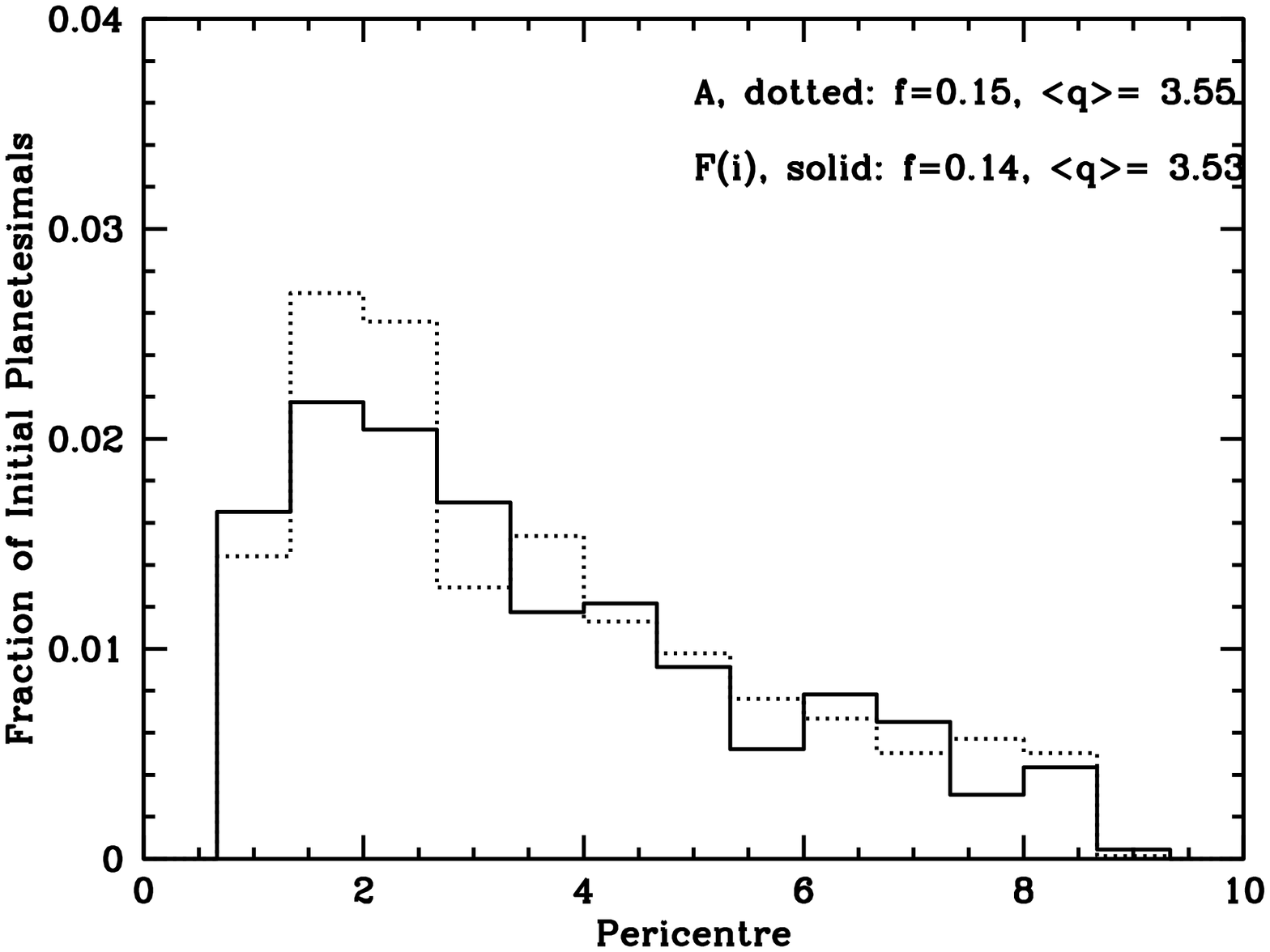}
    }
    \label{FIG:SIMFi}
  }
  \vspace{-0.5cm}
  \subfigure[tight][Simulation F(iii): Additional Planet, Scattered to Outer System]{
    \centerline{
      \psfig{figure=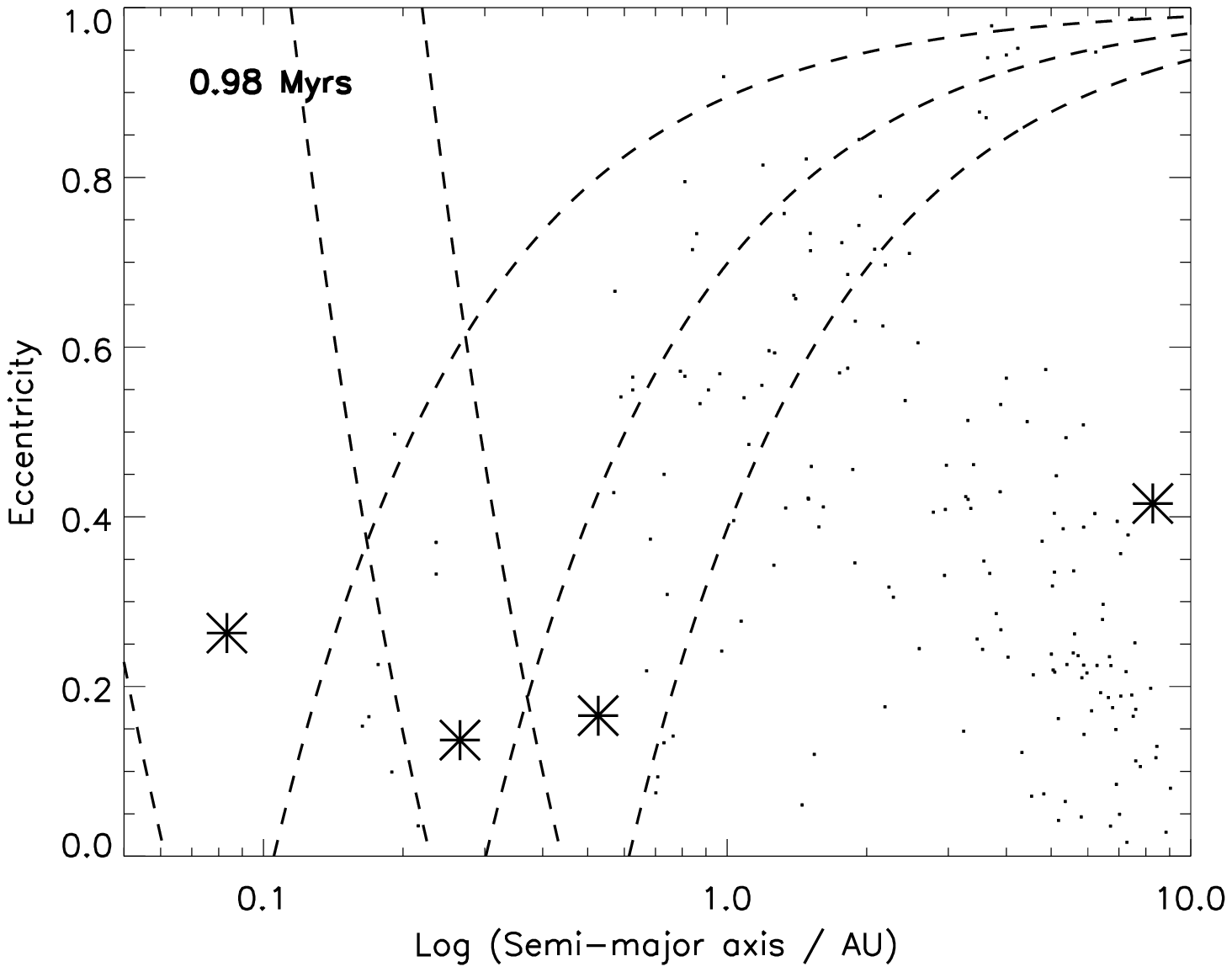,width=0.33\textwidth}
      \includegraphics[trim = 0mm 0mm 0mm 49mm, clip, width=0.33\textwidth]{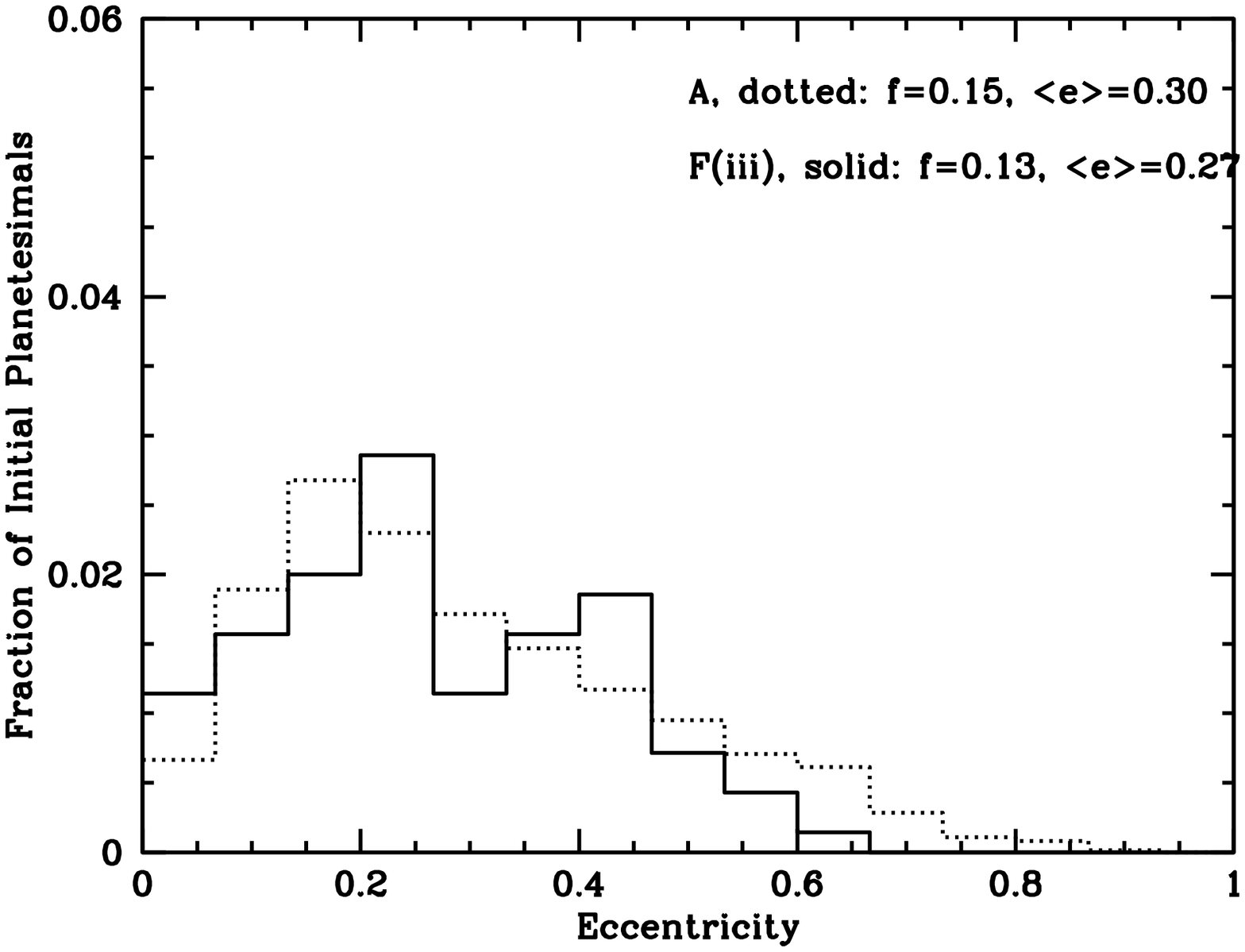}
      \includegraphics[trim = 0mm 0mm 0mm 49mm, clip, width=0.33\textwidth]{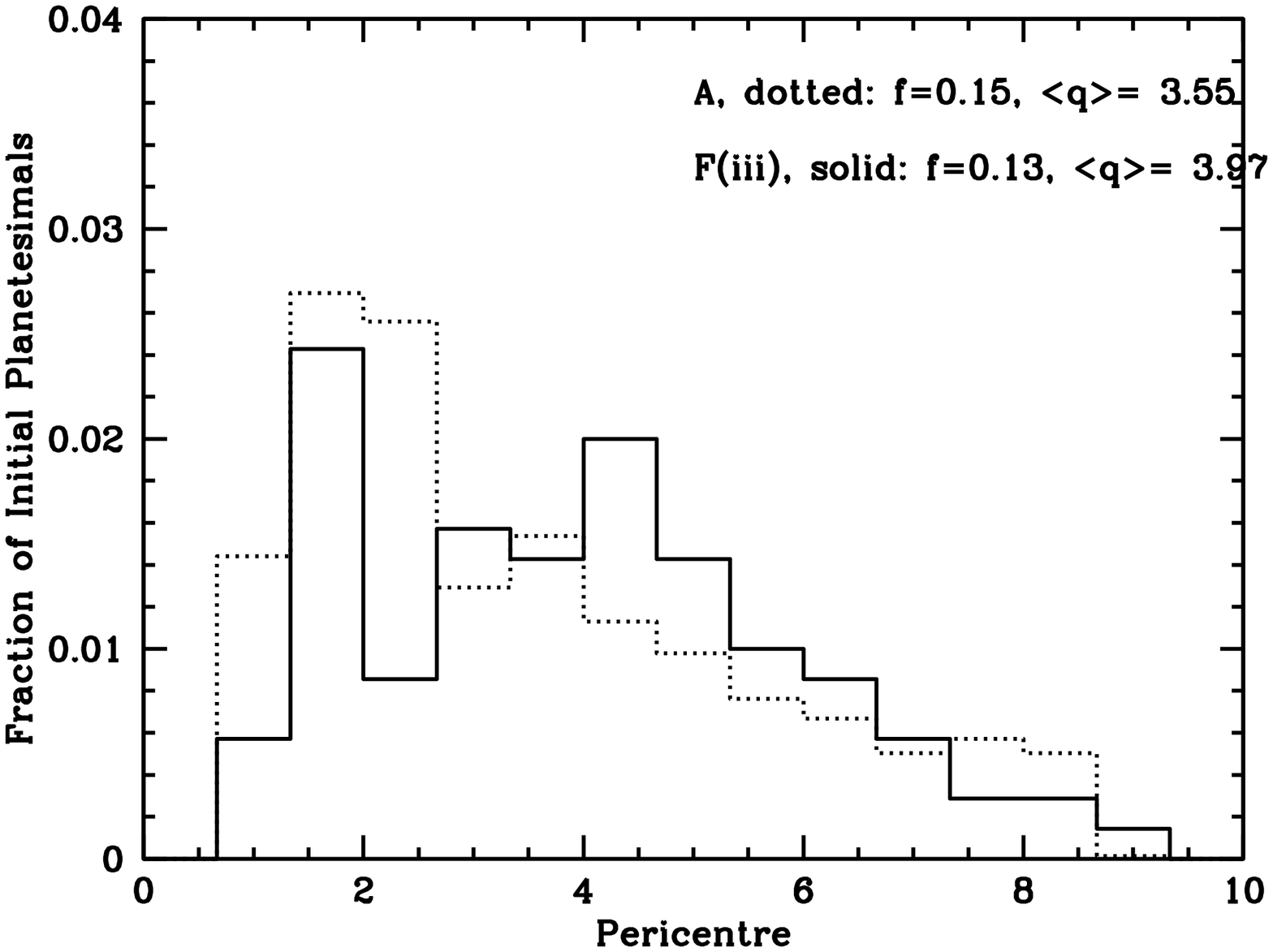}
    }
    \label{FIG:SIMFiii}
  }
  \vspace{-0.5cm}
  \subfigure[tight][Simulation G: Migration After Growth]{
    \centerline{
      \psfig{figure=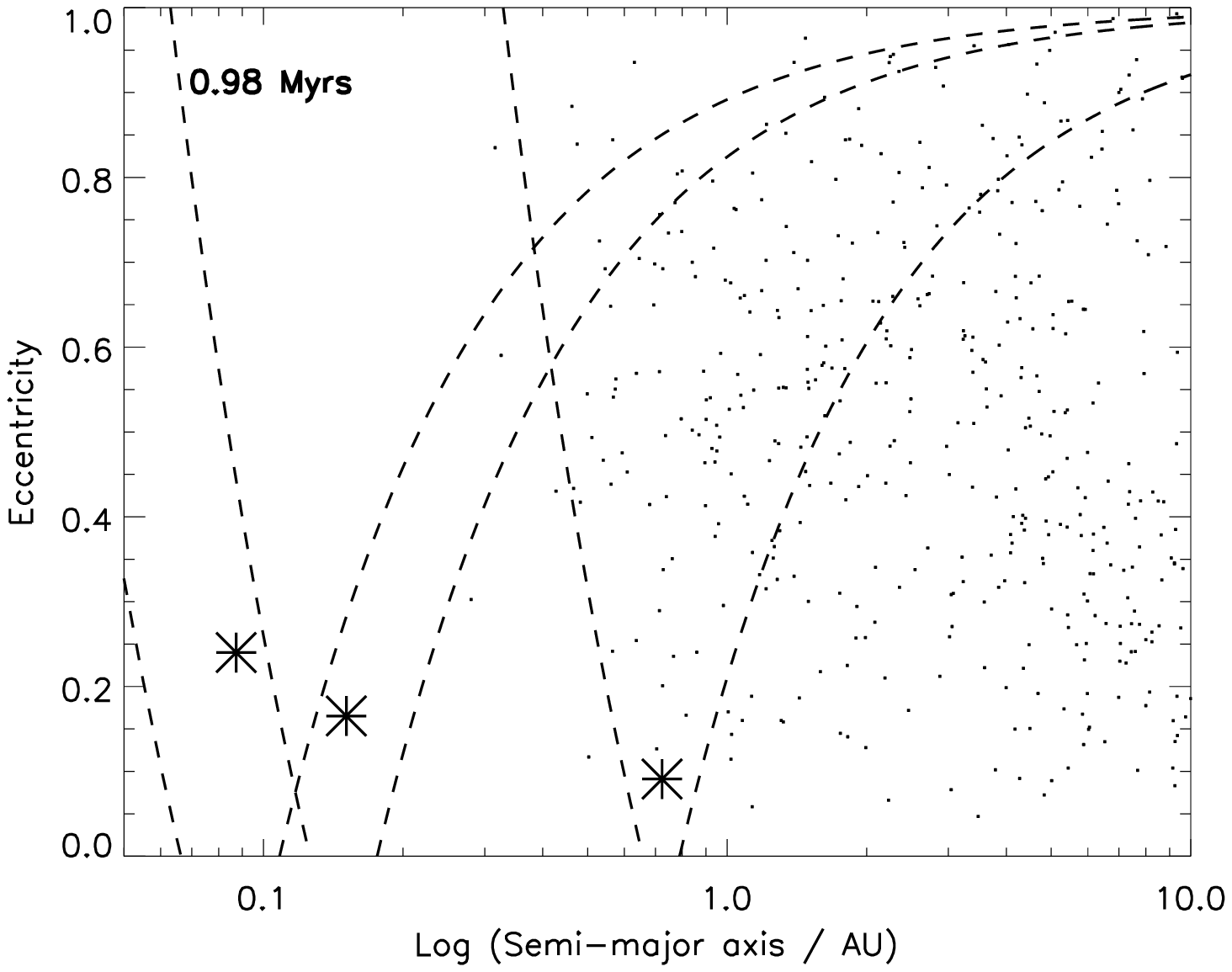,width=0.33\textwidth}
      \includegraphics[trim = 0mm 0mm 0mm 49mm, clip, width=0.33\textwidth]{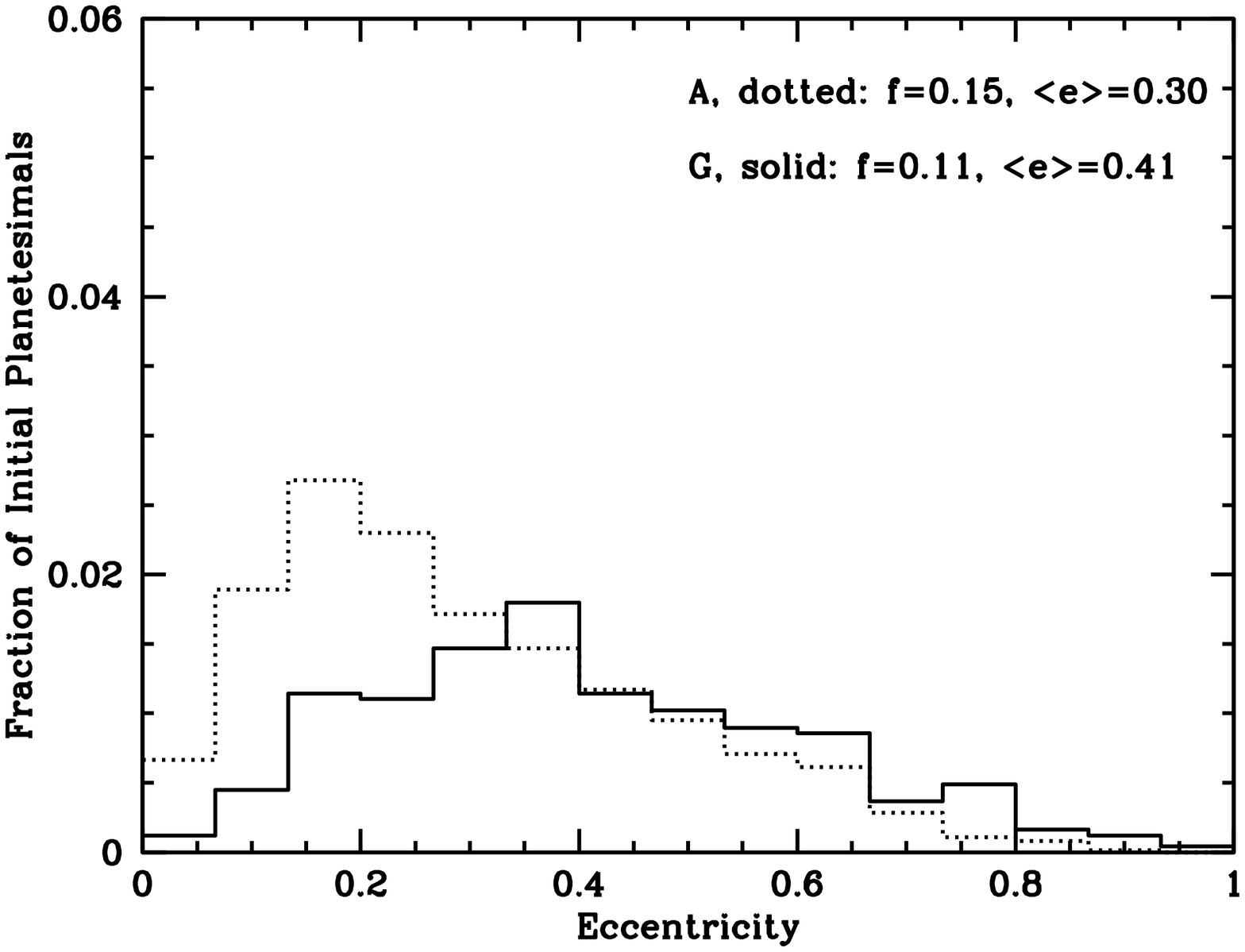}
      \includegraphics[trim = 0mm 0mm 0mm 49mm, clip, width=0.33\textwidth]{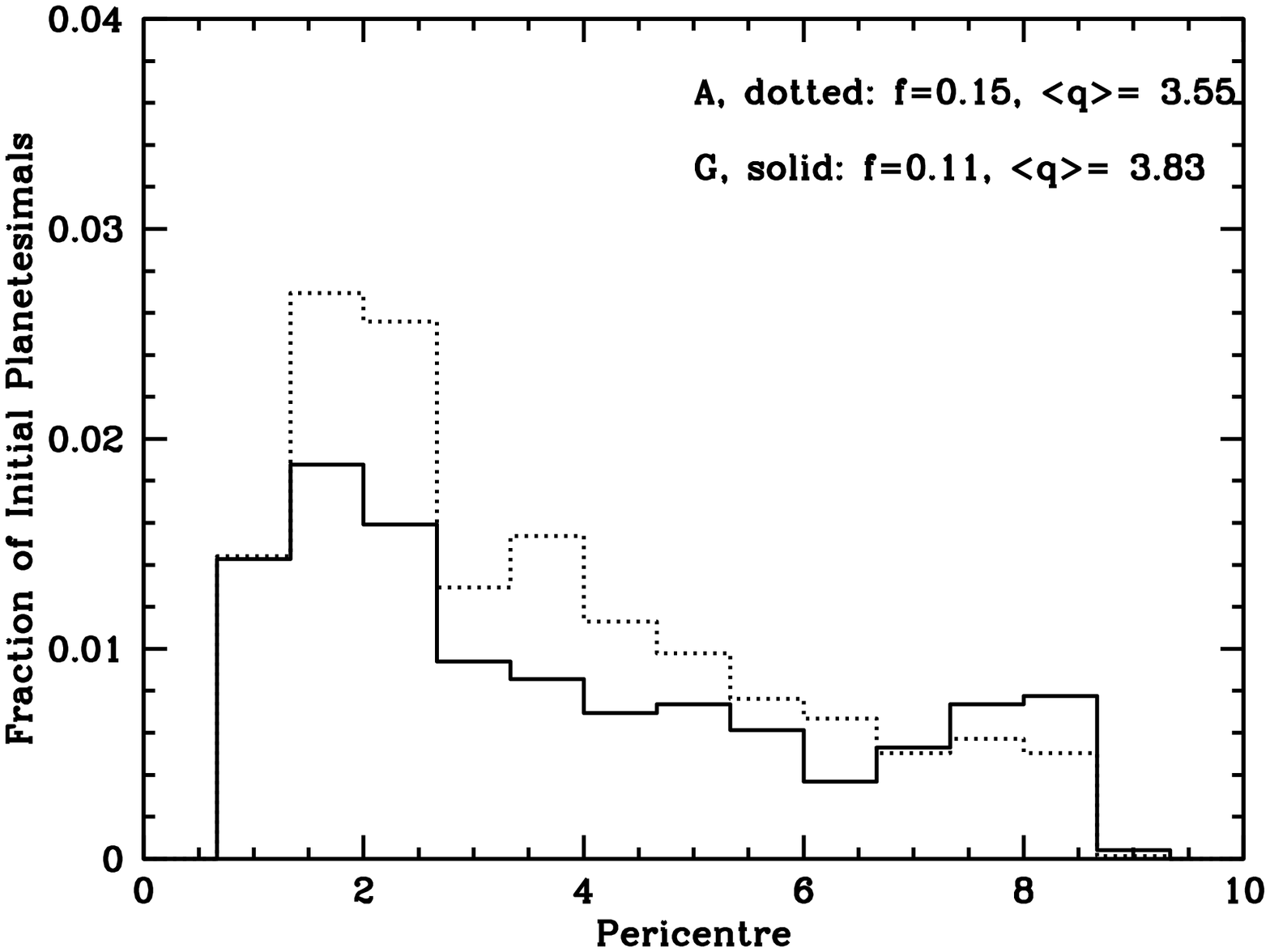}
    }
    \label{FIG:SIMG}
  }
    \vspace{-0.5cm}
  \subfigure[tight][Simulation H: Comparison of simulation H (Gas Drag in operation + Initial Semi-Major Axes altered to 0.5, 1.0 \& 6.0 AU) with the standard simulation, A, and the Gas Drag simulation D1.]{
    \centerline{
      \psfig{figure=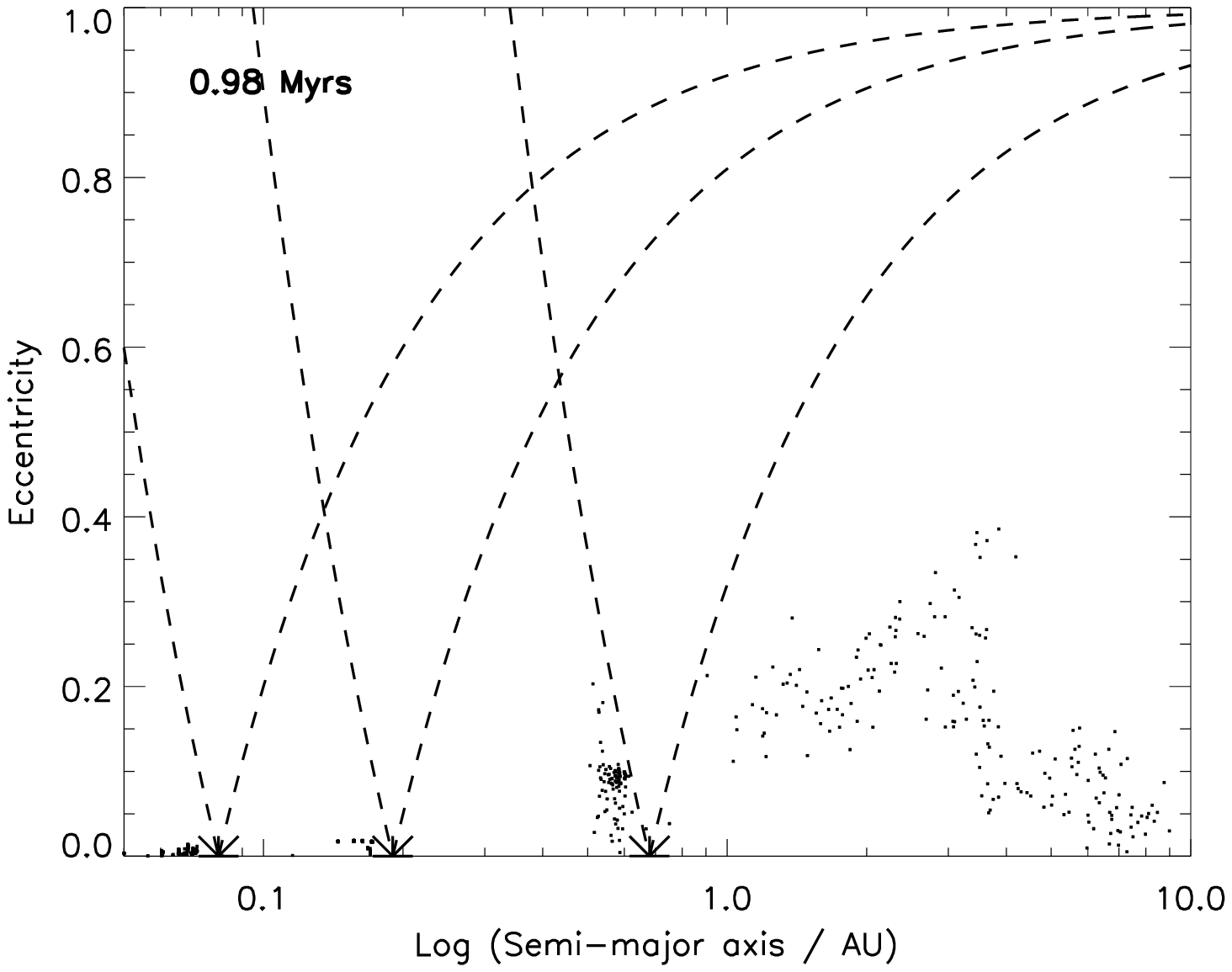,width=0.33\textwidth}
      \includegraphics[trim = 0mm 0mm 0mm 49mm, clip, width=0.33\textwidth]{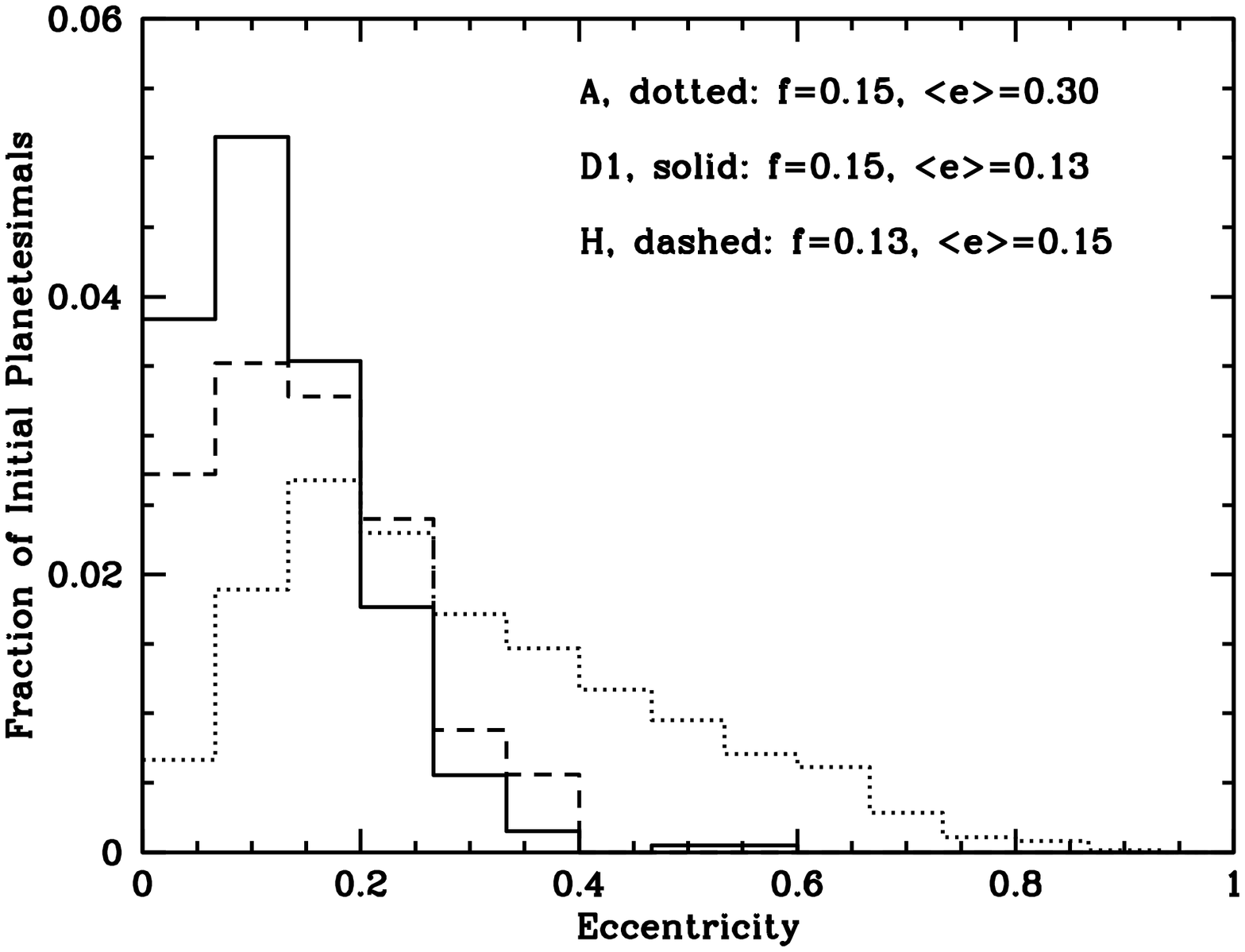}
      \includegraphics[trim = 0mm 0mm 0mm 49mm, clip, width=0.33\textwidth]{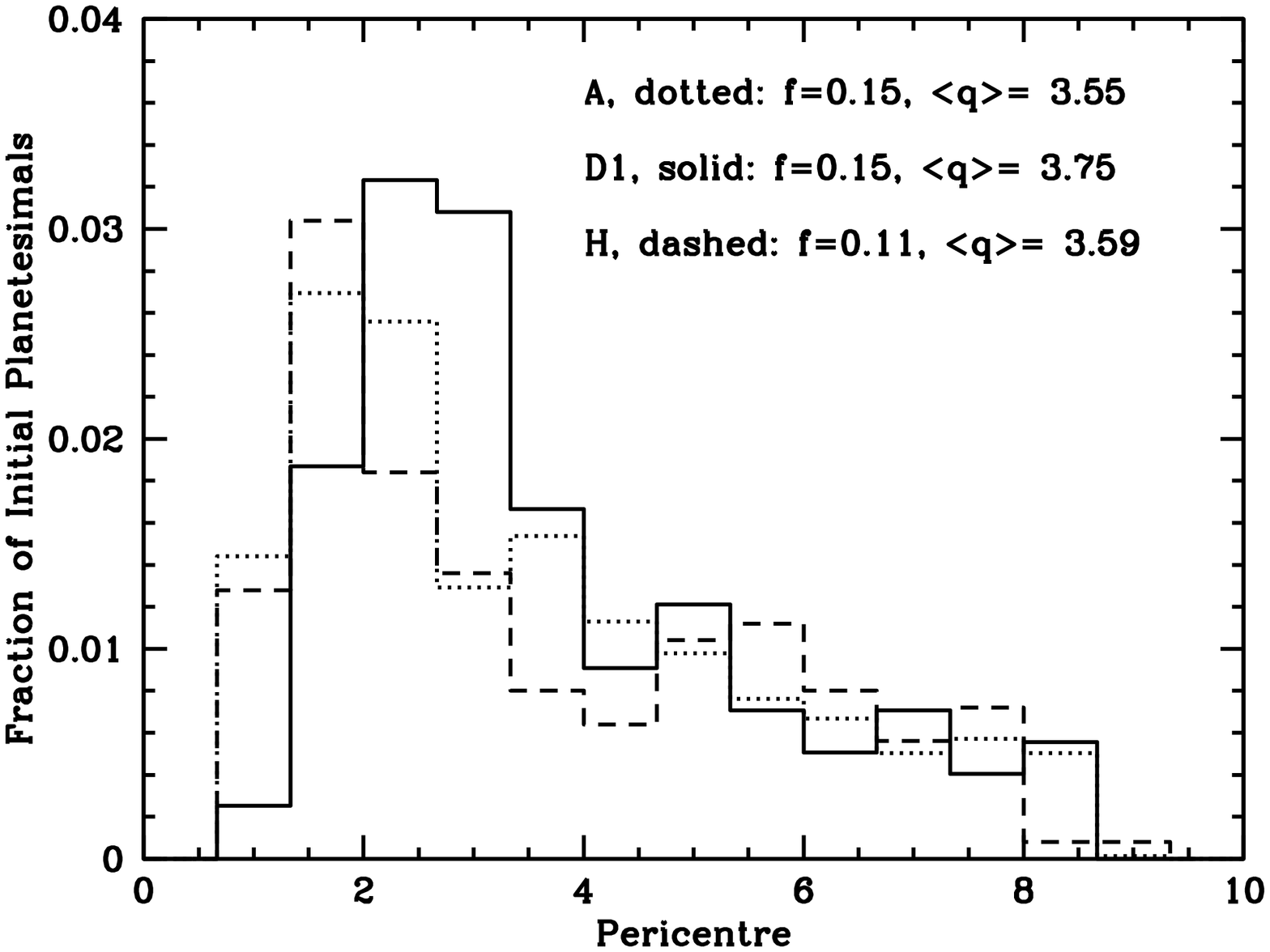}
    }
    \label{FIG:SIMH}
  }
  \caption{
    Comparative results plots for selected simulations. 
%    \\ 
    Left Hand Column = Eccentricity-v-Semi-Major Axis Plots for Planets \& Planetesimals; 
%    \\
    Central Column = Final \emph{Eccentricity} Histograms for planetesimals (only showing planetesimals with $q<1$); 
%    \\
    Right Hand Column = Final \emph{Pericentre} Histograms for planetesimals (only showing planetesimals with $q<1$).
  }
  \label{FIG:SIMGRIDNEW}
\end{figure*}

For a limited number of the simulations, we perform further detailed simulations to calculate the expected planetesimal distribution resulting from the addition of an additional embryo. These additional simulations, referred to as F(i)-F(iv), are detailed in Table \ref{TABLE:Simulations} and a summary of the results is plotted in Figure \ref{FIG:SIMGRIDNEW}.

Irrespective of whether the additional planet collided with one of the other planets (simulations F(i) \& F(ii)), is scattered to the outer system (F(iii)) or is ejected from the system (F(iv)), the population of scattered planetesimals with $q>1$ is very similar to that observed in the standard simulation (set A), having very similar population fractions and very similar eccentricity excitations. N.B. There are fewer particles in Sim F(iii) than in F(i), hence the difference in appearances of plots \ref{FIG:SIMFi} and \ref{FIG:SIMFiii}, despite both having similar fractions of retained particles with $q>1$. The only significant difference appears to be the slight reduction in the number of planetesimals accreting onto the inner planet, the difference essentially being made up by an increase in the number of planetesimals ejected onto parabolic orbits. However, the accretion rate onto the inner planet is still an order of magnitude larger than that onto the other two planets.

If we repeat the insertion of an additional planet, but now apply gas damping to the planets, then the situation becomes very different. We now find that in 98\% of cases the additional planet remains securely trapped between the two planets it starts between, migrating in between them, with any mutual eccentricity excitation being quickly damped away (average planetary eccentricities drop to $<e>\sim 0.0005$). Whilst this scenario clearly will do little or nothing \emph{directly} to stir or excite the outer population of planetesimals, it may provide a simple way to initially trap planetary embryos in the inner system before subsequent perturbations eject them into the outer system, potentially providing a trigger for an LHB-type scenario. However, without long term simulations this remains speculative, especially given that the low planetary eccentricities ($\sim 10^{-3}$) may preclude any significant secular evolution to drive an instability.

\subsection{Migration \emph{after} Growth}\label{MAG}
If the planets carry out the majority of their mass growth whilst they are stationary, then the eccentricity that they self-excite as they migrate may be significant, perhaps even more than that found in section \ref{mass}, as a significant fraction of the eccentricity growth occurs as a result of MMR crossing, and as such is dependent on planetary mass at the time of resonance crossing. To investigate this in simulation set G, we take the extreme case of the planets growing to full mass at their initial locations and then migrating through the planetesimal disk to their final location. 

We find in Fig \ref{FIG:SIMGRIDNEW} and Table \ref{TABLE:Simulations} that the distribution of the remaining planetesimals is such that the percentage of planetesimals with pericentres $>1$ AU is now somewhat reduced compared to set A (11\% versus 15\%), but the average eccentricity is now significantly higher (0.4 versus 0.25). Thus, as might be expected, the results are somewhat similar to those simulations with increased mass and growth rates (E.g. set B, D2 \& E2) in that the average eccentricity has been increased, but this time we also have more efficient clearing of the disk external to 1 AU.

Finally, we note that the high initial masses in this model serve to excite significant eccentricities as the planets migrate, resulting in average eccentricities (over 100 simulations) of 0.24, 0.17 \& 0.09 for the inner, middle and outer planets respectively.

\subsection{Different Initial Semi-Major Axes}
Given that the inner planet suffers a much higher collision rate than the other two planets, we consider a scenario in which the inner planet starts closer to the central star. This means that it migrates through a much smaller portion of the disk and thus has the potential to interact and collide with a smaller number of planetesimals. In addition, a greater number of planetesimals remain in the unperturbed region \emph{exterior} to the inner planet, thus increasing the chance of impacts occurring with the two outer planets. We keep the migration timescale the same as well as $M(t)$, but alter $a(t)$ such that the planets migrate inwards at a constant rate, ${\dot a}(t) = \kappa$, arriving at their final semi-major axes after $0.9Myr$.

When the effects of gas drag on 100km planetesimals are \emph{excluded}, rearranging the initial semi major axes of the planets does \emph{not} have any significant effect, unless the inner planet is started from exceedingly small semi-major axes (due to the power-law nature of the planetesimal disk). When gas drag \emph{is} included, then conducting a variety of simulations in which the initial positions of the three planets are varied gives the results shown in Table \ref{TABLE:SEMIS}. The first three columns give the initial semi-major axes, the next three columns give the resultant accretion rates onto the planets whilst the final column gives the overall average planetary eccentricity. For reference, the top line includes the results from \S \ref{DampPlanets}, where gas drag was included on the standard distribution of planets. The most promising results (with regards to the accretion rates) from the additional simulations seem to occur when the inner two planets (b \& c) are started close together and a relatively small semi-major axes, whilst the outer planet starts much further out: see for example the last line of Table \ref{TABLE:SEMIS}, in which the planets start at 0.5, 1.0 \& 6.0 AU respectively and the subsequent fractions of planetesimals accreting onto the planets are 0.6, 0.8 \& 1.0 respectively.

We note in passing that the eccentricities of the planets in these models (right hand column of Table \ref{TABLE:SEMIS}) are rather harder to tie in to the observed values. The final row of the table (which has the most even distribution of planetesimal collisions) has planetary eccentricities which are around four orders of magnitude lower than the mean observed values. Clearly one of the other models, in which the outer two planets start closer together and hence pass through a significant MMR and thus generate eccentricities of $\sim 0.1$ would be favoured if future RV measurements do confirm the current eccentricities.

\begin{table}
\setlength{\tabcolsep}{4.5pt}
\caption{Details of simulation outcomes for various different initial planetary semi-major axes}
\label{TABLE:SEMIS}
\begin{tabular}{ccccccc}
\hline
\multicolumn{3}{|c|}{Initial Semi-Major Axis} & 
\multicolumn{3}{|c|}{Accretion Rate onto Planet} &
\multicolumn{1}{|c|}{Average } \\
\multicolumn{3}{|c|}{(AU)} & 
\multicolumn{3}{|c|}{(\%)} &
\multicolumn{1}{|c|}{ Planetary } \\
b   & c   & d   &   b   & c   & d   &  Eccentricity \\
\hline
3.0 & 6.3 & 8.0 &   2.7 & 1.0 & 0.2 &   0.001  \\
\hline
3.0 & 3.5 & 4.0 &   1.2 & 0.2 & 0.7 &   0.22   \\ %7250-7299
5.0 & 6.0 & 7.0 &   1.2 & 0.2 & 0.2 &   0.23   \\ %7300-7359
1.0 & 6.3 & 8.0 &   0.6 & 1.6 & 0.2 &   0.09   \\ %7350-7399
0.5 & 6.3 & 8.0 &   3.0 & 2.3 & 0.4 &   0.09   \\ %7400-7459
1.0 & 2.0 & 8.0 &   0.5 & 1.1 & 1.3 &   0.0002 \\ %7450-7499
0.5 & 2.0 & 8.0 &   0.9 & 1.1 & 0.3 &   0.0001 \\ %7500-7559
1.0 & 4.0 & 6.0 &   0.2 & 2.0 & 0.3 &   0.05   \\ %7550-7599
0.5 & 1.0 & 6.0 &   0.6 & 0.8 & 1.0 &   0.0002 \\ %7600-7659 This is also Simulation H
\hline
\end{tabular}
\end{table}

%
%%%%%%%%%%%%%%%%%%%%%%%%%%%%%%%%%%%%%%%%%%%%%%%%%%%%%%%%%%%%%%%%%%%%%%%%%%%%%%%%%%%%%%%%
%
%
%%%%%%%%%%%%%%%%%%%%%%%%%%%%%%%%%%%%%%%%%%%%%%%%%%%%%%%%%%%%%%%%%%%%%%%%%%%%%%%%%%%%%%%%
%

\section{Discussion}\label{Discussion}
\subsection{Planet Formation Models}\label{PFM}
Numerous studies \citep{Kokubo06, 2006ApJ...644.1223R} show that the growth of embryos to isolation mass in a $\Sigma \propto a^{-3/2}$ disk will occur inside-out, that is, the timescale for growth to isolation in such a disk will be $\tau_{iso} \propto a^{2}$ and hence isolation will occur quickest in the inner disk. Therefore a migrating embryo which initially formed at 3 AU will be migrating through a region of space which is \emph{not} occupied by a smooth solid disk, but rather by a series of massive embryos which have already grown to isolation. The core growth of the embryo must then take place via giant impacts, which will clearly have a significant impact on both the planetary semi-major axes and eccentricities which would be predicted at the end of the growth and migration regime.

In addition, we have seen in our work above that significant ``shepherding'' of material can occur in front of migrating protoplanets. This phenomena combined with the stirring up of the planetesimal disk by the passage of the first embryo means that the core growth rate for the first embryo is (at least) an order of magnitude greater than for the subsequent cores, a result that is significantly at odds with the assumptions of semi-analytic models.

We also find that embryos which have grown more massive \emph{prior} to passage through MMR are prefered in the model, as this allows an easier, more natural explanation of the observed planetary eccentricities, although the subsequent gas damping of these eccentricities still presents problems.

These results serve to emphasise that a consistent model for planet formation and evolution is not possible using the current generation of semi-analytic models. In order to make progress, the semi-analytic growth models need to be combined with n-body models to ensure that both the mutual planet-planet interactions as well as the planet-planetesimal interactions are self-consistently modelled and able to be compared with observations.

\subsection{Eccentric Planetesimal Disk}\label{EPD}
We find in \S \ref{PlanetesimalExcite} that the end product of our n-body simulations is the prediction that the excitation and retention of an eccentric planetesimal disk exterior to the three planets in HD 69830 is almost certain to take place (within the model parameters investigated). Significantly, we found that the mass of planetesimals remaining in this disk was large: after the planets had migrated and the gas disk dissipated, there would be $\sim 25 - 45 \Mearth$ of material present in the scattered disk (using the initial disk profile of \citep{Alibert_et_al_06}). 

The presence of this amount of material in the young system (few Myr) does not necessarily guarantee that significant amounts of mass will remain in the system at late times (few Gyr), as the steady-state collisional processing of the disk will remove a large amount of material. The results of \citet{Wyatt_et_al_07a} indicate that low eccentricity planetesimal belts evolve to a mass that is \emph{independent} of initial mass, and in the case of HD69830, if this material were placed in a ring at 1 AU, then after $2Gyr$ the collisional processing would result in just $10^{-5} \Mearth$ of planetesimals remaining, which would be $1,000$ times less than required to explain the observed emission. This conclusion may be mitigated to some extent by the results of \citet{Lohne08} which show a small dependence of dust mass at late times on initial planetesimal mass, but this would probably not be sufficient as there results imply that the initial planetesimal mass would have to be unrealistically large to explain the observed emission in HD 69830.

To work out how our planetesimals evolve we need to consider the collisional evolution of a population with a range of eccentricities and semimajor axes: This has yet to be achieved, however, work in preparation (Wyatt et al) already shows that increasing the eccentricity of a planetesimal belt from nearly circular to 0.9 while keeping the pericentre constant increases the amount of material remaining at late times by 1 to 2 orders of magnitude. Although certainly giving no guarantee, this increases confidence that the distributions we derived might be able to explain the observed emission.

However, our simulations also show that the size distribution of the planetesimals is extremely important, since it is only $> 100km$ planetesimals that remain on highly eccentric orbits following scattering by planet migration and subsequent gas damping. Thus, it is important to consider the fraction of the remaining $25-45\Mearth$ of planetesimals that have radii $> 100km$. The planetesimals would be expected to have a bimodal population of the type seen in the simulations of \citet{Morishima08}, who found that at $10^6$yrs, $\sim 70\%$ of the solid mass would be in the form of a few large embryos, with the rest residing in a population of planetesimals with a number density distribution given by $n(m)dm = m^{-2}dm$. Such a distribution has mass evenly spread at all sizes (on a log-scale), and thus we would anticipate that approximately $10\%$ ($\sim 2.5-4.5\Mearth$) of the surviving planetesmal material would reside in bodies of size 100 - 1000 km, with a further $70\%$ ($\sim 18-32\Mearth$) locked up in the very large embryos. In the \citet{Alibert_et_al_06} model, from the initial $0.5 M_J = 160\Mearth$ of solid material in the disk, only $\sim 27\Mearth$ ($15\%$) is accreted onto the migrating cores, so the contribution of the large embryos to the scattered distribution could be significant. Thus there is likely to be a significant mass fraction in large scattered bodies, which, along with the size sorting of the planetesimals via gas drag, works to ensure that a large proportion of the remnant planetesimal mass would be locked up in large bodies which would be on the most eccentric orbits.

\section{Summary}\label{Summary}
We used the model of \citet{Alibert_et_al_06} as the basis for undertaking a suite of n-body simulations which allow us to consider planetesimal disk formation in HD 69830 in the aftermath of planet formation.

\begin{itemize}

\item We find that the model of \citet{Alibert_et_al_06} in its basic form cannot coherently explain the observed eccentricities in the HD69830 system. Additional mass, either in the form of additional planets or increased masses of the observed planets, would need to be present in the system in order to excite greater eccentricities in the observed planets.

\item Some way of mitigating the effect of gas damping on the planets or stimulating additional eccentricity excitation after the end of the gas damping phase would also seem to be required.

\item The eccentricities of the planets are relatively \emph{un}important in deciding the overall availability of an excited planetesimal population with pericentre(s) outside 1 AU. 

\item We find that the eccentricity damping of planetesimals does \emph{not} act as an insurmountable obstacle to the existence of an excited eccentric disk: We consistently find in \emph{all} simulations that $\sim15\%$ (equivalent to $\sim 25 \Mearth$) or more of the total solid disk material will remain in excited orbits having pericentres, $q>1$ AU. It therefore seems probable that after the planet formation process had concluded, HD 69830 would have been left with a significant swarm of eccentric planetesimals.

\item We cannot yet definitively rule-out any of the IR emission models discussed in \S \ref{IRobs}, but the consistent production and survival of eccentric disks during and after the planet formation process in our n-body models does suggest the eccentric swarm idea is worthy of further investigation.

\item Gas damping of planetesimals does not significantly alter the \emph{number} of available planetesimals, but it does serve to significantly decrease the proportion of these planetesimals which have high eccentricities ($e > 0.5$).

\item Gas damping of planetesimals works to ``size-sort'' the planetesimals, preferentially leaving the larger planetesimals (which contain a large proportion of the mass) occupying the higher eccentricity orbits whilst the low mass objects are efficiently circularised.

\item Further work is needed (and is in progress) to model the lifetime and emission characteristics of extended non-aligned eccentric swarms of planetesimals to understand in more detail whether the reservoirs of planetesimals predicted in this paper could survive long enough to explain the observed IR emission. 

\end{itemize}

\section{Acknowledgments}
The authors acknowledge the High-Performance Computing Centers of the University of Florida and the University of Cambridge for providing computational resources and support. In addition we wish to thank Yann Alibert for supplying suitably formatted data, Philippe Thebault for comments on planetesimal size distributions, Graeme Lufkin, Derek Richardson and Richard Edgar for their time in discussing their previous simulation work and our anonymous referee for suggesting some useful clarifications. MJP acknowledges the UK PPARC/STFC for a research studentship and thanks EF and the University of Florida for their hospitality during his visit.

%\bibliography{references}

\begin{thebibliography}{}
\small
\itemindent -0.48cm

\bibitem[\protect\citeauthoryear{{Adams}, {Laughlin} \& {Bloch}}{{Adams}  et~al.}{2008}]{Adams08}{Adams} F.~C.,  {Laughlin} G.,    {Bloch} A.~M.,  2008, \apj, 683, 1117

\bibitem[\protect\citeauthoryear{{Alibert}, {Baraffe}, {Benz}, {Chabrier}, {Mordasini}, {Lovis}, {Mayor}, {Pepe}, {Bouchy}, {Queloz} \& {Udry}}{Alibert et~al.}{2006}]{Alibert_et_al_06}{Alibert} Y.~{et al.},  2006, \aap, 455, L25

\bibitem[\protect\citeauthoryear{{Alibert}, {Mordasini}, {Benz} \& {Winisdoerffer}}{Alibert et~al.}{2005}]{Alibert_et_al_05}{Alibert} Y.,  {Mordasini} C.,  {Benz} W.,    {Winisdoerffer} C.,  2005, \aap,  434, 343

\bibitem[\protect\citeauthoryear{{Alibert}, {Mousis}, {Mordasini} \& {Benz}}{Alibert et~al.}{2005}]{Alibert_et_al_05b}{Alibert} Y.,  {Mousis} O.,  {Mordasini} C.,    {Benz} W.,  2005, \apjl, 626,  L57

\bibitem[\protect\citeauthoryear{{Beichman}, {Bryden}, {Gautier}, {Stapelfeldt}, {Werner}, {Misselt}, {Rieke}, {Stansberry} \& {Trilling}}{Beichman et~al.}{2005}]{Beichman_et_al_05}{Beichman} C.~A.~{et al.},  2005, \apj, 626, 1061

\bibitem[\protect\citeauthoryear{{Benz}, {Mordasini}, {Alibert} \& {Naef}}{Benz et~al.}{2008}]{Benz08}{Benz} W.,  {Mordasini} C.,  {Alibert} Y.,    {Naef} D.,  2008, Physica Scripta  Volume T, 130, 014022

\bibitem[\protect\citeauthoryear{{Chambers}}{{Chambers}}{1999}]{Chambers99}{Chambers} J.~E.,  1999, \mnras, 304, 793

\bibitem[\protect\citeauthoryear{{Chatterjee}, {Ford} \& {Rasio}}{{Chatterjee}  et~al.}{2007}]{Chatterjee07}{Chatterjee} S.,  {Ford} E.~B.,    {Rasio} F.~A.,  2007, astro-ph/0703166

\bibitem[\protect\citeauthoryear{{Daisaka}, {Tanaka} \& {Ida}}{{Daisaka}  et~al.}{2006}]{Daisaka06}{Daisaka} J.~K.,  {Tanaka} H.,    {Ida} S.,  2006, Icarus, 185, 492

\bibitem[\protect\citeauthoryear{{Edgar} \& {Artymowicz}}{{Edgar} \&  {Artymowicz}}{2004}]{Edgar04}{Edgar} R.,  {Artymowicz} P.,  2004, \mnras, 354, 769

\bibitem[\protect\citeauthoryear{{Fogg} \& {Nelson}}{{Fogg} \&  {Nelson}}{2005}]{2005A&A...441..791F}{Fogg} M.~J.,  {Nelson} R.~P.,  2005, \aap, 441, 791

\bibitem[\protect\citeauthoryear{{Fogg} \& {Nelson}}{{Fogg} \&  {Nelson}}{2007a}]{2007arXiv0710.3730F}{Fogg} M.~J.,  {Nelson} R.~P.,  2007a, ArXiv e-prints, 710

\bibitem[\protect\citeauthoryear{{Fogg} \& {Nelson}}{{Fogg} \&  {Nelson}}{2007b}]{2007A&A...461.1195F}{Fogg} M.~J.,  {Nelson} R.~P.,  2007b, \aap, 461, 1195

\bibitem[\protect\citeauthoryear{{Fogg} \& {Nelson}}{{Fogg} \&  {Nelson}}{2007c}]{2007A&A...472.1003F}{Fogg} M.~J.,  {Nelson} R.~P.,  2007c, \aap, 472, 1003

\bibitem[\protect\citeauthoryear{{Ford}}{{Ford}}{2006}]{Ford06}{Ford} E.~B.,  2006, \apj, 642, 505

\bibitem[\protect\citeauthoryear{{Ford}, {Lystad} \& {Rasio}}{{Ford}  et~al.}{2005}]{Ford05}{Ford} E.~B.,  {Lystad} V.,    {Rasio} F.~A.,  2005, \nat, 434, 873

\bibitem[\protect\citeauthoryear{{Ford} \& {Rasio}}{{Ford} \&  {Rasio}}{2007}]{2007astro.ph..3163F}{Ford} E.~B.,  {Rasio} F.~A.,  2007, astro-ph/0703163

\bibitem[\protect\citeauthoryear{{Fortier}, {Benvenuto} \& {Brunini}}{{Fortier}  et~al.}{2007}]{Fortier07}{Fortier} A.,  {Benvenuto} O.~G.,    {Brunini} A.,  2007, \aap, 473, 311

\bibitem[\protect\citeauthoryear{{Gomes}, {Levison}, {Tsiganis} \& {Morbidelli}}{Gomes et~al.}{2005}]{2005Natur.435..466G}{Gomes} R.,  {Levison} H.~F.,  {Tsiganis} K.,    {Morbidelli} A.,  2005, \nat,  435, 466

\bibitem[\protect\citeauthoryear{{Ida} \& {Lin}}{{Ida} \& {Lin}}{2008}]{IL5}{Ida} S.,  {Lin} D.~N.~C.,  2008, \apj, 673, 487

\bibitem[\protect\citeauthoryear{{Ji}, {Kinoshita}, {Liu} \& {Li}}{{Ji}  et~al.}{2007}]{Ji07}{Ji} J.,  {Kinoshita} H.,  {Liu} L.,    {Li} G.,  2007, \apj, 657, 1092

\bibitem[\protect\citeauthoryear{{Kokubo}, {Kominami} \& {Ida}}{{Kokubo}  et~al.}{2006}]{Kokubo06}{Kokubo} E.,  {Kominami} J.,    {Ida} S.,  2006, \apj, 642, 1131

\bibitem[\protect\citeauthoryear{{Lee} \& {Peale}}{{Lee} \&  {Peale}}{2002}]{Lee02}{Lee} M.~H.,  {Peale} S.~J.,  2002, \apj, 567, 596

\bibitem[\protect\citeauthoryear{{Lisse}, {Beichman}, {Bryden} \& {Wyatt}}{Lisse et~al.}{2007}]{Lisse_et_al_07}{Lisse} C.~M.,  {Beichman} C.~A.,  {Bryden} G.,    {Wyatt} M.~C.,  2007, \apj,  658, 584

\bibitem[\protect\citeauthoryear{{L{\"o}hne}, {Krivov} \& {Rodmann}}{L{\"o}hne et~al.}{2008}]{Lohne08}{L{\"o}hne} T.,  {Krivov} A.~V.,    {Rodmann} J.,  2008, \apj, 673, 1123

\bibitem[\protect\citeauthoryear{{Lovis}, {Mayor}, {Pepe}, {Alibert}, {Benz}, {Bouchy}, {Correia}, {Laskar}, {Mordasini}, {Queloz}, {Santos}, {Udry}, {Bertaux} \& {Sivan}}{Lovis et~al.}{2006}]{Lovis_et_al_06}{Lovis} C.~{et al.},  2006, \nat, 441, 305

\bibitem[\protect\citeauthoryear{{Lufkin}, {Richardson} \& {Mundy}}{{Lufkin}  et~al.}{2006}]{Lufkin06}{Lufkin} G.,  {Richardson} D.~C.,    {Mundy} L.~G.,  2006, \apj, 653, 1464

\bibitem[\protect\citeauthoryear{{Mandell}, {Raymond} \& {Sigurdsson}}{Mandell et~al.}{2007}]{Mandell07}{Mandell} A.~M.,  {Raymond} S.~N.,    {Sigurdsson} S.,  2007, \apj, 660, 823

\bibitem[\protect\citeauthoryear{{Mayor}, {Udry}, {Lovis}, {Pepe}, {Queloz}, {Benz}, {Bertaux}, {Bouchy}, {Mordasini} \& {Segransan}}{Mayor et~al.}{2008}]{Mayor08}{Mayor} M.~{et al.},  2008,  ArXiv e-prints, 806

\bibitem[\protect\citeauthoryear{{Meyer}, {Hillenbrand}, {Backman}, {Beckwith}, {Bouwman}, {Brooke}, {Carpenter}, {Cohen}, {Cortes}, {Crockett}, {Gorti}, {Henning}, {Hines}, {Hollenbach}, {Kim}, {Lunine}, {Malhotra}, {Mamajek}, {Metchev}, {Moro-Martin}, {Morris}, {Najita}, {Padgett}, {Pascucci}, {Rodmann}, {Schlingman}, {Silverstone}, {Soderblom}, {Stauffer}, {Stobie}, {Strom}, {Watson}, {Weidenschilling}, {Wolf} \& {Young}}{Meyer et~al.}{2006}]{Meyer06}{Meyer} M.~R.~{et al.},  2006, \pasp, 118, 1690

\bibitem[\protect\citeauthoryear{{Morishima}, {Schmidt}, {Stadel} \& {Moore}}{Morishima et~al.}{2008}]{Morishima08}{Morishima} R.,  {Schmidt} M.~W.,  {Stadel} J.,    {Moore} B.,  2008, \apj,  685, 1247

\bibitem[\protect\citeauthoryear{{Moro-Mart{\'{\i}}n}, {Carpenter}, {Meyer}, {Hillenbrand}, {Malhotra}, {Hollenbach}, {Najita}, {Henning}, {Kim}, {Bouwman}, {Silverstone}, {Hines}, {Wolf}, {Pascucci}, {Mamajek} \& {Lunine}}{Moro-Mart{\'{\i}}n et~al.}{2007}]{Moro-Martin07}{Moro-Mart{\'{\i}}n} A.~{et al.},  2007, \apj, 658, 1312

\bibitem[\protect\citeauthoryear{{Murray-Clay} \& {Chiang}}{{Murray-Clay} \&  {Chiang}}{2006}]{MurrayClay06}{Murray-Clay} R.~A.,  {Chiang} E.~I.,  2006, \apj, 651, 1194

\bibitem[\protect\citeauthoryear{{Nelson} \& {Papaloizou}}{{Nelson} \&  {Papaloizou}}{2004}]{Nelson04}{Nelson} R.~P.,  {Papaloizou} J.~C.~B.,  2004, \mnras, 350, 849

\bibitem[\protect\citeauthoryear{{Papaloizou}, {Nelson} \& {Masset}}{Papaloizou et~al.}{2001}]{Papaloizou01}{Papaloizou} J.~C.~B.,  {Nelson} R.~P.,    {Masset} F.,  2001, \aap, 366, 263

\bibitem[\protect\citeauthoryear{{Perryman}, {Lindegren}, {Kovalevsky}, {Hoeg}, {Bastian}, {Bernacca}, {Cr{\'e}z{\'e}}, {Donati}, {Grenon}, {van  Leeuwen}, {van der Marel}, {Mignard}, {Murray}, {Le Poole}, {Schrijver}, {Turon}, {Arenou}, {Froeschl{\'e}} \& {Petersen}}{Perryman et~al.}{1997}]{Perryman97}{Perryman} M.~A.~C.~{et al.},  1997, \aap, 323, L49

\bibitem[\protect\citeauthoryear{{Rafikov}}{{Rafikov}}{2004}]{Rafikov04}{Rafikov} R.~R.,  2004, \aj, 128, 1348

\bibitem[\protect\citeauthoryear{{Raymond}, {Barnes} \& {Kaib}}{{Raymond}  et~al.}{2006}]{2006ApJ...644.1223R}{Raymond} S.~N.,  {Barnes} R.,    {Kaib} N.~A.,  2006, \apj, 644, 1223

\bibitem[\protect\citeauthoryear{{Rein} \& {Papaloizou}}{{Rein} \&  {Papaloizou}}{2008}]{Rein08}{Rein} H.,  {Papaloizou} J.~C.~B.,  2008, ArXiv e-prints

\bibitem[\protect\citeauthoryear{{Tanaka}, {Takeuchi} \& {Ward}}{{Tanaka}  et~al.}{2002}]{Tanaka02}{Tanaka} H.,  {Takeuchi} T.,    {Ward} W.~R.,  2002, \apj, 565, 1257

\bibitem[\protect\citeauthoryear{{Wyatt}}{{Wyatt}}{2003}]{Wyatt03}{Wyatt} M.~C.,  2003, \apj, 598, 1321

\bibitem[\protect\citeauthoryear{{Wyatt}, {Smith}, {Greaves}, {Beichman}, {Bryden} \& {Lisse}}{Wyatt et~al.}{2007}]{Wyatt_et_al_07a}{Wyatt} M.~C.,  {Smith} R.,  {Greaves} J.~S.,  {Beichman} C.~A.,  {Bryden} G.,    {Lisse} C.~M.,  2007, \apj, 658, 569

\bibitem[\protect\citeauthoryear{{Zhou}, {Lin} \& {Sun}}{{Zhou}  et~al.}{2007}]{Zhou07}{Zhou} J.-L.,  {Lin} D.~N.~C.,    {Sun} Y.-S.,  2007, \apj, 666, 423

\end{thebibliography}
%\bibliographystyle{mn2e}

\label{lastpage}
\end{document}